\documentclass[nofootinbib,twocolumn,showpacs,preprintnumbers,amsmath,amssymb,floatfix]{revtex4}
\usepackage{graphicx,dcolumn,bm}
\usepackage{multirow}
\usepackage{wasysym}
\usepackage{color}

\RequirePackage{xspace}

\usepackage{relsize}
\def\babar{\mbox{\slshape B\kern-0.1em{\smaller A}\kern-0.1em
    B\kern-0.1em{\smaller A\kern-0.2em R}}}

\def\en         {\ensuremath{e^-}\xspace}   

\def\epem       {\ensuremath{e^+e^-}\xspace}

\def\gaga  {\ensuremath{\gamma\gamma}\xspace}  

\def\q     {\ensuremath{q}\xspace}

\def\qqbar {\ensuremath{q\overline q}\xspace}

\def\piz   {\ensuremath{\pi^0}\xspace}

\def\pip   {\ensuremath{\pi^+}\xspace}
\def\pim   {\ensuremath{\pi^-}\xspace}

\def\pipm  {\ensuremath{\pi^\pm}\xspace}
\def\pimp  {\ensuremath{\pi^\mp}\xspace}

\def\Kbar  {\kern 0.2em\overline{\kern -0.2em K}{}\xspace}

\def\Kz    {\ensuremath{K^0}\xspace}
\def\Kzb   {\ensuremath{\Kbar^0}\xspace}
\def\KzKzb {\ensuremath{\Kz \kern -0.16em \Kzb}\xspace}
\def\Kp    {\ensuremath{K^+}\xspace}
\def\Km    {\ensuremath{K^-}\xspace}
\def\Kpm   {\ensuremath{K^\pm}\xspace}
\def\Kmp   {\ensuremath{K^\mp}\xspace}
\def\KpKm  {\ensuremath{\Kp \kern -0.16em \Km}\xspace}
\def\KS    {\ensuremath{K^0_{\scriptscriptstyle S}}\xspace}

\def\Dbar    {\kern 0.2em\overline{\kern -0.2em D}{}\xspace}

\def\Dz      {\ensuremath{D^0}\xspace}
\def\Dzb     {\ensuremath{\Dbar^0}\xspace}
\def\DzDzb   {\ensuremath{\Dz {\kern -0.16em \Dzb}}\xspace}
\def\Dp      {\ensuremath{D^+}\xspace}
\def\Dm      {\ensuremath{D^-}\xspace}

\def\DpDm    {\ensuremath{\Dp {\kern -0.16em \Dm}}\xspace}

\def\B       {\ensuremath{B}\xspace}
\def\Bbar    {\kern 0.18em\overline{\kern -0.18em B}{}\xspace}

\def\Bz      {\ensuremath{B^0}\xspace}
\def\Bzb     {\ensuremath{\Bbar^0}\xspace}
\def\BzBzb   {\ensuremath{\Bz {\kern -0.16em \Bzb}}\xspace}
\def\Bu      {\ensuremath{B^+}\xspace}
\def\Bub     {\ensuremath{B^-}\xspace}

\def\BpBm    {\ensuremath{\Bu {\kern -0.16em \Bub}}\xspace}

\def\BorBbar    {\kern 0.18em\optbar{\kern -0.18em B}{}\xspace}
\def\DorDbar    {\kern 0.18em\optbar{\kern -0.18em D}{}\xspace}
\def\KorKbar    {\kern 0.18em\optbar{\kern -0.18em K}{}\xspace}

\def\jpsi     {\ensuremath{{J\mskip -3mu/\mskip -2mu\psi\mskip 2mu}}\xspace}

\mathchardef\Upsilon="7107
\def\Y#1S{\ensuremath{\Upsilon{(#1S)}}\xspace}

\mathchardef\Deltares="7101
\mathchardef\Xi="7104
\mathchardef\Lambda="7103
\mathchardef\Sigma="7106
\mathchardef\Omega="710A

\def\Deltabar{\kern 0.25em\overline{\kern -0.25em \Deltares}{}\xspace}
\def\Lbar{\kern 0.2em\overline{\kern -0.2em\Lambda\kern 0.05em}\kern-0.05em{}\xspace}
\def\Sigbar{\kern 0.2em\overline{\kern -0.2em \Sigma}{}\xspace}
\def\Xibar{\kern 0.2em\overline{\kern -0.2em \Xi}{}\xspace}
\def\Obar{\kern 0.2em\overline{\kern -0.2em \Omega}{}\xspace}
\def\Nbar{\kern 0.2em\overline{\kern -0.2em N}{}\xspace}
\def\Xb{\kern 0.2em\overline{\kern -0.2em X}{}\xspace}

\newcommand{\tev}{\ensuremath{\mathrm{\,Te\kern -0.1em V}}\xspace}
\newcommand{\gev}{\ensuremath{\mathrm{\,Ge\kern -0.1em V}}\xspace}
\newcommand{\mev}{\ensuremath{\mathrm{\,Me\kern -0.1em V}}\xspace}
\newcommand{\kev}{\ensuremath{\mathrm{\,ke\kern -0.1em V}}\xspace}
\newcommand{\ev}{\ensuremath{\mathrm{\,e\kern -0.1em V}}\xspace}
\newcommand{\gevc}{\ensuremath{{\mathrm{\,Ge\kern -0.1em V\!/}c}}\xspace}
\newcommand{\mevc}{\ensuremath{{\mathrm{\,Me\kern -0.1em V\!/}c}}\xspace}
\newcommand{\gevcc}{\ensuremath{{\mathrm{\,Ge\kern -0.1em V\!/}c^2}}\xspace}
\newcommand{\mevcc}{\ensuremath{{\mathrm{\,Me\kern -0.1em V\!/}c^2}}\xspace}

\def\mm   {\ensuremath{{\rm \,mm}}\xspace}

\def\nb         {\ensuremath{{\rm \,nb}}\xspace}
\def\invnb {\ensuremath{\mbox{\,nb}^{-1}}\xspace}

\def\mus  {\ensuremath{\rm \,\mus}\xspace}

\def\mus        {\ensuremath{\,\mu{\rm s}}\xspace}    

\def\to                 {\ensuremath{\rightarrow}\xspace}

\def\pep2{PEP-II}

\newcommand{\chisq}{\ensuremath{\chi^2}\xspace}

\def\gsim{{~\raise.15em\hbox{$>$}\kern-.85em
          \lower.35em\hbox{$\sim$}~}\xspace}
\def\lsim{{~\raise.15em\hbox{$<$}\kern-.85em
          \lower.35em\hbox{$\sim$}~}\xspace}

\def\CP                {\ensuremath{C\!P}\xspace}

\xspace

\newcommand{\jprlBase}       {Phys.\ Rev.\ Lett.\xspace}
\newcommand{\jprBase}        {Phys.\ Rev.\xspace}
\newcommand{\jplBase}        {Phys.\ Lett.\xspace}
\newcommand{\nimBaseA}       {Nucl.\ Instrum.\ Methods Phys.\ Res., Sect.\ A\xspace}

\newcommand{\npBase}         {Nucl.\ Phys.\xspace}

\newcommand{\nima}      [1]  {\nimBaseA~{\bf #1}}

\newcommand{\npb}       [1]  {\npBase\ B~{\bf #1}}

\newcommand{\plb}       [1]  {\jplBase\ B~{\bf #1}}

\newcommand{\jprl}      [1]  {\jprlBase\ {\bf #1}}
\newcommand{\jprd}      [1]  {\jprBase\ D~{\bf #1}}

\def\jetset74   {\mbox{\tt Jetset \hspace{-0.5em}7.\hspace{-0.2em}4}\xspace}

\def\k         {\ensuremath{K}}
\def\K         {\ensuremath{K}}
\def\kk        {\ensuremath{\Kp\Km}}
\def\Kst       {\ensuremath{\K^*(892)}\xspace}
\def\kst       {\ensuremath{\K^*(892)}\xspace}
\def\kstz      {\ensuremath{\K^{*0}(892)}\xspace}
\def\kstc      {\ensuremath{\K^{*\pm}(892)}\xspace}

\def\kskpi     {\ensuremath{\KS\Kpm\pimp}\xspace}
\def\kkpiz     {\ensuremath{\Kp\Km\piz}\xspace}
\def\phipiz    {\ensuremath{\phi\piz}\xspace}
\def\kketa     {\ensuremath{\Kp\Km\eta}\xspace}
\def\phieta    {\ensuremath{\phi\eta}\xspace}

\def\Ecm       {\ensuremath{E_{\rm c.m.}}\xspace}
\def\phip       {\ensuremath{\phi(1680)}\xspace}
\def\DP        {Dalitz plot}

\newcommand{\bc}{\begin{center}}
\newcommand{\ec}{\end{center}}

\newcommand{\bi}{\begin{itemize}}
\newcommand{\ei}{\end{itemize}}
\newcommand{\be}{\begin{eqnarray}}
\renewcommand{\en}{\end{eqnarray}}
\newcommand{\ba}{\begin{array}}
\newcommand{\ea}{\end{array}}
\renewcommand{\bc}{\begin{center}}
\renewcommand{\ec}{\end{center}}
\newcommand{\bfi}{\begin{figure}}
\newcommand{\efi}{\end{figure}}

\newcommand{\ov}[1]{\overline{#1}}
\newcommand{\no}{\nonumber}

\def\ifb{\ensuremath{\,\mathrm{fb}^{-1}}}

\def\makeatletter{\catcode `\@=11\relax}
\def\makeatother{ \catcode `\@=12\relax}
\makeatletter
\let\@@comma=\,       \def\,{\ensuremath{\@@comma}}
\let\@@colon=\;       \def\;{\ensuremath{\@@colon}}
\makeatother

\long\def\inst#1{\par\nobreak\kern 4pt\nobreak
    {\it #1}\par\vskip 10pt plus 3pt minus 3pt}

\begin{document}




\title{\boldmath Measurements of $\epem\to \kketa$,
 \kkpiz and \kskpi Cross Sections Using Initial State Radiation Events}

%
\author{B.~Aubert}
\author{M.~Bona}
\author{D.~Boutigny}
\author{Y.~Karyotakis}
\author{J.~P.~Lees}
\author{V.~Poireau}
\author{X.~Prudent}
\author{V.~Tisserand}
\author{A.~Zghiche}
\affiliation{Laboratoire de Physique des Particules, IN2P3/CNRS et Universit\'e de Savoie, F-74941 Annecy-Le-Vieux, France }
\author{J.~Garra~Tico}
\author{E.~Grauges}
\affiliation{Universitat de Barcelona, Facultat de Fisica, Departament ECM, E-08028 Barcelona, Spain }
% ...
\collaboration{The \babar\ Collaboration}
\noaffiliation


\begin{abstract}
This paper reports measurements of processes:
$\epem\to\gamma\kskpi$, $\epem\to\gamma\kkpiz$, 
$\epem\to\gamma\phieta$ and $\epem\to\gamma\phipiz$.
The initial state radiated photon allows to cover the 
hadronic final state in the energy range from thresholds up to $\approx$ 4.6 GeV.
The overall size of the data sample analyzed is  232 \ifb, 
collected by the \babar\ detector running at the PEP-II \epem storage ring.
From the Dalitz plot analysis  of the \kskpi final state, 
moduli and relative phase of the isoscalar and the isovector components 
of the $\epem\to\K\kst$ cross section are determined.
Parameters of $\phi$ and $\rho$ recurrences are also measured, 
using a global fitting procedure which exploits the interconnection 
among amplitudes, moduli and phases 
of the $\epem\to\kskpi,\kkpiz,\phieta$ final states.
The cross section for the OZI-forbidden process $\epem\to\phipiz$, 
and the \jpsi branching fractions to $\K\kst$ and $\kketa$  
are also measured.
\end{abstract}

\pacs{13.66.Bc, 14.40.Cs, 13.25.Gv, 13.25.Jx, 25.70.Ef}
\vfill
\maketitle
%
%
%
\section{INTRODUCTION}
\label{sec:intro}

The \babar\ detector~\cite{babar},  designed  
to study time-dependent \CP-violation in \B decays,
is very well suited  to study hadronic final state
production. At an $\epem$ collider, lower center-of-mass 
(c.m.) energies can be studied using initial state
radiation (ISR).
The possibility of exploiting such processes
 was first outlined in 
Ref.~\cite{baier} and discussed in
Refs.~\cite{arbus,kuehn,ivanch}.
                                
The cross section $\sigma_{\gamma f}$ 
to radiate a photon of energy $E_{\gamma}$
and subsequently produce  
a definite hadronic final state $f$ is related to the 
corresponding \epem production cross section $\sigma_f(s)$ by:
\begin{equation}
\frac{d\sigma_{\gamma f}(s,x)}{dx} = W(s,x)\cdot \sigma_f[s(1-x)]\ ,
\label{eq1}
\end{equation}
where $x=2E_{\gamma}/\sqrt{s}$, $\sqrt{s}$ is the nominal \epem c.m. energy  
and $\Ecm\equiv\sqrt{s(1-x)}$ is the effective c.m. energy at which the 
final state $f$ is produced. This quantity can be determined, 
for instance, by measuring the invariant mass of the 
hadronic final state. The radiator function $W(s,x)$, which describes the 
probability density for photon emission,
can be evaluated  with 
better than 1\% accuracy (see for example Ref.~\cite{phokara}). 
The radiator function falls rapidly with increasing $E_{\gamma}$, 
but has a long tail, 
which produces sizeable event rates at very low \Ecm.   
In the present study we require the ISR photon 
to be detected in the \babar\ electromagnetic calorimeter.  
The angular distribution of the ISR photon peaks along the beam direction, 
but $10\--15\%$~\cite{ivanch} of the photons are emitted within the detector
acceptance. 

\section{THE \babar\ DETECTOR AND DATASET}
\label{sec:detector}
The data used in this analysis, collected with the \babar\ detector
at the \pep2 storage ring, correspond to an 
integrated luminosity of $232\;\ifb$.
Charged-particle momenta are measured in a tracking
system consisting of a five-layer double-sided silicon
vertex tracker (SVT) and a 40-layer central drift chamber
(DCH), immersed in a 1.5  T axial magnetic field.
An internally reflecting ring-imaging Cherenkov detector
(DIRC) with quartz bar radiators provides charged-particle
identification.
A CsI electromagnetic calorimeter (EMC) is used to detect
and identify photons and electrons.
Muons are identified in the instrumented magnetic
flux return system (IFR).

Electron candidates are selected using the ratio of
the shower energy deposited in the EMC to the measured 
momentum, the shower shape, the specific ionization
in the DCH, and the Cherenkov angle measured by the
DIRC.
Muons are identified by the depth of penetration
into the IFR, the IFR cluster geometry, and the
energy deposited in the EMC.
Charged pion and kaon  candidates are selected using 
a likelihood function based on the specific ionization 
in the DCH and SVT, and the Cherenkov angle measured 
in the DIRC.
Photon candidates are defined as  clusters
in the EMC that have a shape consistent with an
electromagnetic shower, without an associated
charged track.

In order to study the detector acceptance and efficiency, a
special package of simulation programs for radiative processes
has been developed. Algorithms for generating hadronic final 
states via ISR are derived from Ref.~\cite{czyz}.
Multiple soft-photon emission from initial-state
charged particles is implemented by means of the
structure-function technique~\cite{kuraev,strfun}, while
extra photon radiation from 
final-state particles is simulated via the PHOTOS package~\cite{PHOTOS}.

The $\epem\to\kskpi\gamma$, $\kkpiz\gamma$, $\phieta\gamma$ and $\phipiz\gamma$
reactions are simulated with the hadronic final states 
generated according to their phase space distributions. Both processes 
$e^+e^-\to P_1P_2P_3$ and $e^+e^-\to VP$~\cite{czyz}, where $P_{(i)}$ 
 and $V$ are pseudoscalar and vector mesons, respectively, are 
used in the simulation.  
To evaluate the backgrounds, a number of different  ISR-produced 
final states, such as $4\pi$, $\K\K\pi\pi$, and $\phi\piz\piz$, are simulated as well.

The quark-antiquark continuum production 
and hadronization (\qqbar background in the following, where $\q=u,d,s,c$) and $\tau\tau$ events are 
simulated using JETSET~\cite{jetset} and \hbox{KORALB}~\cite{koralb}, respectively.

The generated events undergo a
detector simulation based on GEANT4~\cite{GEANT4} and
are analyzed as experimental data.


\section{EVENT SELECTION AND KINEMATICS}
\label{sec:selection}

A  preliminary set of selection criteria are applied to 
the data regardless of the final state under study.
We consider only  
charged tracks emitted at a polar angle 
between 0.45 and 2.40 radians 
and that extrapolate within 0.25 cm of the beam axis 
and within 3 cm of the nominal collision point along the axis. 
We retain photon candidates with energy above 30 \mev 
and emitted at a polar angle between 0.35 and 2.40 radians.
The highest-energy photon, assumed to be the ISR photon, 
is required to have an energy greater than 3~\gev.  

We then apply channel specific selection criteria and 
subject each candidate event to a one-constraint (1C) kinematic fit, 
both to select the signal and to estimate the background contamination.
The fit uses as input the measured quantities (momenta and angles)  
of the selected charged tracks and photons, 
together with the associated error matrices.
However, because of the excellent resolution
of the DCH, the three-momentum vector of the ISR photon is better determined
through the momentum conservation than through the measurement in the EMC, 
thus reducing the number of constraints to one. 
For final states  containing a $\piz$ or an $\eta$, 
we perform the kinematic fit
for every possible pair of detected photons, without 
applying a mass constraint, and retain the pair producing the lowest \chisq. 
The fitting procedure predicts the ISR photon direction, 
which is then required to be consistent   
with the observed ISR photon within 15 mrad.
In order to reduce multi-photon events, we also require, 
that no more than 400~\mev
of additional neutral energy, besides the ISR and \piz ($\eta$) 
decay photons, be detected in the EMC. 
This cut has no effect on signal events.
\\ \indent 
We select as $ \pi^0 \, ( \eta ) $ candidates $\gamma \gamma $ pairs
whose masses lie in the range $0.110\-- 0.160$ ($0.520\-- 0.580$)~\gevcc.
In calculating the invariant masses of these pairs, we
use the photon momenta resulting from the 1C fits described above.
This improves the  $\piz$, $\eta$ mass resolution by $ \approx 20\% $.
To get rid of the contamination due to additional soft photons
produced by machine background or by interactions in the detector material,
we require both photons used for \piz or $\eta$ candidates to have  
an energy greater than 60~\mev and at least one cluster containing more
 than  100~\mev.\\ \indent 
As $ K^0_S $ candidates, we select pairs of charged tracks forming
a vertex with a good $\chi^2$ at least 2~\mm away from the
event primary vertex, and whose $ \pi^+\pi^- $
invariant mass lies within 15 MeV/$c^2$ of the nominal
$K^0_S$ mass. 
We also require that $\theta_{\KS}$, the angle between the $ K^0_S $ 
momentum vector and the line connecting the $ K^0_S $ vertex to the
primary vertex position, satisfy the condition cos($\theta_{\KS}$) $>0.99$.
For the purposes of the 1C fit described earlier,
the $ K^0_S $ candidate is treated as a single particle
and its covariance matrix is used as the corresponding
track covariance matrix.\\
\indent Charged kaon and pion candidates are selected using 
likelihood ratios for the hypotheses
 $ e $, $\mu $, $\pi$, $ \K $, and $ p $ based upon specific ionization
measurements in the SVT and DCH, and Cherenkov light observed
in the DIRC.
The algorithm used here for \Kpm identification is over 80\% efficient
in the momentum range of interest, with  $\pi$, $\mu$, 
$p$, $e\to\k$ misidentification rates well below 2\%.


%
\begin{figure}[h!]\vspace{-3mm}
\includegraphics[width=.9\columnwidth]{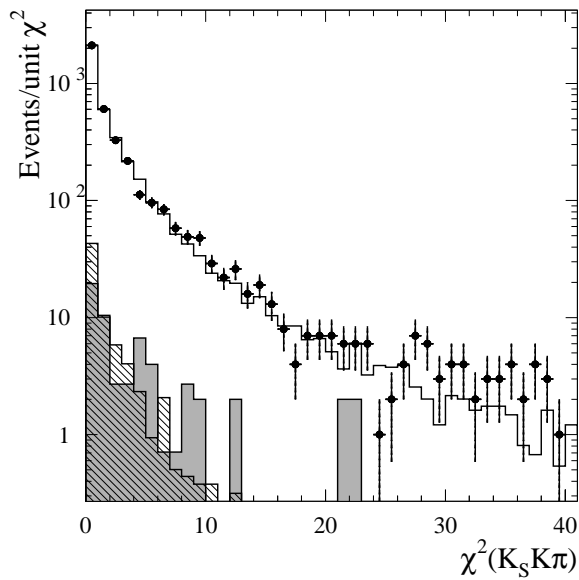}\vspace{-3mm}
\caption{The $\chi^2$ distributions for data (points), signal MC (open histogram) 
normalized to the first bin of the data, background contribution due to 
\qqbar (hatched histogram) and different ISR-produced final states (gray histogram).}
\label{fig:chi2_kskpi}\vspace{-3mm}
\end{figure}
\section{The \kskpi\ FINAL STATE}
\label{sec:kskpi}
\subsection{Event selection}
\label{sec:kskpi-selection}

We require exactly four charged tracks with zero net charge, two of them 
coming from the primary vertex of the event and identified respectively as a 
kaon and a pion, and the other two forming a \KS.\\
\indent
Figure~\ref{fig:chi2_kskpi} shows the \chisq distribution 
of the selected data sample, together with  
three different MC simulations: signal events, background contribution due to
$\epem\to q\bar{q}$, and ISR-produced processes $\Kp\Km\pip\pim+2(\pip\pim)$ 
that might mimic this final state.
A total of 3860 events are selected within the \chisq signal region, 
$i.e.\,\chisq<20$.\\
The background due to mis-reconstructed \KS is small, as can be inferred
from the \KS mass distribution shown in Fig.~\ref{fig:ks_mass}, and
it is almost exclusively given by 
the reactions $\epem\to\gamma\Kp\Km\pi^+\pi^-$ and 
$\epem\to\gamma\pi^+\pi^-\pi^+\pi^-$.
These processes differ from the signal only by one $\K-\pi$ exchange,
and produce a \chisq distribution peaking at low values.
The direct simulation of these two sources of background yields to 
 respectively $44\pm15$ and $10\pm5$ events in the \chisq signal region,
where the errors are taken from the uncertainties of the
corresponding production cross sections~\cite{bib:4pi}.\\

\begin{figure}[htb]
\includegraphics[width=.9\columnwidth,height=.9\columnwidth]{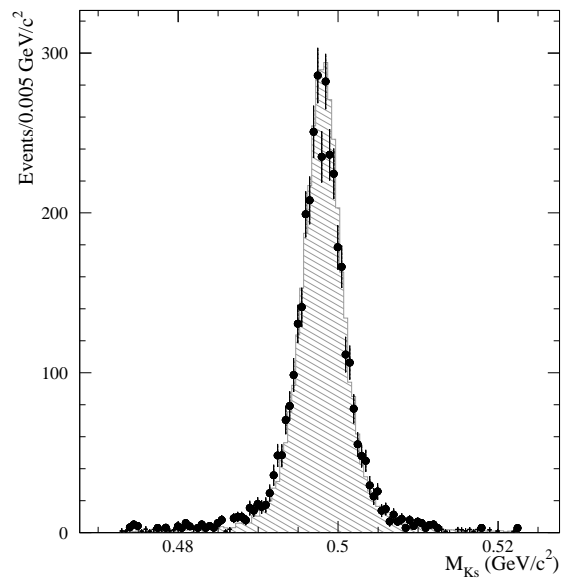}
\caption{Reconstructed \KS mass for data (points) and signal 
  MC simulation (histogram).}
  \label{fig:ks_mass}
\end{figure}  

The cross sections for the background processes in 
conti\-nuum $q\bar{q}$ production are poorly known; 
this type of background is almost exclusively produced
by processes like $\epem\to\kskpi\piz$ with the \piz decaying into two 
very energetically asymmetric photons and/or photons 
merged in a single EMC cluster that fake the ISR photon.
This class of events, as a consequence, produces a sharp peak at the \piz 
invariant mass when the fake ISR photon is combined with another photon
in the event.
\begin{figure}[htb]
\begin{tabular}{cc}
\includegraphics[width=\columnwidth]{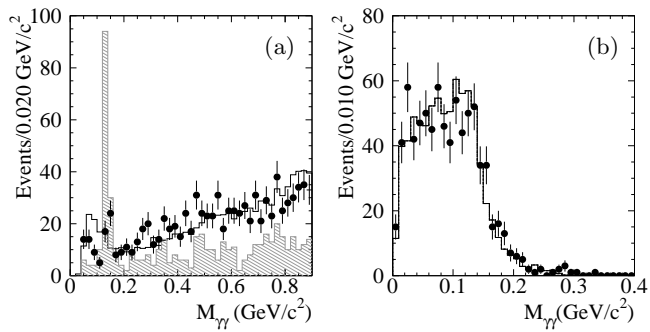}
\end{tabular}
\caption{(a) invariant mass of the ISR photon and any other 
  photon in the event. Distributions for data (points with error bars), 
  signal MC (open histogram) and \qqbar MC (shaded histogram) samples are shown. 
  (b) $\gamma\gamma$ mass distribution for additional photons in data 
  (points) compared with signal MC simulation.} 
\label{fig:kskpi_isr}
\end{figure}

Figure~\ref{fig:kskpi_isr}(a) depicts   
the invariant mass combination of the candidate ISR photon with any other 
photon in the event. 
A sharp peak at the \piz mass is evident for the \qqbar MC sample, while 
the signal and the data distributions are both essentially flat in that region.
We perform a fit to the data distribution, combining with weights 
the distributions of \qqbar background and signal. The difference between the shapes of
the signal and \qqbar distributions allows to obtain unambiguously the weights from the fit.
We find a weight for the \qqbar of about $1/9$, which translates into 
$72\pm42$ \qqbar background events.\\
\indent
Any remaining background due to other ISR processes should results in 
a relatively flat \chisq distribution, 
as a consequence of the mis-reconstructed final state. 
We check the consistency of the \chisq 
distribution by normalizing the data and the simulated \kskpi sample
to the first bin of the data distribution (see Fig.~\ref{fig:chi2_kskpi}), 
after subtracting the already estimated background components.
The difference between data and normalized  signal 
distributions gives an estimate of additional
background events consistent with zero.

\begin{figure}[htb]
\includegraphics[width=\columnwidth,height=.9\columnwidth]{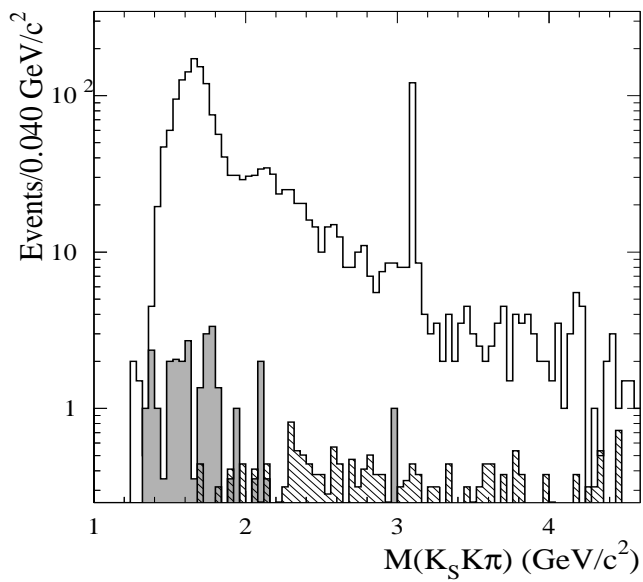}
\caption{The invariant mass distribution of the \kskpi system (open histogram).
  The hatched and gray histograms show the \qqbar and ISR backgrounds, respectively.} 
  \label{fig:kskpi_mass}
\end{figure}  

As a cross-check, we show in  Fig.~\ref{fig:kskpi_isr}(b)
 the invariant mass distribution
of any pair of detected photons in the event, not including the ISR photon.
The very good agreement between the data and the simulated signal sample 
and the absence of a peak at the \piz mass, confirm that background 
contribution from final states with extra \piz's is negligible.\\ 
\indent
The \kskpi invariant mass distribution for the final
sample is shown in Fig.~\ref{fig:kskpi_mass}, 
together with the estimated    
\qqbar and ISR backgrounds.
The final signal yield after background subtraction
consists of 3714 events.

\begin{figure}[htb]
\includegraphics[width=\columnwidth,height=.9\columnwidth]{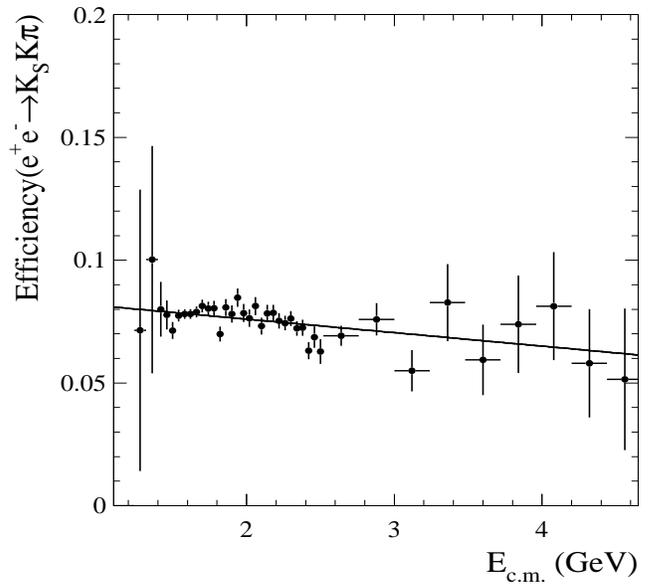}
\caption{Detection and reconstruction efficiency as a function 
  of the c.m. energy for the \kskpi final state. 
The solid line is the result of a linear fit to the data.}
\label{fig:kskpi_eff}
\end{figure}

\subsection{\boldmath Cross section for $\epem\to \kskpi$}
\label{sec:xskskpi}

The cross section for the $\epem \to \kskpi$ process as a function of the 
effective c.m. energy, \Ecm, is evaluated from:
\begin{equation}
  \sigma_{\epem\to\kskpi}(\Ecm)
    = \frac{dN_{\KS K\pi\gamma}(\Ecm)}
         {d{\cal L}(\Ecm)
          \cdot\epsilon(\Ecm)} ~,
\end{equation}
where \Ecm is the invariant mass of the reconstructed
$\KS K^{\pm}\pi^{\mp}$ system, 
$dN_{\KS K\pi\gamma}$ is the number of selected
$\KS K^{\pm}\pi^{\mp}$ events after background subtraction in the interval 
$d\Ecm$, and $\epsilon(\Ecm)$ is the corresponding detection
efficiency obtained from the signal MC simulation.
The differential luminosity, $d{\cal L}(\Ecm)$, 
in each interval  $d\Ecm$, is evaluated  
from ISR $\mu\mu\gamma$ events with the photon in the same fiducial range
used for simulation; the procedure is described in 
Refs.~\cite{isr4pi,bad1163}. From data-simulation comparison, 
we conservatively  estimate a systematic uncertainty  of 3.0\% on $d{\cal L}$. 
This error also includes the uncertainties on radiative corrections,
estimated from a MC simulation to be $\approx$ 1\%.\\
\indent 
The efficiency, that accounts also for the $\KS\to\pip\pim$ decay rate,
has been corrected for differences between data and simulation:
the angular distributions are different from the simulated distributions
because the latter, based on a phase space model, do not account for 
the presence of intermediate resonant states. 
However, the angular acceptance is essentially uniform for ISR events, 
so the uncertainty in estimated efficiencies due to imprecise 
modeling of the process dynamics is, at most,  3.0\%.
We study several control samples to determine any difference on track 
  reconstruction efficiency between data and simulation. 
  We find that the MC efficiency is overestimated by 
  0.8\% per charged track, with an uncertainty of $\pm0.5\%$. 
Analogously, we determine a correction of 2.6\% on the \KS 
  reconstruction efficiency, with an associated systematic 
  uncertainty of 1.1\%.
The uncertainties on \qqbar and ISR background estimates lead 
to systematic errors of 1.1\% and 0.5\%, respectively.

The main systematic uncertainties assigned to the measurement 
are summarized in Table~\ref{tab:error-kskpi}. 
The corrections applied to the measured cross sections sum to 
$+4.2\%$, while the estimated total systematic error is $\pm 5.1\%$.

\begin{figure}[htb]
\includegraphics[width=\columnwidth]{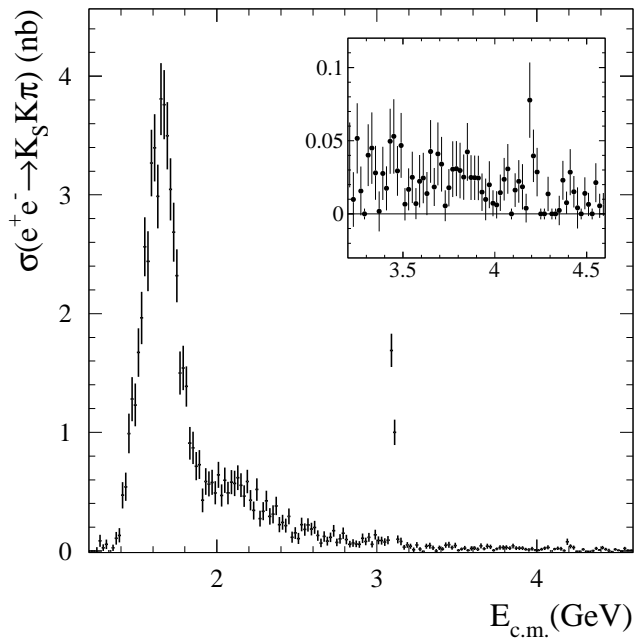}
\caption{The $\epem\to\KS\Kpm\pimp$ cross section.
Inset: expanded view in the mass range
$3.2 < \Ecm <4.6 ~\gev$.}
\label{fig:sigma_kskpi_all}
\end{figure}
\begin{figure}[htb]
\includegraphics[width=\columnwidth,height=.9\columnwidth]{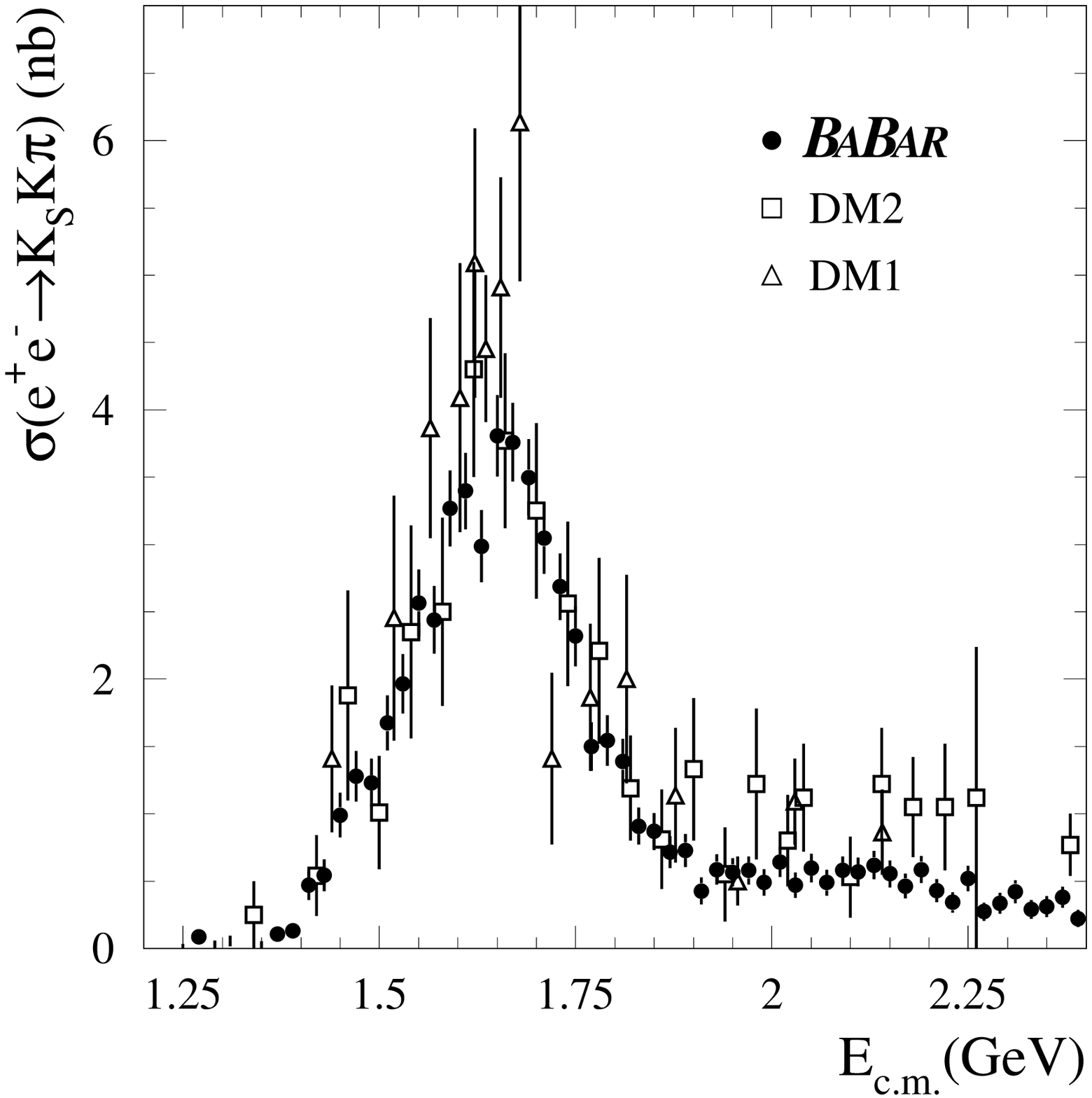}
\caption{Comparison of the $\epem\to \kskpi$ cross section measured 
  in \babar\ with previous experiments, in the mass range
  $1.3 < \Ecm <2.4 ~\gev$.}
\label{fig:sigma_kskpi_12}
\end{figure}

\begin{table}[htb]
\caption{Summary of corrections and systematic errors applied to  
the  $\epem\to\kskpi$ cross section measurement.}
\vspace{-2mm}
\label{tab:error-kskpi}
\begin{center}
\begin{tabular}{l c c} 
\hline\hline
Source                         & Correction            & Systematic error\\
\hline
ISR luminosity                 &  -                    & $3.0\%$ \\
\qqbar background                & -                     & $1.1\%$ \\
ISR background                 & -                     & $0.5\%$ \\
Track reconstruction efficiency& $+1.6\%$              & $1.0\%$ \\
PID efficiency                 & -                     & $2.0\%$ \\
\KS reconstruction efficiency  & $+2.6\%$              & $1.1\%$ \\
${\cal B}(\KS\!\to\!\pip\pim)$~\cite{pdg} & -          & $0.7\%$ \\
MC model                       & -                     & $3.0\%$ \\
\hline
\multirow{2}{*}{Total}         & \multirow{2}{*}{$+4.2\%$} & \multirow{2}{*}{$5.1\%$} \\
                               &                       & \\
\hline\hline
\end{tabular}
\end{center}
\end{table}

\begin{table*}[htb]
\caption{Measurement of the $\epem\to\kskpi$ 
cross section as a function of \Ecm. Errors are statistical only.}
\label{tab:kskpi}
\begin{ruledtabular}
\begin{tabular}{cc cc cc cc}
$\Ecm (\gev)$ & $\sigma$ (nb)& 
$\Ecm (\gev)$ & $\sigma$ (nb)& 
$\Ecm (\gev)$ & $\sigma$ (nb)& 
$\Ecm (\gev)$ & $\sigma$ (nb)\\
\hline
 1.24$-$1.28 &  0.037 $\pm$  0.034 &  1.96$-$1.98 &  0.578 $\pm$  0.105 &  2.62$-$2.64 &  0.131 $\pm$  0.044 &  3.30$-$3.34 &  0.043 $\pm$  0.016 \\
 1.28$-$1.32 &  0.042 $\pm$  0.024 &  1.98$-$2.00 &  0.491 $\pm$  0.098 &  2.64$-$2.66 &  0.068 $\pm$  0.036 &  3.34$-$3.38 &  0.015 $\pm$  0.011 \\
 1.32$-$1.36 &  0.014 $\pm$  0.014 &  2.00$-$2.02 &  0.642 $\pm$  0.109 &  2.66$-$2.68 &  0.122 $\pm$  0.044 &  3.38$-$3.42 &  0.023 $\pm$  0.012 \\
 1.36$-$1.38 &  0.107 $\pm$  0.054 &  2.02$-$2.04 &  0.469 $\pm$  0.093 &  2.68$-$2.70 &  0.083 $\pm$  0.037 &  3.42$-$3.46 &  0.051 $\pm$  0.017 \\
 1.38$-$1.40 &  0.132 $\pm$  0.059 &  2.04$-$2.06 &  0.597 $\pm$  0.107 &  2.70$-$2.72 &  0.116 $\pm$  0.040 &  3.46$-$3.50 &  0.038 $\pm$  0.014 \\
 1.40$-$1.42 &  0.470 $\pm$  0.111 &  2.06$-$2.08 &  0.488 $\pm$  0.095 &  2.72$-$2.74 &  0.170 $\pm$  0.050 &  3.50$-$3.54 &  0.012 $\pm$  0.009 \\
 1.42$-$1.44 &  0.542 $\pm$  0.118 &  2.08$-$2.10 &  0.581 $\pm$  0.102 &  2.74$-$2.76 &  0.073 $\pm$  0.033 &  3.54$-$3.58 &  0.016 $\pm$  0.010 \\
 1.44$-$1.46 &  0.990 $\pm$  0.167 &  2.10$-$2.12 &  0.569 $\pm$  0.107 &  2.76$-$2.78 &  0.102 $\pm$  0.041 &  3.58$-$3.62 &  0.023 $\pm$  0.012 \\
 1.46$-$1.48 &  1.279 $\pm$  0.187 &  2.12$-$2.14 &  0.619 $\pm$  0.104 &  2.78$-$2.80 &  0.151 $\pm$  0.048 &  3.62$-$3.66 &  0.028 $\pm$  0.013 \\
 1.48$-$1.50 &  1.231 $\pm$  0.180 &  2.14$-$2.16 &  0.555 $\pm$  0.099 &  2.80$-$2.82 &  0.097 $\pm$  0.041 &  3.66$-$3.70 &  0.030 $\pm$  0.013 \\
 1.50$-$1.52 &  1.673 $\pm$  0.205 &  2.16$-$2.18 &  0.463 $\pm$  0.091 &  2.82$-$2.84 &  0.058 $\pm$  0.029 &  3.70$-$3.74 &  0.020 $\pm$  0.011 \\
 1.52$-$1.54 &  1.964 $\pm$  0.220 &  2.18$-$2.20 &  0.586 $\pm$  0.100 &  2.84$-$2.86 &  0.067 $\pm$  0.032 &  3.74$-$3.78 &  0.024 $\pm$  0.011 \\
 1.54$-$1.56 &  2.564 $\pm$  0.249 &  2.20$-$2.22 &  0.428 $\pm$  0.085 &  2.86$-$2.88 &  0.060 $\pm$  0.028 &  3.78$-$3.82 &  0.030 $\pm$  0.013 \\
 1.56$-$1.58 &  2.440 $\pm$  0.253 &  2.22$-$2.24 &  0.342 $\pm$  0.076 &  2.88$-$2.90 &  0.056 $\pm$  0.028 &  3.82$-$3.86 &  0.034 $\pm$  0.013 \\
 1.58$-$1.60 &  3.269 $\pm$  0.282 &  2.24$-$2.26 &  0.518 $\pm$  0.095 &  2.90$-$2.92 &  0.108 $\pm$  0.040 &  3.86$-$3.90 &  0.025 $\pm$  0.011 \\
 1.60$-$1.62 &  3.397 $\pm$  0.283 &  2.26$-$2.28 &  0.272 $\pm$  0.067 &  2.92$-$2.94 &  0.081 $\pm$  0.032 &  3.90$-$3.94 &  0.020 $\pm$  0.010 \\
 1.62$-$1.64 &  2.987 $\pm$  0.268 &  2.28$-$2.30 &  0.336 $\pm$  0.077 &  2.94$-$2.96 &  0.116 $\pm$  0.038 &  3.94$-$3.98 &  0.015 $\pm$  0.011 \\
 1.64$-$1.66 &  3.807 $\pm$  0.302 &  2.30$-$2.32 &  0.422 $\pm$  0.086 &  2.96$-$2.98 &  0.058 $\pm$  0.027 &  3.98$-$4.02 &  0.007 $\pm$  0.006 \\
 1.66$-$1.68 &  3.760 $\pm$  0.291 &  2.32$-$2.34 &  0.291 $\pm$  0.069 &  2.98$-$3.00 &  0.138 $\pm$  0.041 &  4.02$-$4.06 &  0.019 $\pm$  0.010 \\
 1.68$-$1.70 &  3.498 $\pm$  0.284 &  2.34$-$2.36 &  0.310 $\pm$  0.076 &  3.00$-$3.02 &  0.090 $\pm$  0.033 &  4.06$-$4.10 &  0.015 $\pm$  0.008 \\
 1.70$-$1.72 &  3.046 $\pm$  0.262 &  2.36$-$2.38 &  0.381 $\pm$  0.077 &  3.02$-$3.04 &  0.088 $\pm$  0.033 &  4.10$-$4.14 &  0.019 $\pm$  0.009 \\
 1.72$-$1.74 &  2.686 $\pm$  0.247 &  2.38$-$2.40 &  0.222 $\pm$  0.064 &  3.04$-$3.06 &  0.075 $\pm$  0.031 &  4.14$-$4.18 &  0.011 $\pm$  0.009 \\
 1.74$-$1.76 &  2.319 $\pm$  0.223 &  2.40$-$2.42 &  0.242 $\pm$  0.064 &  3.06$-$3.08 &  0.091 $\pm$  0.036 &  4.18$-$4.22 &  0.059 $\pm$  0.016 \\
 1.76$-$1.78 &  1.500 $\pm$  0.181 &  2.42$-$2.44 &  0.215 $\pm$  0.059 &  3.08$-$3.10 &  1.690 $\pm$  0.140 &  4.22$-$4.26 &  0.013 $\pm$  0.009 \\
 1.78$-$1.80 &  1.543 $\pm$  0.187 &  2.44$-$2.46 &  0.291 $\pm$  0.067 &  3.10$-$3.12 &  0.999 $\pm$  0.106 &  4.26$-$4.30 &  0.006 $\pm$  0.006 \\
 1.80$-$1.82 &  1.387 $\pm$  0.170 &  2.46$-$2.48 &  0.116 $\pm$  0.047 &  3.12$-$3.14 &  0.098 $\pm$  0.033 &  4.30$-$4.38 &  0.005 $\pm$  0.004 \\
 1.82$-$1.84 &  0.908 $\pm$  0.136 &  2.48$-$2.50 &  0.153 $\pm$  0.049 &  3.14$-$3.16 &  0.081 $\pm$  0.032 &  4.38$-$4.46 &  0.014 $\pm$  0.006 \\
 1.84$-$1.86 &  0.869 $\pm$  0.137 &  2.50$-$2.52 &  0.106 $\pm$  0.047 &  3.16$-$3.18 &  0.031 $\pm$  0.019 &  4.46$-$4.54 &  0.003 $\pm$  0.004 \\
 1.86$-$1.88 &  0.714 $\pm$  0.119 &  2.52$-$2.54 &  0.223 $\pm$  0.058 &  3.18$-$3.20 &  0.051 $\pm$  0.025 &  4.54$-$4.62 &  0.011 $\pm$  0.005 \\
 1.88$-$1.90 &  0.729 $\pm$  0.122 &  2.54$-$2.56 &  0.181 $\pm$  0.051 &  3.20$-$3.22 &  0.040 $\pm$  0.022 &  4.62$-$4.70 &  0.003 $\pm$  0.004 \\
 1.90$-$1.92 &  0.427 $\pm$  0.099 &  2.56$-$2.58 &  0.219 $\pm$  0.057 &  3.22$-$3.24 &  0.010 $\pm$  0.019 &  & \\
 1.92$-$1.94 &  0.586 $\pm$  0.111 &  2.58$-$2.60 &  0.184 $\pm$  0.053 &  3.24$-$3.26 &  0.052 $\pm$  0.024 &  & \\
 1.94$-$1.96 &  0.565 $\pm$  0.105 &  2.60$-$2.62 &  0.200 $\pm$  0.054 &  3.26$-$3.30 &  0.007 $\pm$  0.009 &  & \\
\end{tabular}
\end{ruledtabular}
\end{table*}

For a typical ISR reaction, the selection efficiency is expected to have a 
smooth behavior as a function of \Ecm. Therefore, we fit the efficiency
values obtained from MC as  shown in Fig.~\ref{fig:kskpi_eff},
and use the fit result, together with the estimated corrections,
 to obtain the cross section plotted in Fig.~\ref{fig:sigma_kskpi_all}.
The cross section has a maximum  of about 3.8~nb around 1.65\gev, and 
a long tail.
The only  clear structure observed at higher energies is the \jpsi.
In the inset of the same figure the cross section at
energies higher than the $J/\psi$ is shown: there is no evidence of the
recently discovered $Y(4260)$ resonance~\cite{y4260}, 
we obtain an upper
limit for $\Gamma^{Y(4260)}_{ee}\mathcal{B}^{Y(4260)}_{\kskpi}$ of $0.5$ \ev at
90\% C.L.. 
An excess of events is observed at $\approx$
4.2 GeV: there are 15 events in a 40 MeV energy window, to be
compared to an expectation of 5 events, corresponding to 
a probability of 2.2$\times10^{-4}$ to be a statistical fluctuation of
the cross section.\\
\indent The values of the measured cross section are reported in
Table~\ref{tab:kskpi} with the corresponding  statistical errors.
Figure~\ref{fig:sigma_kskpi_12} shows a compilation  
of the measured cross section as obtained by previous direct 
$\epem$ experiments (data are available only up to 2.4~\gev~\cite{dm2}).
The data show reasonable consistency; 
our measurements, however, have much smaller statistical 
and systematic errors. 


\begin{figure}[htb]
\includegraphics[width=\columnwidth]{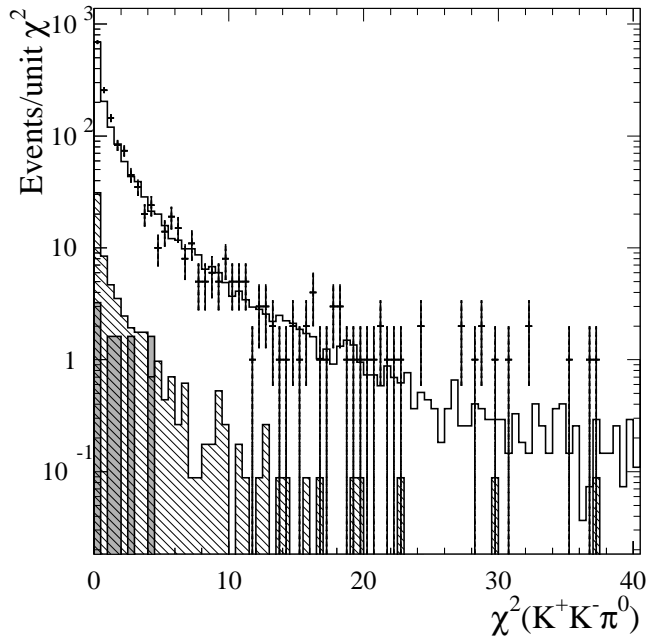}
\caption{The $\chi^2$ distributions for data (points),
  signal MC simulation (open histogram) 
  normalized to the first bin of the data, background contributions due 
  to \qqbar (hatched histogram) and to $\epem\to\omega\gamma\to\pi^+\pi^-\piz\gamma$ 
  MC simulation (gray histogram).}
\label{fig:chi2_kkpi0_60_feb}
\end{figure}

\section{The \kkpiz FINAL STATE}
\label{sec:kkpiz}

We select \kkpiz candidate events by requiring two
charged tracks identified as kaons, and two photons 
forming a \piz candidate, in addition to the ISR photon.
In order to avoid any possible contamination from events with
intermediate $\phi\to\Kp\Km$ production, we require $M_{\K\K}>1.045~\gevcc$,
where $M_{\K\K}$ is the invariant mass of the $\Kp\Km$ system.
As previously described, we perform the kinematic fit for every possible
pair of selected photons and we retain    
for each event the $\gamma\gamma$ combination giving the lowest \chisq. 
Figure~\ref{fig:chi2_kkpi0_60_feb} shows the \chisq distribution 
of the selected  data sample, 
and the normalized distributions of signal MC simulation and different background 
sources (\qqbar and $\epem\to\omega\gamma\to\pi^+\pi^-\piz\gamma$).
The selection procedure  results in 1524 events  with a $\chisq < 20$.\\ 
\begin{figure}[htb]
\includegraphics[width=\columnwidth]{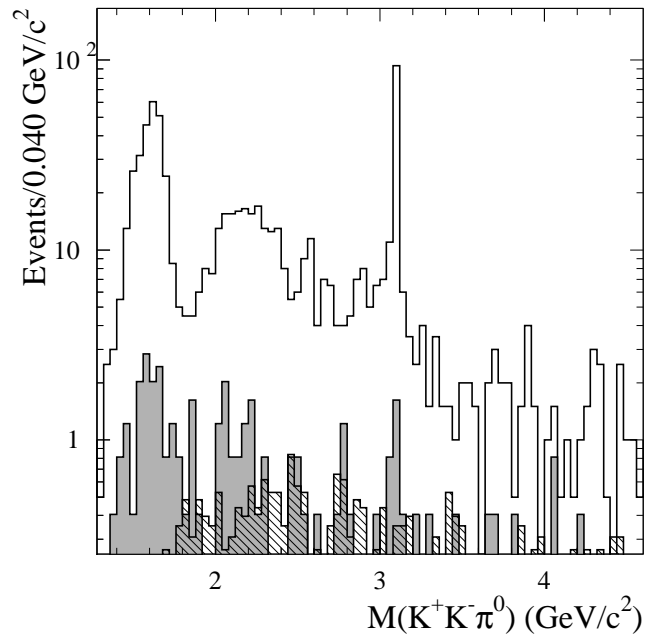}
\caption{The invariant mass distribution of the \kkpiz system 
  (open histogram). 
  The hatched and gray histograms show the \qqbar and ISR backgrounds,
  respectively. }
\label{fig:nevents_kkpi0_feb_1}
\end{figure}
\indent 
Applying a  procedure analogous to the one described in the
previous section for the \kskpi analysis, 
we determine $62\pm 35$ \qqbar and $84\pm27$ 
ISR-background events from different ISR processes.
The mass spectrum of the selected sample is 
shown in  Fig.~\ref{fig:nevents_kkpi0_feb_1}, together with
the estimated \qqbar and ISR backgrounds.
After background subtraction, 1378 signal events are retained.\\
%
%
\begin{figure}[htb]
\includegraphics[width=\columnwidth]{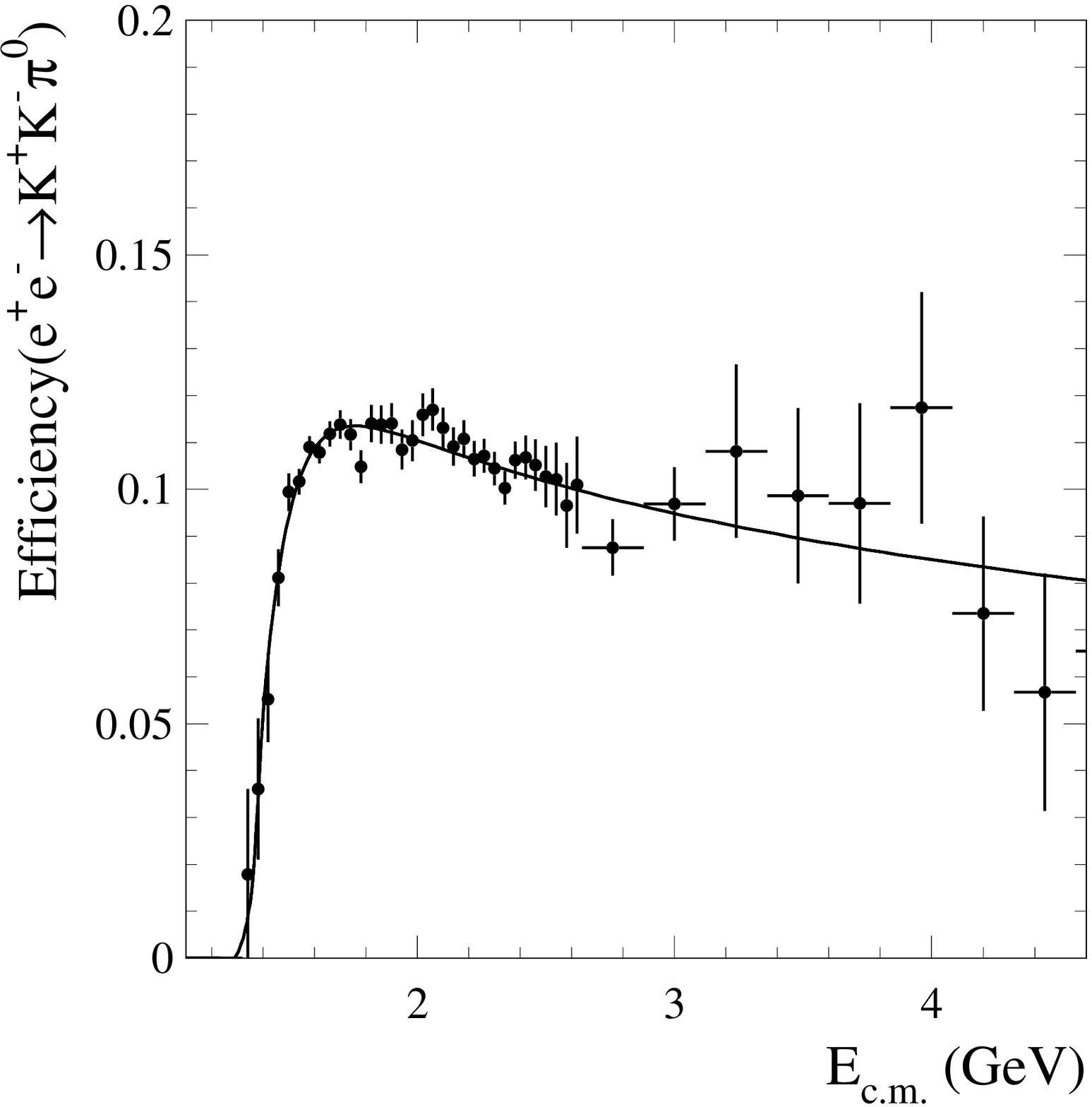}
\caption{Detection and reconstruction efficiency as a 
  function of the c.m. energy for \kkpiz final state.
The solid line is the result of a fit to the data with the function  $a\left( 1 - (x/b)^p\cdot e^{\xi (b-x)}\right)$.}
\label{fig:eff-kkpiz}
\end{figure}
\begin{figure}[htb]
\includegraphics[width=\columnwidth]{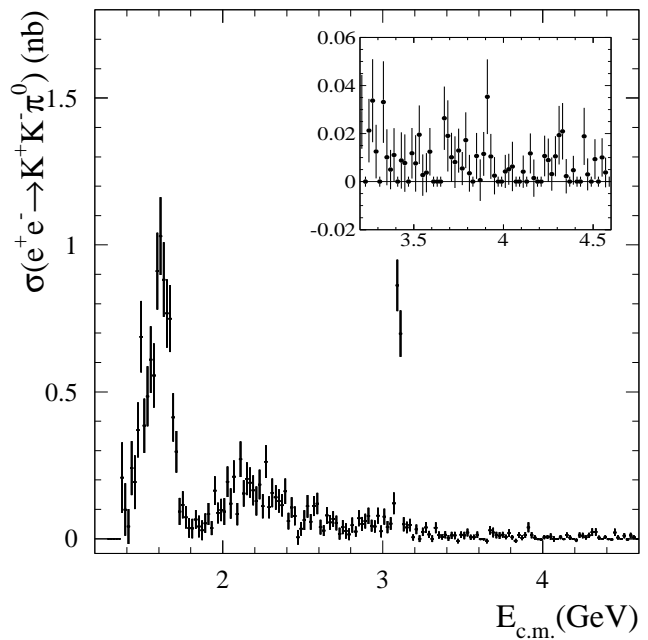}
\caption{The $\epem\to\kkpiz$ cross section. Inset: expanded view in the
$3.2 < \Ecm <4.6 \;\gev$ mass range.}
\label{fig:sigma_kkpi0_all_feb}
\end{figure}
\begin{figure}[htb]
\includegraphics[width=\columnwidth]{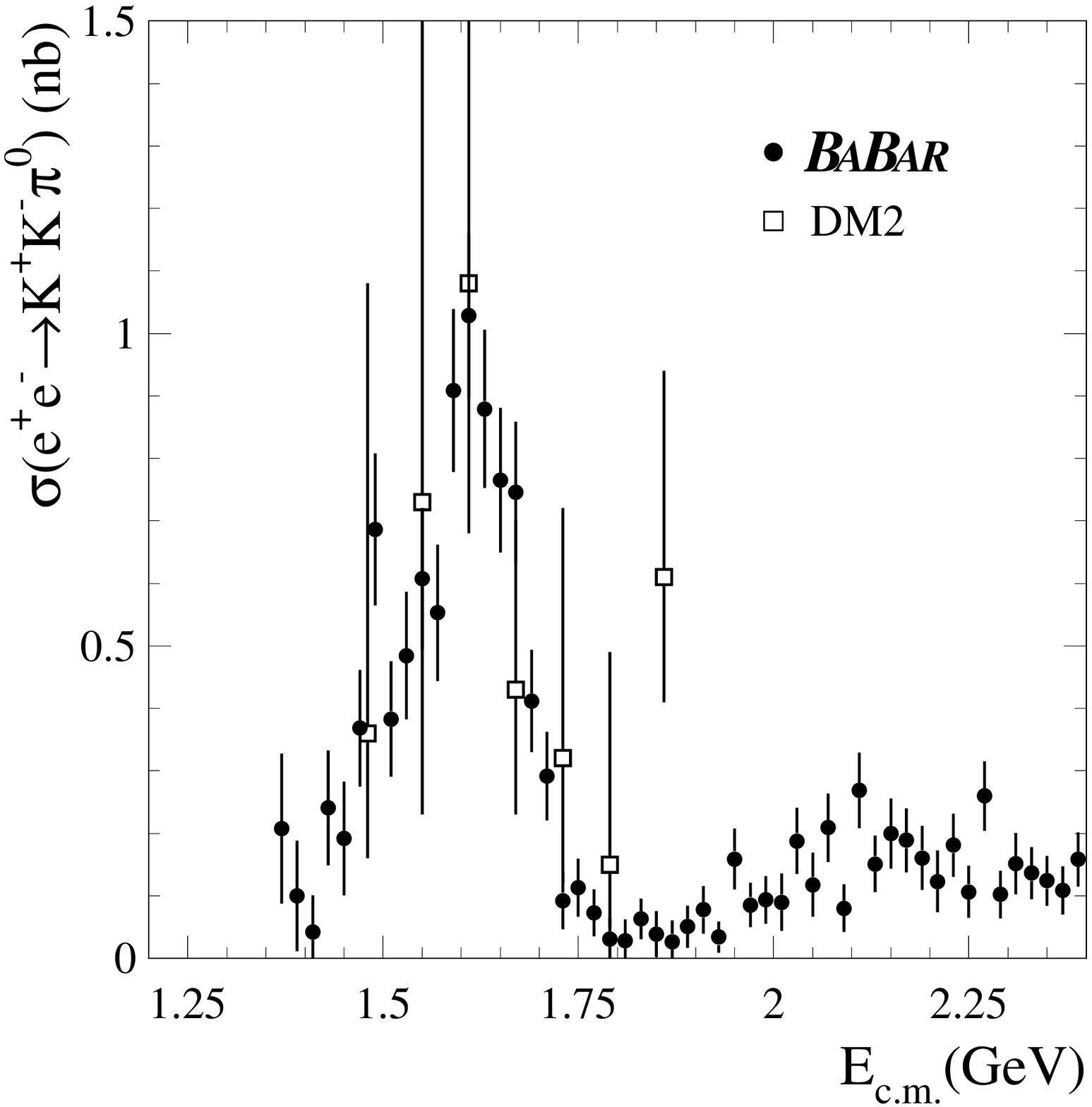}
\caption{Comparison of the $\epem\to \kkpiz$ cross section measured 
  in \babar\ with previous experiments.}
\label{fig:sigma_kkpi0_12_feb}
\end{figure}
%
As in the \kskpi channel, 
we measure  the $\epem\to\kkpiz$ cross section 
as a function of the effective c.m. energy, in 20~\mev energy bins. 
The reconstruction and selection efficiency for each energy interval
is shown in Fig.~\ref{fig:eff-kkpiz}, while the corresponding cross section
values are shown in Fig.~\ref{fig:sigma_kkpi0_all_feb} 
and reported in Table~\ref{kkpi0_tab} (variable width bin size).
Figure~\ref{fig:sigma_kkpi0_12_feb} shows a comparison with the 
DM2 measurement~\cite{dm2}.\\
\indent
The cross section values shown in Fig.~\ref{fig:sigma_kkpi0_all_feb}
include only statistical errors. 
The inset of the same figure is an enlarged view of the 
$3.3-4.6$ \gev energy region, there is no evidence of the 
$Y(4260)$, and we estimate an upper limit for
$\Gamma^{Y(4260)}_{ee}\mathcal{B}^{Y(4260)}_{\kkpiz}$ of 
$0.6$ \ev at 90\% C.L..
The systematic errors are summarized in Table~\ref{tab:error-kkpi0}, along 
with the corrections applied to the cross section measurements~\cite{isr4pi,bad1163}.
The correction and  systematic error associated with the MC/data difference
in the \piz reconstruction efficiency 
have been studied using the ISR-produced high statistics 
channel $\omega\piz\to\pip\pim\piz\piz$~\cite{bad1163}.
It has been found that the MC efficiency  is $3.0\pm3.0\%$ higher than 
the data.  
The corrections applied to the measured cross sections
sum to +4.6\%, while the estimated total systematic error 
is $\pm 6.5\%$.

\begin{table*}[htb!]
\caption{Measurement of the $\epem\to \kkpiz$ cross section as a function of \Ecm. Errors are statistical only.}
\label{kkpi0_tab}
\begin{ruledtabular}
\begin{tabular}{ cc cc cc cc}
$\Ecm (\gev)$ & $\sigma$ (nb)& 
$\Ecm (\gev)$ & $\sigma$ (nb)& 
$\Ecm (\gev)$ & $\sigma$ (nb)& 
$\Ecm (\gev)$ & $\sigma$ (nb)\\
\hline
 1.36$-$1.38 &  0.208 $\pm$  0.120 &  1.96$-$1.98 &  0.088 $\pm$  0.035 &  2.56$-$2.58 &  0.111 $\pm$  0.034 &  3.16$-$3.18 &  0.047 $\pm$  0.021 \\
 1.38$-$1.40 &  0.100 $\pm$  0.089 &  1.98$-$2.00 &  0.098 $\pm$  0.038 &  2.58$-$2.60 &  0.120 $\pm$  0.036 &  3.18$-$3.20 &  0.005 $\pm$  0.010 \\
 1.40$-$1.42 &  0.042 $\pm$  0.059 &  2.00$-$2.02 &  0.092 $\pm$  0.046 &  2.60$-$2.62 &  0.040 $\pm$  0.025 &  3.20$-$3.24 &  0.016 $\pm$  0.010 \\
 1.42$-$1.44 &  0.241 $\pm$  0.092 &  2.02$-$2.04 &  0.194 $\pm$  0.052 &  2.62$-$2.64 &  0.028 $\pm$  0.018 &  3.24$-$3.28 &  0.031 $\pm$  0.012 \\
 1.44$-$1.46 &  0.193 $\pm$  0.091 &  2.04$-$2.06 &  0.120 $\pm$  0.051 &  2.64$-$2.66 &  0.079 $\pm$  0.028 &  3.28$-$3.32 &  0.007 $\pm$  0.008 \\
 1.46$-$1.48 &  0.370 $\pm$  0.093 &  2.06$-$2.08 &  0.212 $\pm$  0.055 &  2.66$-$2.68 &  0.056 $\pm$  0.025 &  3.32$-$3.36 &  0.025 $\pm$  0.011 \\
 1.48$-$1.50 &  0.688 $\pm$  0.122 &  2.08$-$2.10 &  0.084 $\pm$  0.038 &  2.68$-$2.70 &  0.067 $\pm$  0.026 &  3.36$-$3.40 &  0.010 $\pm$  0.007 \\
 1.50$-$1.52 &  0.385 $\pm$  0.092 &  2.10$-$2.12 &  0.271 $\pm$  0.060 &  2.70$-$2.72 &  0.053 $\pm$  0.025 &  3.40$-$3.44 &  0.007 $\pm$  0.008 \\
 1.52$-$1.54 &  0.485 $\pm$  0.102 &  2.12$-$2.14 &  0.154 $\pm$  0.045 &  2.72$-$2.74 &  0.017 $\pm$  0.020 &  3.44$-$3.48 &  0.002 $\pm$  0.007 \\
 1.54$-$1.56 &  0.610 $\pm$  0.113 &  2.14$-$2.16 &  0.204 $\pm$  0.056 &  2.74$-$2.76 &  0.040 $\pm$  0.023 &  3.48$-$3.52 &  0.012 $\pm$  0.008 \\
 1.56$-$1.58 &  0.556 $\pm$  0.109 &  2.16$-$2.18 &  0.191 $\pm$  0.051 &  2.76$-$2.78 &  0.026 $\pm$  0.026 &  3.52$-$3.56 &  0.013 $\pm$  0.008 \\
 1.58$-$1.60 &  0.911 $\pm$  0.130 &  2.18$-$2.20 &  0.165 $\pm$  0.051 &  2.78$-$2.80 &  0.015 $\pm$  0.020 &  3.56$-$3.62 &  0.006 $\pm$  0.005 \\
 1.60$-$1.62 &  1.030 $\pm$  0.132 &  2.20$-$2.22 &  0.128 $\pm$  0.049 &  2.80$-$2.82 &  0.046 $\pm$  0.025 &  3.62$-$3.68 &  0.006 $\pm$  0.006 \\
 1.62$-$1.64 &  0.882 $\pm$  0.127 &  2.22$-$2.24 &  0.185 $\pm$  0.050 &  2.82$-$2.84 &  0.025 $\pm$  0.017 &  3.68$-$3.74 &  0.015 $\pm$  0.007 \\
 1.64$-$1.66 &  0.767 $\pm$  0.116 &  2.24$-$2.26 &  0.111 $\pm$  0.041 &  2.84$-$2.86 &  0.070 $\pm$  0.027 &  3.74$-$3.80 &  0.013 $\pm$  0.006 \\
 1.66$-$1.68 &  0.749 $\pm$  0.114 &  2.26$-$2.28 &  0.262 $\pm$  0.056 &  2.86$-$2.88 &  0.050 $\pm$  0.024 &  3.80$-$3.86 &  0.004 $\pm$  0.005 \\
 1.68$-$1.70 &  0.413 $\pm$  0.082 &  2.28$-$2.30 &  0.107 $\pm$  0.037 &  2.88$-$2.90 &  0.060 $\pm$  0.025 &  3.86$-$3.92 &  0.018 $\pm$  0.007 \\
 1.70$-$1.72 &  0.296 $\pm$  0.071 &  2.30$-$2.32 &  0.156 $\pm$  0.049 &  2.90$-$2.92 &  0.078 $\pm$  0.028 &  3.92$-$3.98 &  0.005 $\pm$  0.004 \\
 1.72$-$1.74 &  0.093 $\pm$  0.046 &  2.32$-$2.34 &  0.141 $\pm$  0.041 &  2.92$-$2.94 &  0.044 $\pm$  0.020 &  3.98$-$4.04 &  0.003 $\pm$  0.004 \\
 1.74$-$1.76 &  0.116 $\pm$  0.046 &  2.34$-$2.36 &  0.128 $\pm$  0.040 &  2.94$-$2.96 &  0.042 $\pm$  0.021 &  4.04$-$4.10 &  0.002 $\pm$  0.005 \\
 1.76$-$1.78 &  0.073 $\pm$  0.038 &  2.36$-$2.38 &  0.114 $\pm$  0.038 &  2.96$-$2.98 &  0.079 $\pm$  0.027 &  4.10$-$4.16 &  0.005 $\pm$  0.004 \\
 1.78$-$1.80 &  0.037 $\pm$  0.033 &  2.38$-$2.40 &  0.161 $\pm$  0.043 &  2.98$-$3.00 &  0.025 $\pm$  0.020 &  4.16$-$4.24 &  0.002 $\pm$  0.003 \\
 1.80$-$1.82 &  0.035 $\pm$  0.033 &  2.40$-$2.42 &  0.060 $\pm$  0.028 &  3.00$-$3.02 &  0.075 $\pm$  0.027 &  4.24$-$4.32 &  0.012 $\pm$  0.005 \\
 1.82$-$1.84 &  0.066 $\pm$  0.032 &  2.42$-$2.44 &  0.107 $\pm$  0.035 &  3.02$-$3.04 &  0.040 $\pm$  0.020 &  4.32$-$4.40 &  0.008 $\pm$  0.004 \\
 1.84$-$1.86 &  0.042 $\pm$  0.036 &  2.44$-$2.46 &  0.078 $\pm$  0.032 &  3.04$-$3.06 &  0.052 $\pm$  0.021 &  4.40$-$4.48 &  0.006 $\pm$  0.004 \\
 1.86$-$1.88 &  0.029 $\pm$  0.033 &  2.46$-$2.48 &  0.005 $\pm$  0.025 &  3.06$-$3.08 &  0.121 $\pm$  0.036 &  4.48$-$4.56 &  0.005 $\pm$  0.003 \\
 1.88$-$1.90 &  0.053 $\pm$  0.033 &  2.48$-$2.50 &  0.034 $\pm$  0.023 &  3.08$-$3.10 &  0.862 $\pm$  0.088 &  4.56$-$4.64 &  0.002 $\pm$  0.002 \\
 1.90$-$1.92 &  0.084 $\pm$  0.037 &  2.50$-$2.52 &  0.063 $\pm$  0.033 &  3.10$-$3.12 &  0.698 $\pm$  0.079 &  4.64$-$4.72 &  0.001 $\pm$  0.002 \\
 1.92$-$1.94 &  0.036 $\pm$  0.024 &  2.52$-$2.54 &  0.110 $\pm$  0.038 &  3.12$-$3.14 &  0.049 $\pm$  0.024 &  & \\
 1.94$-$1.96 &  0.164 $\pm$  0.048 &  2.54$-$2.56 &  0.058 $\pm$  0.026 &  3.14$-$3.16 &  0.039 $\pm$  0.019 &  & \\
\end{tabular}
\end{ruledtabular}
\end{table*}

\begin{table}[h!]
\caption{Summary of corrections and systematic errors applied to
the $\epem\to K^+K^-\pi^0$ cross section measurement.}
\vspace{-2mm}
\label{tab:error-kkpi0}
\begin{center}
\begin{tabular}{l c c} 
\hline
\hline
Source & Correction  & Systematic error\\
\hline
ISR luminosity                 & -                     & $3.0\%$ \\
\qqbar background                & -                     & $2.5\%$ \\
ISR background                 & -                     & $2.0\%$ \\
Track reconstruction efficiency& $+1.6\%$              & $1.0\%$ \\
PID efficiency                 & -                     & $2.0\%$ \\
\piz reconstruction efficiency & $+3.0\%$              & $3.0\%$ \\
MC model                       & -                     & $3.0\%$ \\
\hline
\multirow{2}{*}{Total}         & \multirow{2}{*}{$+4.6\%$} & \multirow{2}{*}{$6.5\%$} \\
           &   & \\
\hline
\hline
\end{tabular}
\end{center}
\end{table}


\section{The {\boldmath Dalitz Plot analysis for \kskpi and \kkpiz channels}}
\label{sec:dalitz}
Figures~\ref{dal-kkpi0} and~\ref{dal-kskpi} show the
Dalitz plot at c.m. energies below 3.1 \gev
for the \kkpiz and \kskpi final states, respectively.
We observe that the intermediate states
$KK^*(892)$ and $KK^*_2(1430)$, carrying the allowed combinations of quantum 
numbers, give the main contributions to the $KK\pi$ 
production, while $KK^*_1(1410)$ contributes only weakly
due to the small value of $\Gamma(K^*_1(1410)\to K\pi)$.\\
\indent
The final state \kkpiz can be produced through the intermediate states 
$K^{*\pm}(892)K^{\mp}$ or $K_2^{*\pm}(1430)K^{\mp}$, and we expect a 
\DP\ with a symmetric population density with respect to the exchange 
$K^+\pi^0(K^{*+})\leftrightarrow K^-\pi^0(K^{*-})$, 
see Fig.~\ref{dal-kkpi0}.\\
\indent
The final state \kskpi is obtained via the decays of
$K^{*\pm}(892)K^{\mp}$ and $K^{*0}(892)\KS$, or
$K^{*\pm}_2(1430)K^{\mp}$ and $K_2^{*0}(1430)\KS$, {\it i.e.},  
both charged $K^{*\pm}(\KS \pi^\pm)$ and neutral $K^{*0}(K^\mp\pi^\pm)$
can be produced.
The population density of the Dalitz plot is expected to be asymmetric
in this case.
This effect is clearly seen in Fig.~\ref{dal-kskpi}.\\
\indent
A detailed explanation of the analysis techniques and the results obtained 
is given in the following subsection.

\begin{figure}[htb!]
\bc
\includegraphics[width=.9\columnwidth]{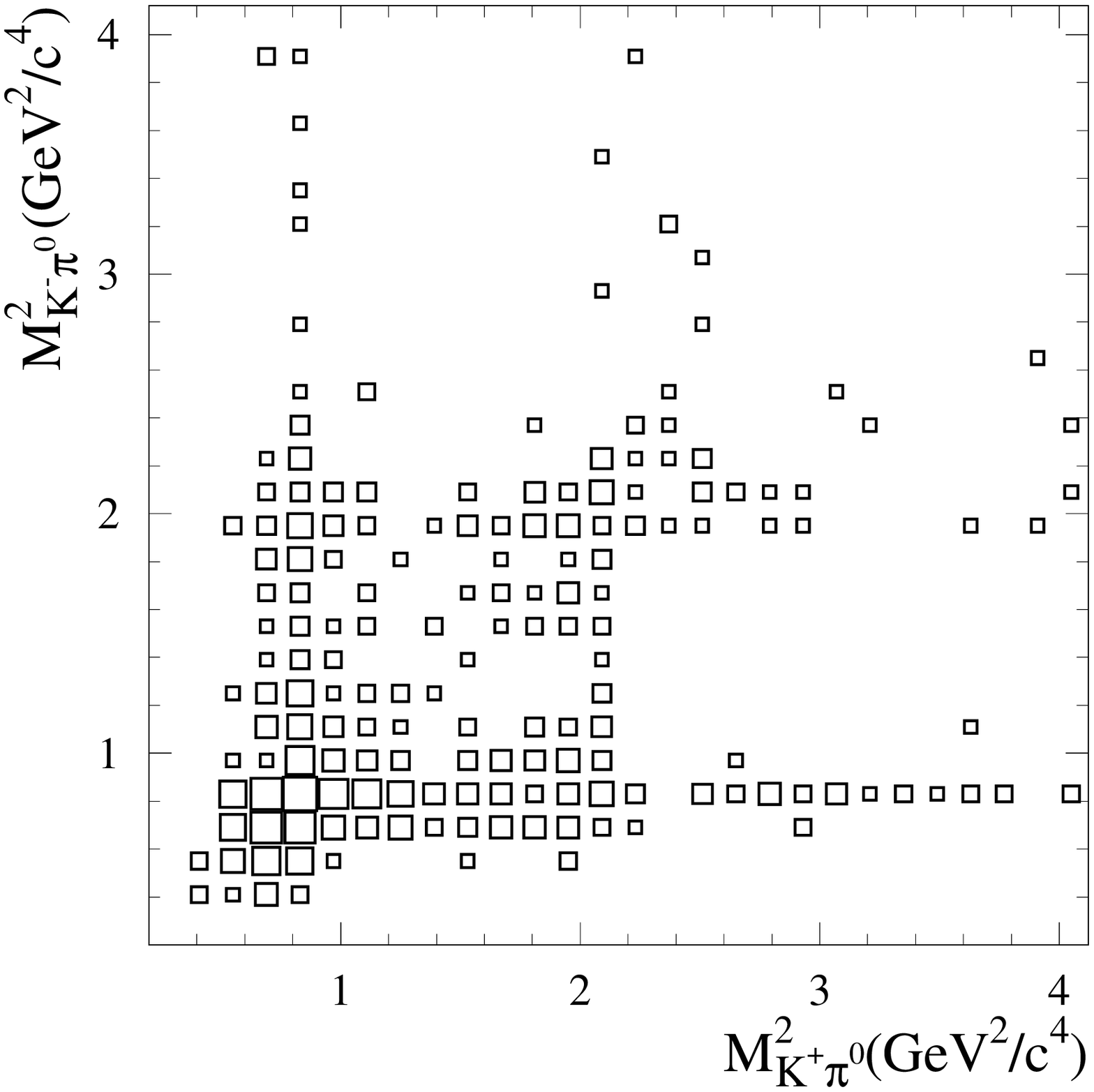}
\caption{\label{dal-kkpi0}
The Dalitz plot distribution for the \kkpiz final state.}
\ec
\end{figure}
\begin{figure}[htb!]
\bc
\includegraphics[width=.9\columnwidth]{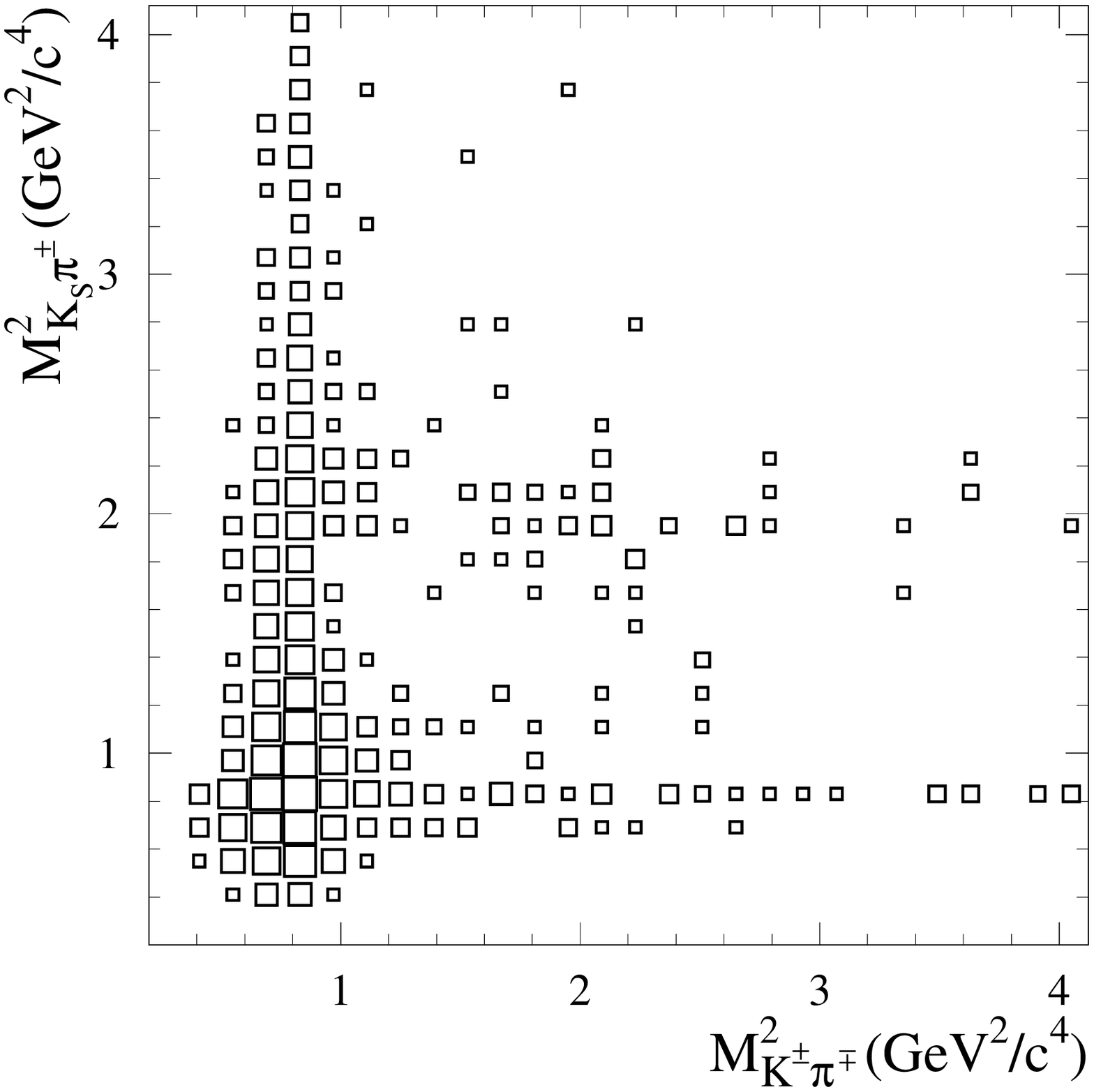}
\caption{\label{dal-kskpi}
  The Dalitz plot distribution for the \kskpi final state.}
\ec
\end{figure}
%

\subsection{The Dalitz method} 

The Dalitz plot population has been fit
assuming $KK^*$ two-body processes as intermediate states and a continuum
contribution coming from tails of resonances coupled with $KK^*$, but 
lying below their production threshold~\cite{dionisi} (see Sec.~\ref{sec:fitting}).
The density function is written in terms of 3-vector isospin amplitudes 
$\vec{A}_{I,K^*}$, that, in turn,
are products of a Breit Wigner distribution ($BW$), the $K^*$ propagator, 
and angular distribution probabilities, 
described by Zemach tensors~\cite{zemach}. In this case, since the
spectator particle is always a pseudoscalar meson (the kaon), the 
Zemach tensors are simply 3-vectors and the amplitudes are:
\begin{eqnarray}
\vec{A}_{I,K^*} \!\!\!&=&\!\!\! C_{I,K^*}\vec{T}_{K^*}(\vec{p}_\pi-\vec{p}_K,\vec{p}_{\ov{K}})\cdot BW_{K^*}[(p_\pi+p_K)^2]+
\no\\
&&\label{zemach-eqn}\\
\!\!\!&&\!\!\! +C_{I,\ov{K}^*}\vec{T}_{\ov{K}^*}(\vec{p}_{\ov{K}}-\vec{p}_\pi,\vec{p}_K)\cdot BW_{\ov{K}^*}[(p_{\ov{K}}+p_\pi)^2],
\no
\end{eqnarray}
where $p_\alpha$  is the 4-momentum of the generic final meson $\alpha$ and
$\vec{T}_{K^*}$ is the Zemach tensor related to the $K^*$. The two terms of
Eq.~\ref{zemach-eqn} correspond to the two possible $K\pi$ and $\ov{K}\pi$ combinations 
in which the $K^*$ and $\ov{K}^*$ can exist.
\indent The coefficient $C_{\alpha}^I$, which accounts for the total isospin $I$
and the charge of the intermediate kaon $\alpha$, has the form
\begin{equation}
C_{I,K^*}=\left\{
\begin{array}{ll}
(-1)^I &\mbox{for charged }K^*\\
1 &\mbox{for neutral }K^*\\
\end{array}\right. .
\end{equation}
%
%
\begin{table}[htb!]
\bc
\caption{Values of the Zemach tensor $\vec{T}_{K^*}$~\cite{dionisi} in the two cases 
under consideration.}
\label{zemach}
\begin{ruledtabular}
\begin{tabular}{cccc} 
Meson $K^*$ & $J^P$ & $L$  & $\vec{T}_{K^*}(\vec{p}_{1}-\vec{p}_2,\vec{p}_3)$  \\ 
\hline
$K^*(892)$ & $1^-$ & P  & $(\vec{p}_{1}-\vec{p}_2)\times\vec{p}_3$\\
$K^*_2(1430)$ & $2^+$ & D  & $[(\vec{p}_{1}-\vec{p}_2)\cdot\vec{p}_3](\vec{p}_{1}-\vec{p}_2)\times\vec{p}_3$\\
\end{tabular}
\end{ruledtabular}
\ec
%
\end{table}
\begin{figure}[htb]
\bc
\includegraphics[width=\columnwidth]{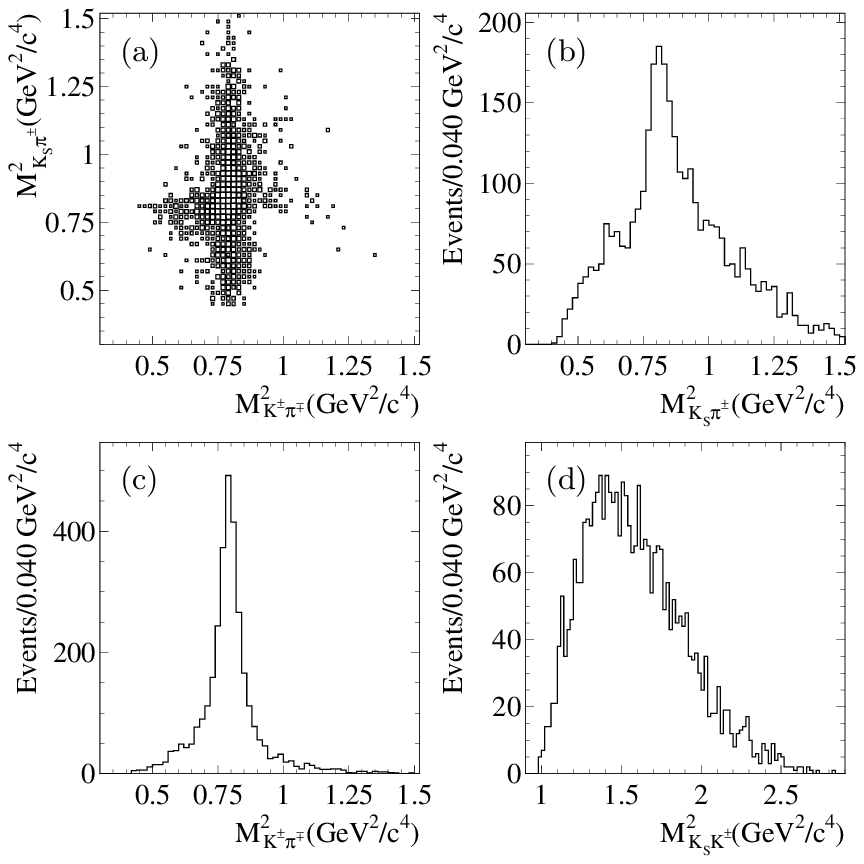}
\caption{\label{kskpi-low}
  (a) Dalitz plot distributions for \kskpi events with an invariant mass 
  $m_{\KS\K\pi} < 2.0\;\gevcc$. 
  (b) \KS\pipm projection, with the broad charged $K^*(892)$ peak,
  (c) \Kpm\pimp projection, with the narrow neutral $K^*(892)$ peak.
  (d) \KS\Kpm projection.}
\ec
\end{figure}
The Zemach tensors for $K^*(892)$ and the $K_2^*(1430)$ are shown  in
Table~\ref{zemach}.\\
The Dalitz density has been fit in each bin of \Ecm with the function
\begin{eqnarray}
&&\mathcal{D}[(p_\pi+p_K)^2,(p_{\ov{K}}+p_\pi)^2,\Ecm]=\\
&&\mathop{\sum\limits_{I,I'=0,1}}_{\begin{minipage}{20mm}\tiny ${K^*,{K^*}'=K^*(892),K^*_2(1430)}$
\end{minipage}}
f_{I,K^*} f_{I',{K^*}'}^*\vec{A}_{I,K^*}\cdot\vec{A}_{I',{K^*}'}^*\, ,\nonumber
\end{eqnarray}
where the $\vec{A}_{I,K^*}$ are the known amplitudes of Eq.~\ref{zemach-eqn},
which depend on the $K^*$ ($K\pi$) invariant mass and the total energy \Ecm,  
and the complex coefficients $f_{I,K^*}$, depending only on \Ecm, and
proportional to the isospin component of the corresponding $K^*K$
channel, are the free parameters. The complex values of the
$f_{I,K^*}$ parameters have been used to extract 
moduli and relative phase of the isospin components.
The analysis described in the following applies only to the \kskpi Dalitz plot, 
since only in this case, with  both charged and neutral $K^*$ contributing,
we can separately extract the isoscalar and isovector components. 
Because the $KK^*(892)$ and $KK^*_2(1430)$ represent the dominant contribution to 
the $KK\pi$ production at c.m. energies below and above $2.0\;\gev$, respectively,
these two regions have been considered separately.

\begin{figure}[htb]
\bc
\includegraphics[width=.8\columnwidth,height=.7\columnwidth]{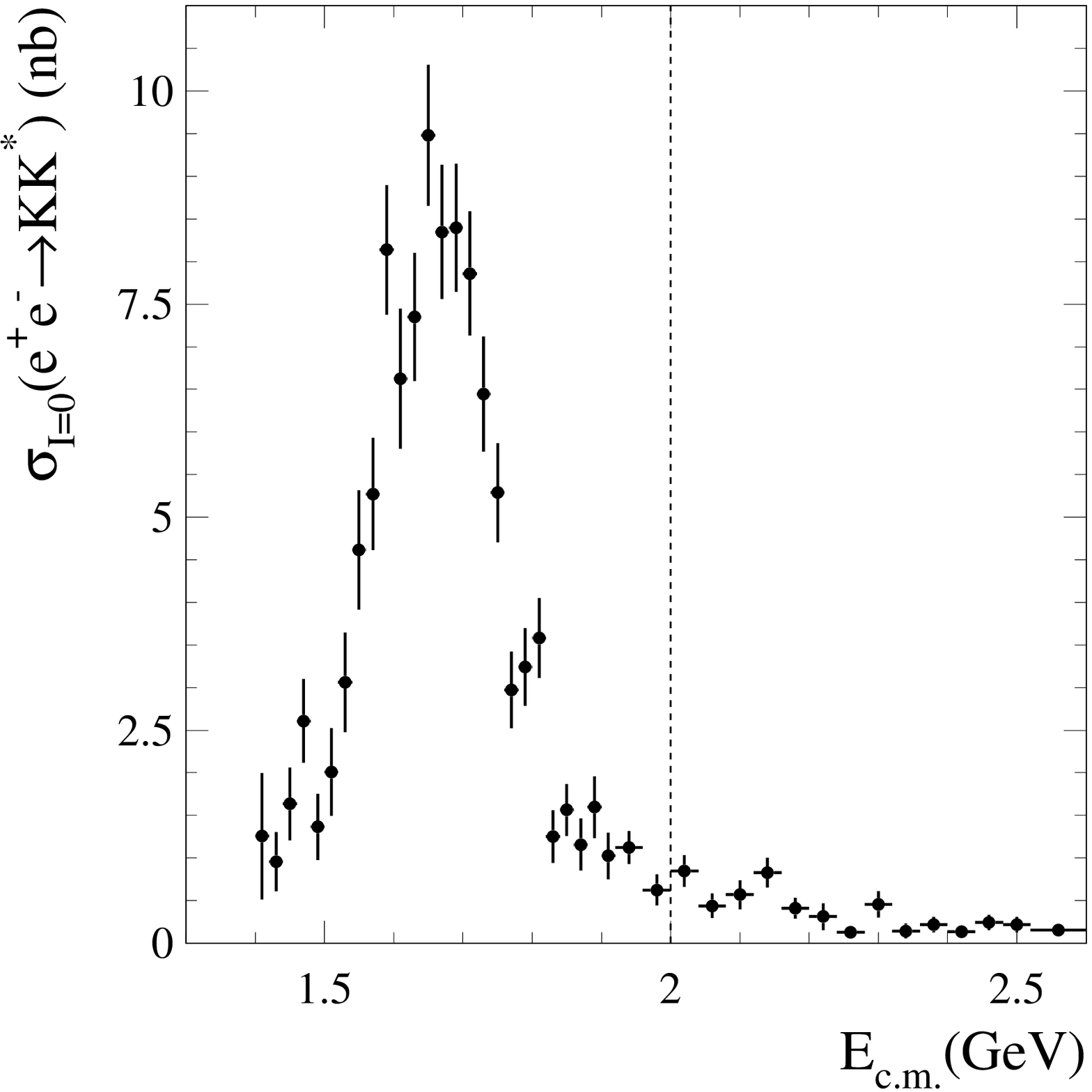}
\caption{\label{kskpi-i0}
Isoscalar component of the cross section for the process
$\epem\to K^*(892)\K$. The vertical line indicates the separation between the
two regions used in the analysis.}
\ec
\end{figure}
\begin{figure}[htb]
\bc
\includegraphics[width=.8\columnwidth,height=.7\columnwidth]{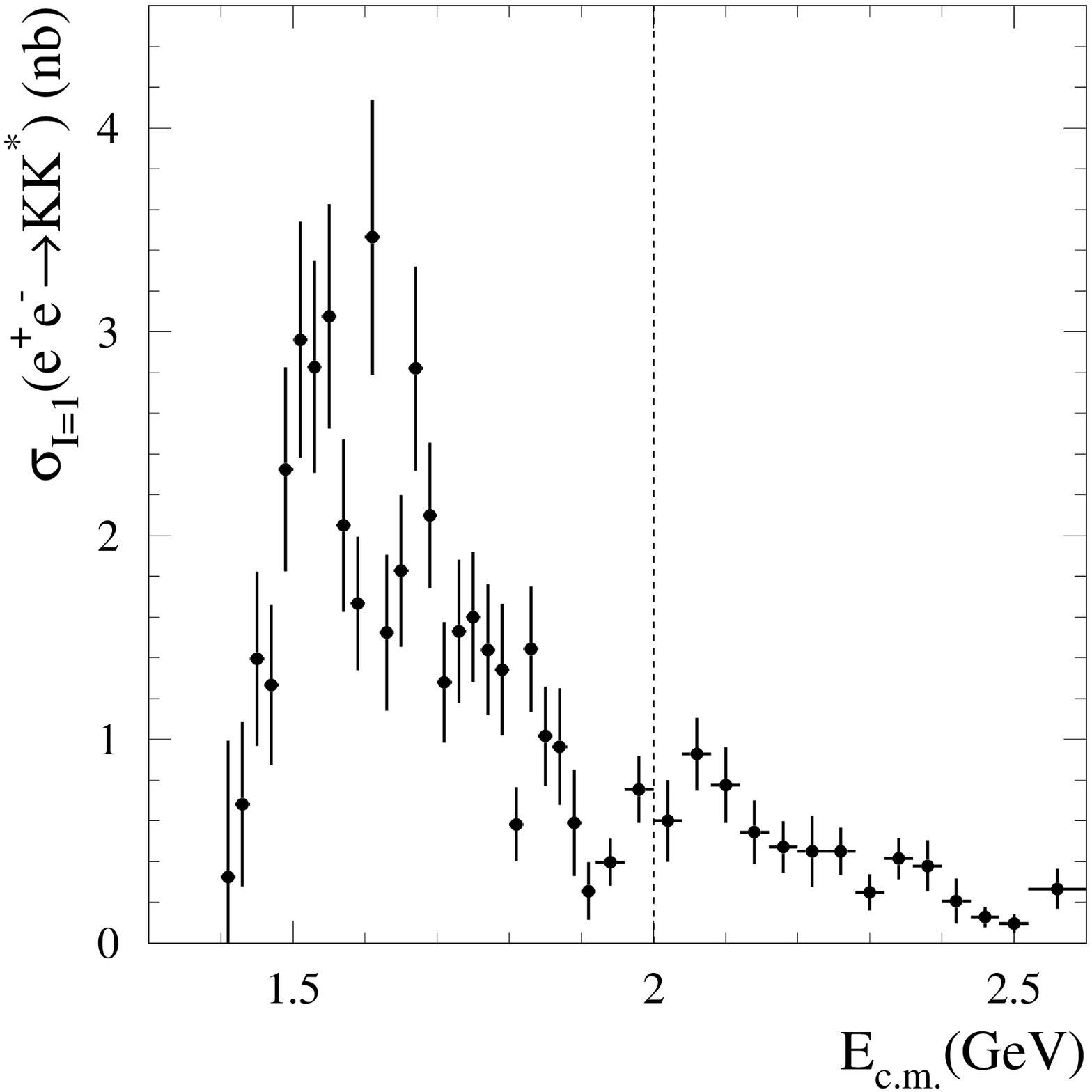}
\caption{\label{kskpi-i1}
Isovector component of the cross section for the process
$\epem\to K^*(892)\K$. The vertical line indicates the separation between the
two regions used in the analysis.}
\ec
\end{figure}
\begin{figure}[htb]
\bc
\includegraphics[width=.8\columnwidth,height=.7\columnwidth]{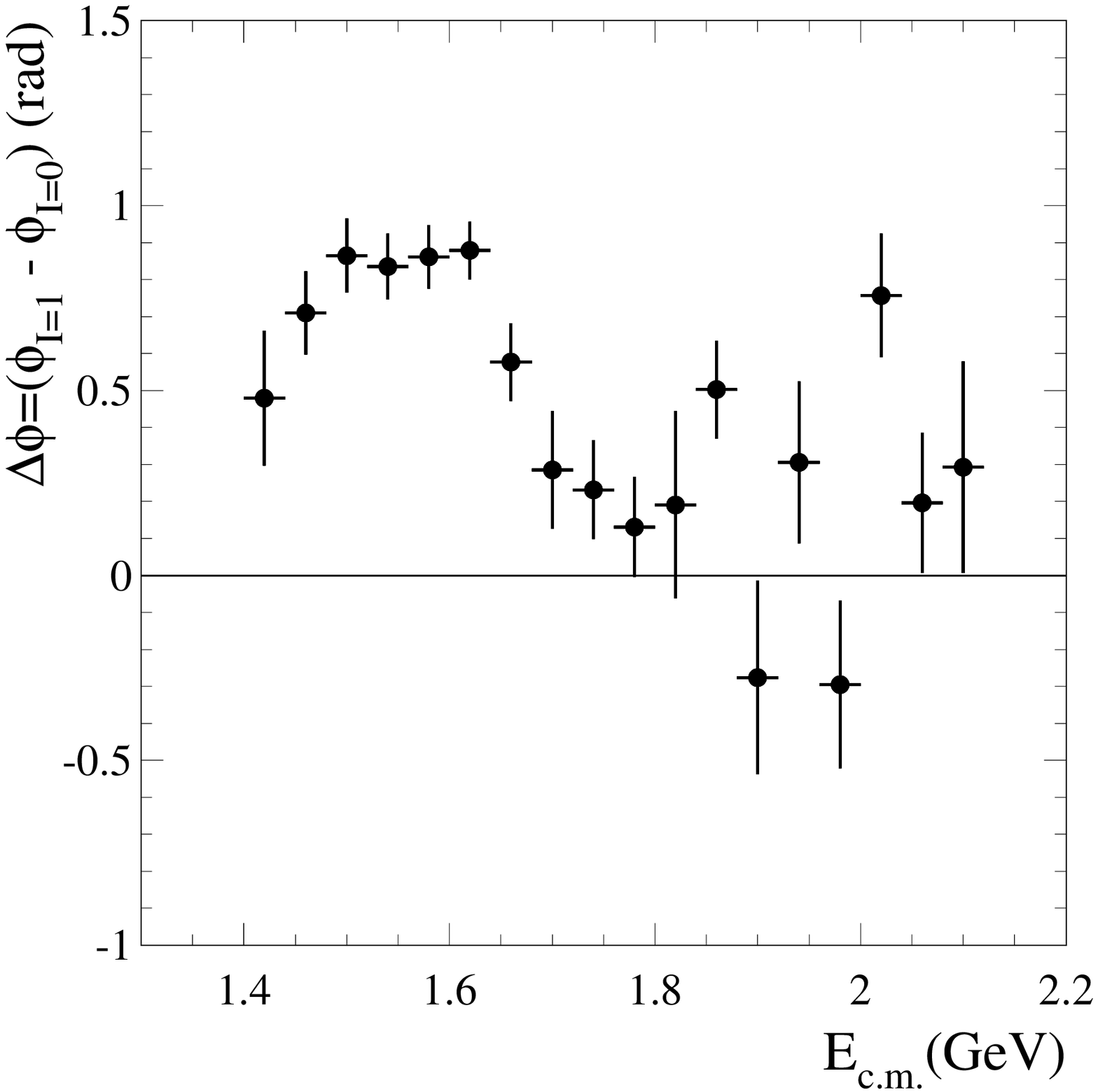}
\caption{\label{kskpi-dph}
  Relative phase between the isoscalar and isovector component of the 
  $\epem\to\K^*(892)\K$ cross section.}
\ec
\end{figure}

Figure~\ref{kskpi-low} shows, in the low-energy region,
the \kskpi Dalitz plot and the three possible projections.
The dominance of the $\K^*(892)$ in each $\K\pi$ channel appears evident.
As a result, in the following we will consider  
the process $\epem\to\K^*(892)K$ as the dominant one.
The different shapes observed for the charged (Fig.~\ref{kskpi-low}(b)) and 
neutral $\K^*(892)$ (Fig.~\ref{kskpi-low}(c)), are due
to interference effects between the isoscalar and isovector components
(see Sec.~\ref{subsec:kkstar}).

\begin{table*}[htb]
\caption{Isoscalar component of the cross section as a function of \Ecm for the process
$e^+e^-\to \kst K$. Errors are statistical only.}
\label{kskpi_890is}
\begin{ruledtabular}
\begin{tabular}{cccccccc}
$\Ecm (\gev)$ & $\sigma$ (nb)& 
$\Ecm (\gev)$ & $\sigma$ (nb)& 
$\Ecm (\gev)$ & $\sigma$ (nb)& 
$\Ecm (\gev)$ & $\sigma$ (nb)\\
\hline
 1.40$-$1.42 &  1.26 $\pm$  0.75 &  1.66$-$1.68 &  8.34 $\pm$  0.78 &  1.92$-$1.96 &  1.11 $\pm$  0.18 &  2.44$-$2.48 &  0.24 $\pm$  0.09 \\
 1.42$-$1.44 &  0.96 $\pm$  0.36 &  1.68$-$1.70 &  8.40 $\pm$  0.75 &  1.96$-$2.00 &  0.63 $\pm$  0.18 &  2.48$-$2.52 &  0.21 $\pm$  0.09 \\
 1.44$-$1.46 &  1.62 $\pm$  0.42 &  1.70$-$1.72 &  7.86 $\pm$  0.72 &  2.00$-$2.04 &  0.84 $\pm$  0.18 &  2.52$-$2.60 &  0.15 $\pm$  0.06 \\
 1.46$-$1.48 &  2.61 $\pm$  0.48 &  1.72$-$1.74 &  6.45 $\pm$  0.69 &  2.04$-$2.08 &  0.45 $\pm$  0.15 &  2.60$-$2.68 &  0.18 $\pm$  0.06 \\
 1.48$-$1.50 &  1.38 $\pm$  0.39 &  1.74$-$1.76 &  5.28 $\pm$  0.57 &  2.08$-$2.12 &  0.57 $\pm$  0.18 &  2.68$-$2.76 &  0.12 $\pm$  0.03 \\
 1.50$-$1.52 &  2.01 $\pm$  0.51 &  1.76$-$1.78 &  2.97 $\pm$  0.45 &  2.12$-$2.16 &  0.84 $\pm$  0.18 &  2.76$-$2.84 &  0.09 $\pm$  0.03 \\
 1.52$-$1.54 &  3.06 $\pm$  0.57 &  1.78$-$1.80 &  3.24 $\pm$  0.45 &  2.16$-$2.20 &  0.42 $\pm$  0.12 &  2.84$-$2.92 &  0.09 $\pm$  0.06 \\
 1.54$-$1.56 &  4.62 $\pm$  0.69 &  1.80$-$1.82 &  3.57 $\pm$  0.48 &  2.20$-$2.24 &  0.30 $\pm$  0.15 &  2.92$-$3.00 &  0.15 $\pm$  0.03 \\
 1.56$-$1.58 &  5.28 $\pm$  0.66 &  1.82$-$1.84 &  1.26 $\pm$  0.30 &  2.24$-$2.28 &  0.12 $\pm$  0.06 &  3.00$-$3.08 &  0.09 $\pm$  0.03 \\
 1.58$-$1.60 &  8.13 $\pm$  0.75 &  1.84$-$1.86 &  1.56 $\pm$  0.30 &  2.28$-$2.32 &  0.45 $\pm$  0.15 &  3.08$-$3.16 &  1.95 $\pm$  0.15 \\
 1.60$-$1.62 &  6.63 $\pm$  0.81 &  1.86$-$1.88 &  1.17 $\pm$  0.30 &  2.32$-$2.36 &  0.15 $\pm$  0.09 & & \\ 
 1.62$-$1.64 &  7.35 $\pm$  0.75 &  1.88$-$1.90 &  1.59 $\pm$  0.36 &  2.36$-$2.40 &  0.21 $\pm$  0.09 & & \\
 1.64$-$1.66 &  9.48 $\pm$  0.84 &  1.90$-$1.92 &  1.02 $\pm$  0.27 &  2.40$-$2.44 &  0.15 $\pm$  0.06 & & \\
\end{tabular}
\end{ruledtabular}
\end{table*}

\begin{table*}[htb]
\caption{Isovector component of the cross section as a function of \Ecm for the process
$e^+e^-\to\kst K$. Errors are statistical only.}
\label{kskpi_890iv}
\begin{ruledtabular}
\begin{tabular}{cccccccc}
$\Ecm (\gev)$ & $\sigma$ (nb)& 
$\Ecm (\gev)$ & $\sigma$ (nb)& 
$\Ecm (\gev)$ & $\sigma$ (nb)& 
$\Ecm (\gev)$ & $\sigma$ (nb)\\
\hline
 1.40$-$1.42 &  0.32 $\pm$  0.67 &  1.64$-$1.66 &  1.83 $\pm$  0.37 &  1.88$-$1.90 &  0.59 $\pm$  0.26 &  2.28$-$2.32 &  0.25 $\pm$  0.09 \\
 1.42$-$1.44 &  0.68 $\pm$  0.40 &  1.66$-$1.68 &  2.82 $\pm$  0.50 &  1.90$-$1.92 &  0.26 $\pm$  0.14 &  2.32$-$2.36 &  0.41 $\pm$  0.10 \\
 1.44$-$1.46 &  1.40 $\pm$  0.43 &  1.68$-$1.70 &  2.10 $\pm$  0.36 &  1.92$-$1.94 &  0.51 $\pm$  0.17 &  2.36$-$2.40 &  0.38 $\pm$  0.13 \\
 1.46$-$1.48 &  1.27 $\pm$  0.39 &  1.70$-$1.72 &  1.28 $\pm$  0.30 &  1.92$-$1.96 &  0.40 $\pm$  0.12 &  2.40$-$2.44 &  0.21 $\pm$  0.11 \\
 1.48$-$1.50 &  2.33 $\pm$  0.50 &  1.72$-$1.74 &  1.53 $\pm$  0.35 &  1.96$-$2.00 &  0.75 $\pm$  0.16 &  2.44$-$2.48 &  0.13 $\pm$  0.05 \\
 1.50$-$1.52 &  2.96 $\pm$  0.58 &  1.74$-$1.76 &  1.60 $\pm$  0.32 &  2.00$-$2.04 &  0.60 $\pm$  0.20 &  2.48$-$2.52 &  0.10 $\pm$  0.05 \\
 1.52$-$1.54 &  2.83 $\pm$  0.52 &  1.76$-$1.78 &  1.44 $\pm$  0.32 &  2.04$-$2.08 &  0.93 $\pm$  0.18 &  2.52$-$2.60 &  0.27 $\pm$  0.10 \\
 1.54$-$1.56 &  3.08 $\pm$  0.55 &  1.78$-$1.80 &  1.34 $\pm$  0.32 &  2.08$-$2.12 &  0.78 $\pm$  0.18 &  2.60$-$2.68 &  0.05 $\pm$  0.03 \\
 1.56$-$1.58 &  2.05 $\pm$  0.42 &  1.80$-$1.82 &  0.58 $\pm$  0.18 &  2.12$-$2.16 &  0.55 $\pm$  0.16 &  2.68$-$2.76 &  0.11 $\pm$  0.04 \\
 1.58$-$1.60 &  1.67 $\pm$  0.33 &  1.82$-$1.84 &  1.44 $\pm$  0.31 &  2.16$-$2.20 &  0.47 $\pm$  0.13 &  2.76$-$2.84 &  0.11 $\pm$  0.03 \\
 1.60$-$1.62 &  3.46 $\pm$  0.68 &  1.84$-$1.86 &  1.02 $\pm$  0.24 &  2.20$-$2.24 &  0.45 $\pm$  0.17 &  2.84$-$2.92 &  0.04 $\pm$  0.01 \\
 1.62$-$1.64 &  1.52 $\pm$  0.38 &  1.86$-$1.88 &  0.96 $\pm$  0.29 &  2.24$-$2.28 &  0.45 $\pm$  0.12 &  2.92$-$3.08 &  0.03 $\pm$  0.01 \\
\end{tabular}
\end{ruledtabular}
\end{table*}

Figures~\ref{kskpi-i0} and~\ref{kskpi-i1} show the isoscalar and isovector 
cross sections; numerical  values are reported  in Tables~\ref{kskpi_890is} 
and~\ref{kskpi_890iv} respectively. Figure~\ref{kskpi-dph} shows the
relative phase between the two amplitudes.
In addition to the errors reported in Table~\ref{tab:error-kskpi} we must
consider the 15\% systematic uncertainty associated with the Dalitz plot-fit
procedure. This quantity has been evaluated by studying the effects
of different parameterizations for the $BW$ distributions which describe 
the intermediate $K^*$ resonances in the amplitudes of Eq.~\ref{zemach-eqn}.
The isoscalar contribution, see Fig.~\ref{kskpi-i0}, is dominant 
with respect to the isovector, (Fig.~\ref{kskpi-i1}) and clearly shows a
resonant behavior. Moreover, the isovector cross section 
seems to be compatible
with a pure phase space shape. However, using the complete information, 
including the relative phase, shown in Fig.~\ref{kskpi-dph}, we also find that 
the isovector component has a resonant structure (for details see Sec.~\ref{subsec:kkstar}).

\begin{figure}[htb]
\bc
\includegraphics[width=\columnwidth]{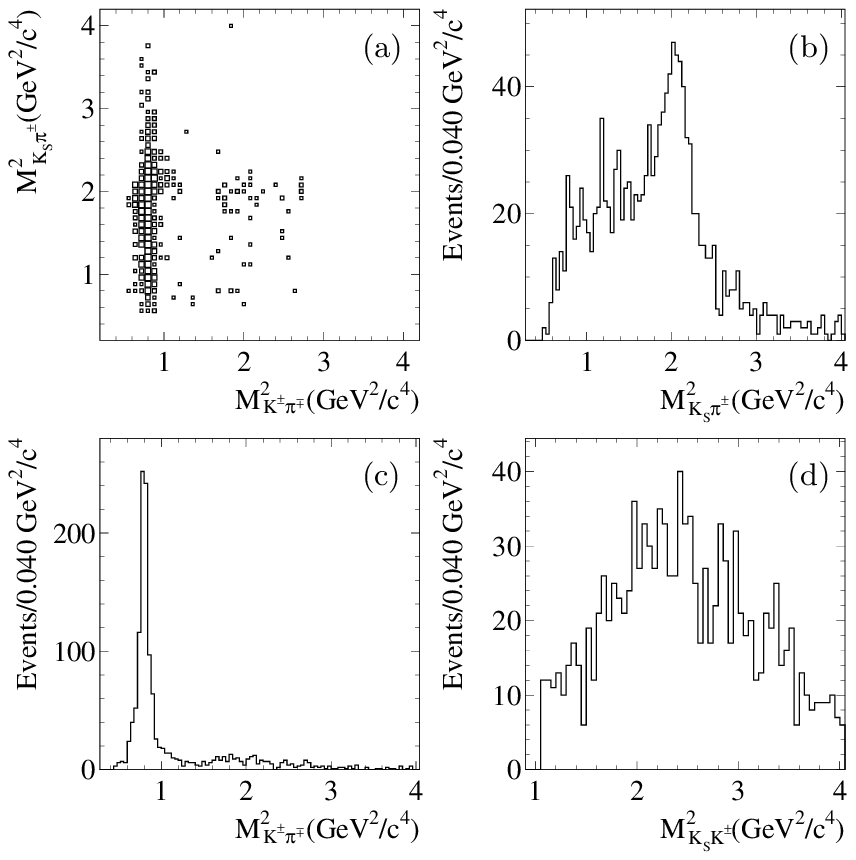}
\caption{\label{kskpi-hi}
  (a) Dalitz plot distributions for \kskpi events with an invariant mass 
  $m_{\KS\K\pi} > 2.0\;\gevcc$. 
  (b) \KS\pipm projection, with the broad charged $\K_2^*(1430)$ peak,
  (c) \Kpm\pimp projection, with the narrow neutral $\K^*(892)$ peak.
  (d) \KS\Kpm projection.}
\ec
\end{figure}
\begin{figure}[htb]
\bc
\includegraphics[width=\columnwidth]{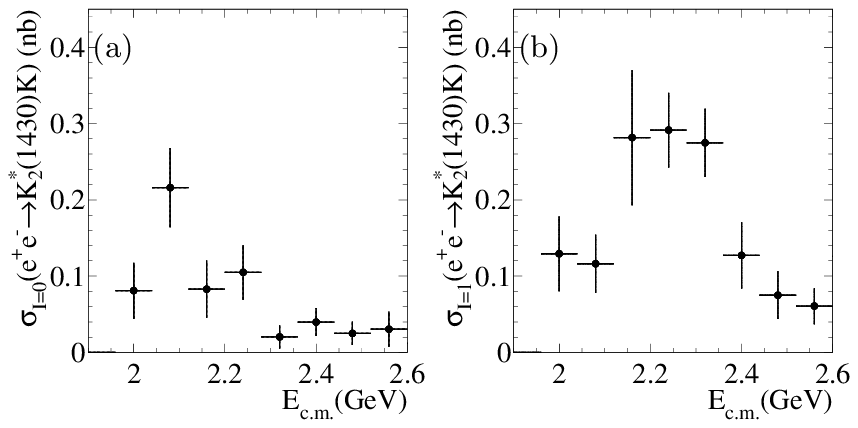}
\caption{\label{kskpi-comp4}
  Isoscalar (a) and isovector (b) components of the cross section for the process
  $\epem\to\K^*_2(1430)\K$.}
\ec
\end{figure}
Figure~\ref{kskpi-hi} shows the Dalitz plot and the projections
obtained in the c.m. energy region above $2.0\;\gev$. In this case, 
in addition to the $\K\K^*(892)$, we have evidence that  
the $\K\K^*_2(1430)$ intermediate state also contributes. 
In particular, as theoretically predicted~\cite{carlsmith}, 
we observe that while 
the neutral $\K^*(892)$ dominates the \Kpm\pimp projection (Fig.~\ref{kskpi-hi}(c)),
the charged $\K_2^*(1430)$ is the main contribution in the \KS\pipm 
projection (Fig.~\ref{kskpi-hi}(b)). This can be interpreted as a different
spatial structure of these mesons: $\K^*(892)$ is a vector, 
while $\K^*_2(1430)$ is a tensor. 
\indent
The Dalitz plot analysis in this energy range 
is performed with four isospin amplitudes, an 
isoscalar and an isovector component for each 
intermediate two-body channel. 
Figures~\ref{kskpi-comp4}(a) and \ref{kskpi-comp4}(b) show the isoscalar and 
isovector components of the $\K^*_2(1430)\K$ channel. The corresponding
values are reported in Tables~\ref{kskpi_1430is} 
and~\ref{kskpi_1430iv} respectively. 
The structure  observed at $\Ecm\simeq 2.1\;\gev$ 
in the isoscalar and $\Ecm\simeq 2.25\;\gev$ in the isovector components
may be interpreted as pure threshold effects. Indeed,
the transition form factors, obtained by taking the square root of
the cross sections divided by the appropriate phase space 
(a pseudoscalar and a tensor meson in the final state), 
have a steep yet smooth behavior, as shown in Fig.~\ref{k1430tff-i0i1}.
The simplest interpretation calls for transition form factors that collect  
the main contributions from low-energy resonances lying below threshold.
Further study and higher statistics would help to clarify this item.

\section{The \jpsi region}
\label{sec:jpsi}

Clean \jpsi signals are observed in almost all the final states under study, 
from which we extract the corresponding decay rates.
Figure~\ref{ksk-jpsi}(a) shows the inclusive distribution in the \jpsi mass region for
the \kskpi data sample.
A fit with a Gaussian for the \jpsi peak and a linear function for the 
continuum, gives $\sigma(M_{\KS K\pi})=6.5~\mevcc$ and less than 1~\mevcc
difference from the PDG~\cite{pdg} value for the \jpsi mass.
To account also for possible radiative or experimental effects
we determine the yields under the \jpsi peak by fitting with a
double-Gaussian.
As shown by the \DP\ analysis, the $\epem\to\K\K\pi$ reactions proceed 
mainly through 
formation of the intermediate \K\Kst state. We could therefore extract 
the $\jpsi\to\Kst\K$
decay rates, where the \kst is charged ($\kstc\to\Kpm\piz$) for the \kkpiz 
final state  and neutral $\kstz\to\Kpm\pimp$ or charged ($\kstc\to\KS\pipm$) 
for the \kskpi final state.  
In order to isolate the different \kst contributions, we require $E_\K> 1.4\;\gev$
in the $\K\K\pi$ c.m. frame for the kaon not coming from the \kst decay,
as expected  for a two-body \jpsi decay.
A check of the \DP\ provides an estimate of the residual contamination from wrong 
$\K\pi$ combinations, which is found to be less than 3.0\%.
From the fit to  the mass distributions obtained for the  
different $\kst\K$ final states, shown in Fig.~\ref{ksk-jpsi}(b-d), 
we derive the number of events reported in Table~\ref{tab:jpsi}, 
together with the corresponding \jpsi branching fractions multiplied by
the $ \jpsi \rightarrow \epem$  width.
The latter quantity has been obtained through
\begin{equation}
  \mathcal{B}^{J/\psi}_{f}\cdot\Gamma^{J/\psi}_{ee}
  = \frac{N(J/\psi\to f)\cdot m_{J/\psi}^2}%
           {6\pi^2\cdot d{\cal L}/dE\cdot\epsilon\cdot C}~,
\end{equation}
where $\mathcal{B}^{J/\psi}_{f}$ is the branching fraction for
$J/\psi\to f$,
$d{\cal L}/dE = 65.6~\invnb/\mev$ is the ISR luminosity
integrated at the \jpsi mass, 
$\epsilon$ 
is the detection efficiency obtained from simulation,
and  $C \equiv (hc/2\pi)^2= 3.894\times 10^{11}~\nb\mev^2$ is a
conversion constant.
The systematic errors have been estimated by considering, in addition to the values 
reported in Tables~\ref{tab:error-kskpi} and \ref{tab:error-kkpi0}, 
the statistical error on the reconstruction efficiency.
Assuming the world average value of the $\jpsi \rightarrow\epem$ width, 
$\Gamma^\jpsi_{ee} =5.55\pm0.14~\kev$~\cite{pdg}, we extract the
decay rates shown in the last column of Table~\ref{tab:jpsi}. 
Here and in the following the first uncertainty reported is statistical 
and the second is systematic.
From these results and the known \kst decay rates, we 
can derive the $\jpsi\to\kstz\overline{K}^{0},\K^{*+}(892)\Km$ 
branching fractions,  summarized in Table~\ref{tab:jpsi2}.

\begin{table*}[htb]
\caption{Isoscalar component of the cross section as a function of \Ecm for the process
$e^+e^-\to K_2^*(1430)K$. Errors are statistical only.}
\label{kskpi_1430is}
\begin{ruledtabular}
\begin{tabular}{cccccccc}
$\Ecm (\gev)$ & $\sigma$ (nb)& 
$\Ecm (\gev)$ & $\sigma$ (nb)& 
$\Ecm (\gev)$ & $\sigma$ (nb)& 
$\Ecm (\gev)$ & $\sigma$ (nb)\\
\hline
 1.92$-$1.96 &  0.069 $\pm$  0.036 &  2.16$-$2.20 &  0.252 $\pm$  0.100 &  2.40$-$2.44 &  0.014 $\pm$  0.007 &  2.76$-$2.84 &  0.009 $\pm$  0.009 \\
 1.96$-$2.00 &  0.048 $\pm$  0.033 &  2.20$-$2.24 &  0.115 $\pm$  0.060 &  2.44$-$2.48 &  0.128 $\pm$  0.066 &  2.84$-$2.92 &  0.022 $\pm$  0.012 \\
 2.00$-$2.04 &  0.083 $\pm$  0.051 &  2.24$-$2.28 &  0.059 $\pm$  0.037 &  2.48$-$2.52 &  0.008 $\pm$  0.004 &  2.92$-$3.00 &  0.065 $\pm$  0.019 \\
 2.04$-$2.08 &  0.261 $\pm$  0.088 &  2.28$-$2.32 &  0.034 $\pm$  0.023 &  2.52$-$2.60 &  0.030 $\pm$  0.031 &  3.00$-$3.08 &  0.044 $\pm$  0.025 \\
 2.08$-$2.12 &  0.172 $\pm$  0.089 &  2.32$-$2.36 &  0.045 $\pm$  0.028 &  2.60$-$2.68 &  0.012 $\pm$  0.012 &  & \\
 2.12$-$2.16 &  0.018 $\pm$  0.018 &  2.36$-$2.40 &  0.063 $\pm$  0.055 &  2.68$-$2.76 &  0.003 $\pm$  0.003 &  & \\
\end{tabular}
\end{ruledtabular}
\end{table*}

\begin{table*}[htb]
\caption{Isovector component of the cross section as a function of \Ecm for the process
$e^+e^-\to K_2^*(1430)K$. Errors are statistical only.}
\label{kskpi_1430iv}
\begin{ruledtabular}
\begin{tabular}{cccccccc}
$\Ecm (\gev)$ & $\sigma$ (nb)& 
$\Ecm (\gev)$ & $\sigma$ (nb)& 
$\Ecm (\gev)$ & $\sigma$ (nb)& 
$\Ecm (\gev)$ & $\sigma$ (nb)\\
\hline
 1.96$-$2.00 &  0.10 $\pm$  0.07 &  2.20$-$2.24 &  0.21 $\pm$  0.07 &  2.44$-$2.48 &  0.06 $\pm$  0.03 &  2.84$-$2.92 &  0.04 $\pm$  0.02 \\
 2.00$-$2.04 &  0.23 $\pm$  0.09 &  2.24$-$2.28 &  0.42 $\pm$  0.10 &  2.48$-$2.52 &  0.02 $\pm$  0.01 &  2.92$-$3.00 &  0.05 $\pm$  0.02 \\
 2.04$-$2.08 &  0.08 $\pm$  0.05 &  2.28$-$2.32 &  0.32 $\pm$  0.11 &  2.52$-$2.60 &  0.06 $\pm$  0.03 &  3.00$-$3.08 &  0.03 $\pm$  0.01 \\
 2.08$-$2.12 &  0.19 $\pm$  0.09 &  2.32$-$2.36 &  0.15 $\pm$  0.07 &  2.60$-$2.68 &  0.13 $\pm$  0.03 &  & \\
 2.12$-$2.16 &  0.33 $\pm$  0.10 &  2.36$-$2.40 &  0.09 $\pm$  0.05 &  2.68$-$2.76 &  0.08 $\pm$  0.03 &  & \\
 2.16$-$2.20 &  0.27 $\pm$  0.10 &  2.40$-$2.44 &  0.18 $\pm$  0.07 &  2.76$-$2.84 &  0.03 $\pm$  0.02 &  & \\
\end{tabular}
\end{ruledtabular}
\end{table*}
\begin{figure}[htb]
\includegraphics[width=.9\columnwidth]{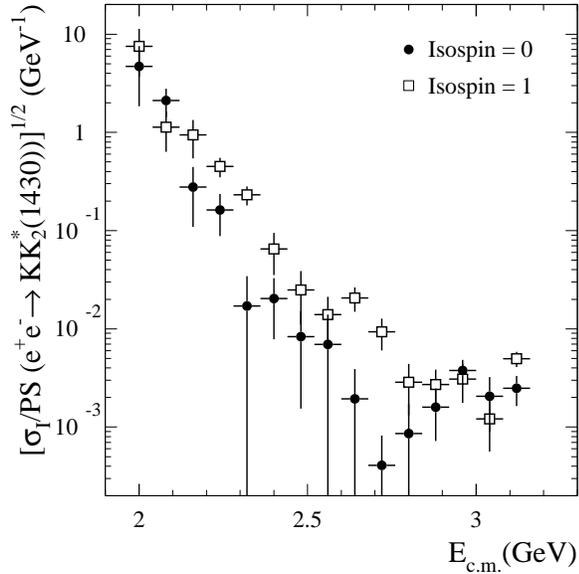}
\caption{\label{k1430tff-i0i1}%
The isospin amplitudes (defined as the square root of the isospin components divided by the 
phase space) for the \kskpi channel, obtained by fitting the Dalitz plot
in the region $\Ecm > 2.0\;\gev$, in the case of  $K^*_2(1430)K$ intermediate
state.}
\end{figure}
%
%
\begin{figure}[htb]
\bc
\includegraphics[width=\columnwidth]{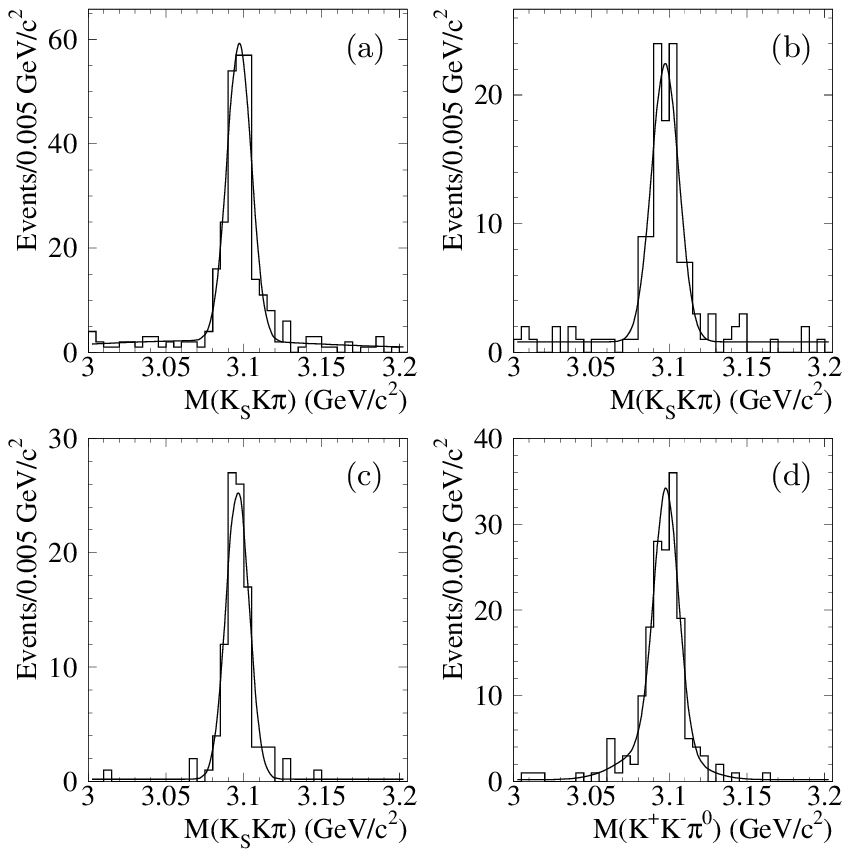}
\caption{\label{ksk-jpsi}
 (a) \kskpi mass distribution in the \jpsi mass region. 
 (b-d) Mass distributions of selected events for: 
  (b) $\kstz\KS$ $(\kstz\to\Kpm\pimp)$;  c) $\kstc\Kmp$ $(\kstc\to\KS\pipm)$ and
  (d) $\kstc\Kmp$ $(\kstc\to\Kpm\piz)$. The curves represent
  the fits obtained using a sum of two Gaussians and a linear function.} 
\ec
\end{figure}
\begin{table}[htb]
\caption{\label{tab:jpsi}%
Branching ratios of the $\jpsi\to KK\pi$ decays. 
The value of $\Gamma_{ee}^\jpsi$ is taken from Ref.~\cite{pdg}.}
\begin{ruledtabular}
\begin{tabular}{cccc}
Final state, $f$          &   Yield          &
$\mathcal{B}^{J/\psi}_{f}\Gamma_{ee}^\jpsi (\ev)$ & $\mathcal{B}^{J/\psi}_{f}\cdot 10^{3}$\\ 
\hline 
$(K^{\pm}\pi^{\mp})K_S$   & $94 \!\pm\! 9$   &
$8.85 \pm 0.85 \pm 0.50$                          & $1.59\!\pm\!0.16 \!\pm\!0.09$     \\
$(K_S\pi^{\pm})K^{\mp}$   & $89 \!\pm\! 9$   &
$8.38 \pm0.85 \pm0.50$                            & $1.51 \!\pm\!0.16 \!\pm\!0.09$    \\
$(K^\pm\pi^0) K^\mp$      & $155 \!\pm\! 12$ & 
$10.96 \!\pm\!0.85 \!\pm\!0.70$                   & $1.97 \!\pm\!0.16 \!\pm\!0.13$    \\
\end{tabular}
\end{ruledtabular}
\end{table}

\begin{table}[htb]
\caption{\label{tab:jpsi2}%
Branching ratios of the $\jpsi\to K^*K$ decay ($\Gamma_{ee}^\jpsi$
from Ref.~\cite{pdg} and c.c. stands for charge conjugate).}
\begin{ruledtabular}
\begin{tabular}{cccc}
\multirow{2}{*}{Final state, $f$} & \multirow{2}{*}{$\mathcal{B}^{J/\psi}_{f}\Gamma_{ee}^\jpsi (\ev)$} & 
\multirow{2}{*}{$\mathcal{B}^{J/\psi}_{f}\!\!\cdot{10^{3}}$} &  $\mathcal{B}^{J/\psi}_{f}\cdot{10^{3}}$ \\ 
& & & (Ref.~\cite{pdg})\\
\hline
$(K^{*+}K^{-})$+c.c.             & $29.0\!\pm\!1.7\!\pm\!1.3$                                 & 
$5.2 \!\pm\!0.3 \!\pm\!0.2$             & $5.0 \!\pm\! 0.4$ \\
$(K^{*0}\overline{K}^{0})$+c.c.  & $26.6\!\pm\!2.5\!\pm\!1.5$                                 & 
$4.8 \!\pm\!0.5 \!\pm\!0.3$             & $4.2 \!\pm\! 0.4$ \\
\end{tabular}
\end{ruledtabular}
\end{table}


\section{The \phipiz FINAL STATE}
\label{sec:phipiz}

\begin{table*}
\caption{Measurement of the $\epem\to\phipiz$ cross section as a function of \Ecm. 
  Errors are statistical only.}
\label{phipi0_tab}
\begin{ruledtabular}
\begin{tabular}{ c c c c c c c c}
$\Ecm (\gev)$ & $\sigma$ (nb)& 
$\Ecm (\gev)$ & $\sigma$ (nb)& 
$\Ecm (\gev)$ & $\sigma$ (nb)& 
$\Ecm (\gev)$ & $\sigma$ (nb)\\
\hline
 1.20$-$1.30 &  0.014 $\pm$  0.016 &  1.55$-$1.60 &  0.191 $\pm$  0.060 &  1.75$-$1.80 &  0.015 $\pm$  0.015 &  1.95$-$2.10 &  0.002 $\pm$  0.004 \\
 1.30$-$1.40 &  0.025 $\pm$  0.018 &  1.60$-$1.65 &  0.087 $\pm$  0.039 &  1.80$-$1.85 &  0.046 $\pm$  0.027 &  2.10$-$2.30 &  0.006 $\pm$  0.005 \\
 1.40$-$1.50 &  0.033 $\pm$  0.022 &  1.65$-$1.70 &  0.072 $\pm$  0.035 &  1.85$-$1.90 &  0.093 $\pm$  0.038 &  2.30$-$2.60 &  0.002 $\pm$  0.002 \\
 1.50$-$1.55 &  0.073 $\pm$  0.038 &  1.70$-$1.75 &  0.100 $\pm$  0.041 &  1.90$-$1.95 &  0.089 $\pm$  0.036 &  & \\
\end{tabular}
\end{ruledtabular}
\end{table*}


The event selection for \phipiz production closely follows the procedure 
described  for \kkpiz with non-resonant \kk pairs.
We tighten a few cuts in order to remove random photon combinations 
that could fake a \piz.
In particular we lowered to 300~\mev the cut on 
the additional detected neutral energy, and we require
that the direction of the observed ISR photon matches that predicted 
by the one-constraint fit within 12~mrad.

Figure~\ref{fig:2d_phipi0_20_feb} shows the \kk\ mass, $M_{KK}$, 
versus the $\gamma\gamma$ mass, $M_{\gamma\gamma}$. 
The plot  is divided into nine equally sized regions, with the central one, 
which corresponds to $1.010 < M_{KK} < 1.030~\gevcc$ 
and $0.110 < M_{\gamma\gamma} < 0.160~\gevcc$, being the signal region.
We extract the number of signal $S$ and background $B$ events 
contained in the central cell, by solving the following linear equation system:
\begin{eqnarray}
  N_0 &=& S + B = 76 \\ \nonumber
  N_1 &=& 0.135\cdot S +2\cdot B = 51 \, ,
\end{eqnarray}
where $N_0$ and $N_1$ are the yields 
in the central cell and the sum of yields in adjacent cells 
({\it i.e.} the cells numbered as 2, 4, 6 and 8 in Fig.~\ref{fig:2d_phipi0_20_feb}), 
respectively. 
The fractions of signal events contained in the central and adjacent cells have been determined 
from MC simulation, while the factor of 2 multiplying $B$ in the second equation is 
a consequence of the fact that the background is linear with respect to both
$\gamma\gamma$ and $\K\K$ masses.
We find 54 signal and 22 background events.

\begin{figure}[htb!]
\includegraphics[width=.9\columnwidth]{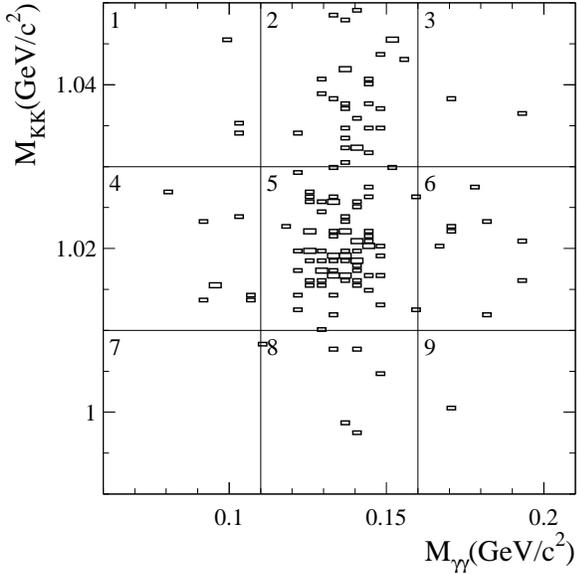}
\caption{The 2-dimensional \Kp\Km, $\gamma\gamma$ invariant mass distribution around 
the $\phi$ and \piz masses. The central cell identifies the signal region.}  
\label{fig:2d_phipi0_20_feb}
\end{figure}

\begin{figure}[htb!]
\includegraphics[width=\columnwidth]{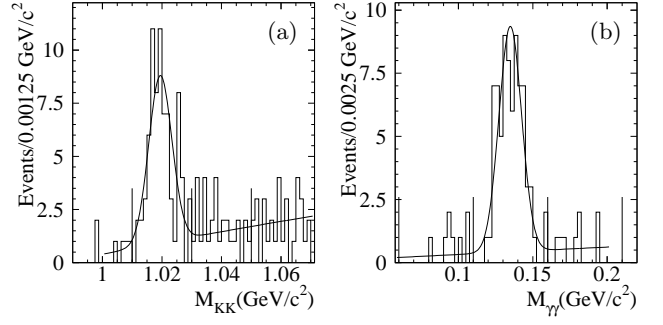}
\caption{
The \Kp\Km (a) and $\gamma\gamma$ (b) invariant mass distributions.
Signal and side-band regions are shown as vertical lines in the plots.
}  
\label{fig:metaphi_phipi0_20_feb}
\end{figure}

The simulation of specific reactions with a real $\phi$ and/or \piz in the   
final state, like $\epem\to\phi\gamma$, $\epem\to\phi\piz\piz\gamma$,  
$\epem\to\K\K\piz\piz\gamma$ (non-resonant $\K\K$ production),
and  $q\bar{q}$, $\tau\tau$ production from continuum, shows that their
contribution is  negligible. 
We can thus conclude that the only sources of background are represented by
the non-resonant $\K\K\piz\gamma$ final state,
and by a number of ISR processes with a true $\phi$ 
and a random $\gamma\gamma$ combination.
Figure~\ref{fig:metaphi_phipi0_20_feb}(a) shows the $\phi$ signal 
in the \kk\ invariant mass, after the cut around the \piz mass is applied; 
viceversa, Fig.~\ref{fig:metaphi_phipi0_20_feb}(b) shows the 
\piz signal once a $\phi$ is selected. 
Fitting these distributions with a Gaussian and a linear background 
we estimate the two background sources to be $15\pm 2$ and $7\pm3$ events, 
respectively.

\begin{figure}[h!]
\includegraphics[width=.9\columnwidth]{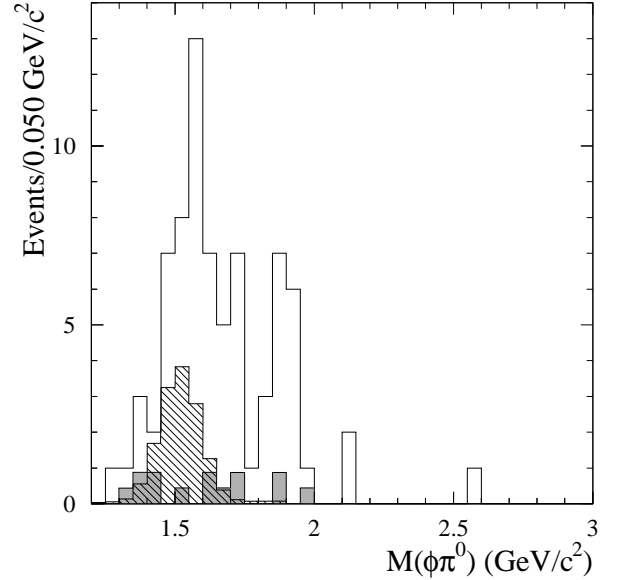}
\caption{The invariant mass distribution of the \phipiz system (open histogram).
 The hatched and gray histograms show the background from non-$\phi$ $\kkpiz$ 
and from $\gamma\gamma$ combinatorial.}
\label{fig:phipi0_mass}
\end{figure}

The  mass spectra for the data and the estimated backgrounds are 
shown in Fig.~\ref{fig:phipi0_mass}.
The mass spectrum for the $\K\K\piz$ background is obtained from the
simulated sample, while the spectrum for the other ISR backgrounds 
is taken directly from the $\piz$ sidebands.


%
The $\epem\to \phi \pi^0$ process is simulated considering only 
$\phi\!\to\!\kk$ decays, so that the 
 cross section as a function of the effective c.m. energy is given by
\begin{equation}
  \sigma_{\epem\!\!\!\to\!\phipiz}(\!\Ecm\!)
  = \frac{dN_{\phi\pi^0\gamma}(\!\Ecm\!)}
         {d{\cal L}(\!\Ecm\!) \
          \epsilon(\!\Ecm\!)
          {\cal B}(\phi\!\to\!\kk)}\ ,
\label{eq:phipi0}
\end{equation}
where \Ecm is the invariant mass of 
the $\phi \pi^0$ system; $dN_{\phi\pi^0\gamma}$ is the number of selected
$\phi\pi^0$ events after background subtraction in the interval 
$d\Ecm$, $\epsilon(\Ecm)$ is the detection
efficiency obtained from MC simulation (see Fig.~\ref{fig:phipi0_eff}),
and ${\cal B}(\phi\!\to\!\kk)$ is the world average value 
of the $\phi\!\to\!\kk$ decay rate~\cite{pdg}.
The energy behavior of the cross section is shown in 
Fig.~\ref{fig:phipi0_cs} and reported in Table~\ref{phipi0_tab}.
The cross section has a maximum of about 0.2~nb around
1.6~\gevcc.

\begin{figure}[h!]
\includegraphics[width=.9\columnwidth]{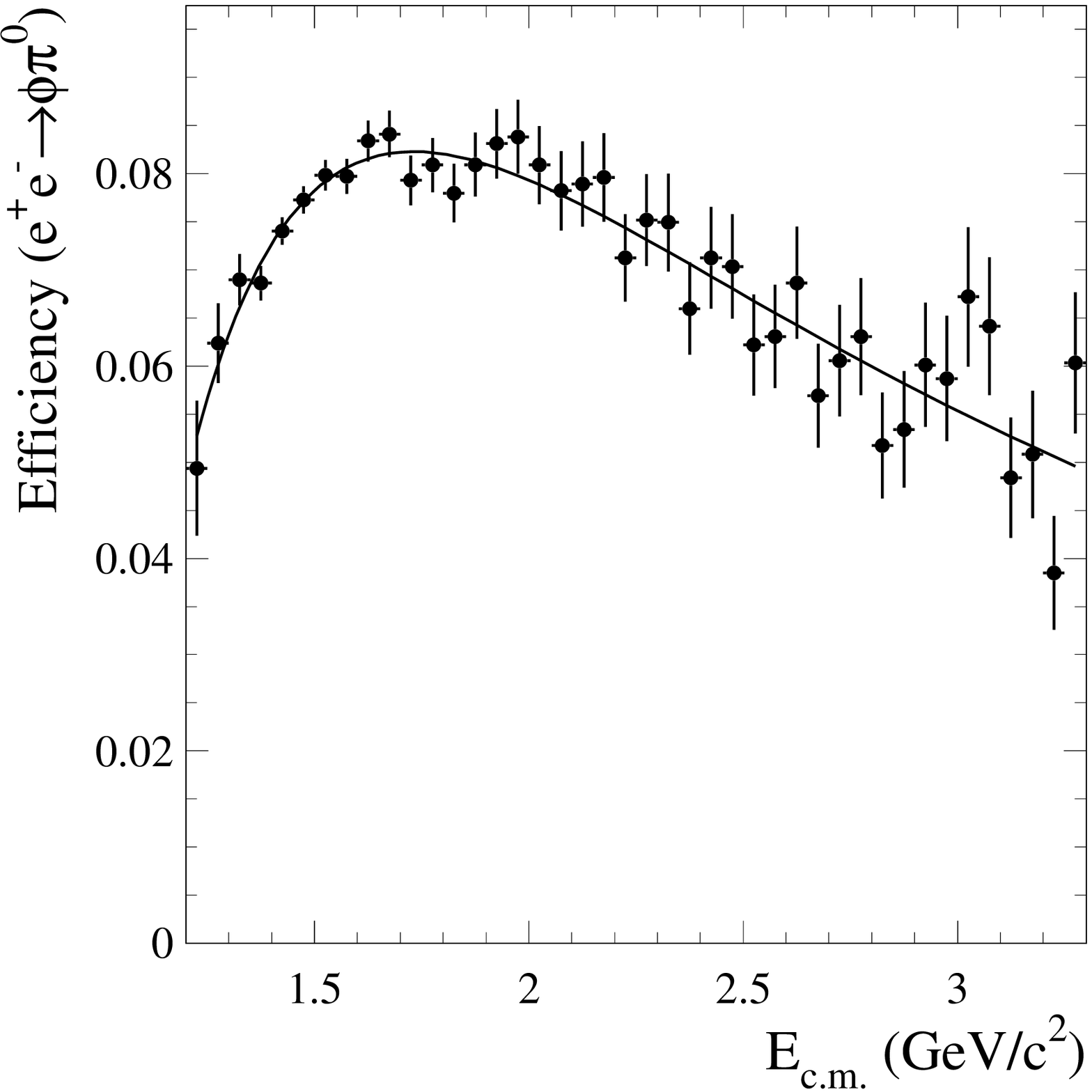}
\caption{Detection and reconstruction efficiency as a function 
  of the c.m. energy for the \phipiz final state.
  The solid line is the result of a fit to the data with the function 
  $a\left( 1 - (x/b)^p\cdot e^{\xi (b-x)}\right)$, where $x=\Ecm$.}
\label{fig:phipi0_eff}
\end{figure}

\begin{figure}[h!]
\includegraphics[width=.9\columnwidth]{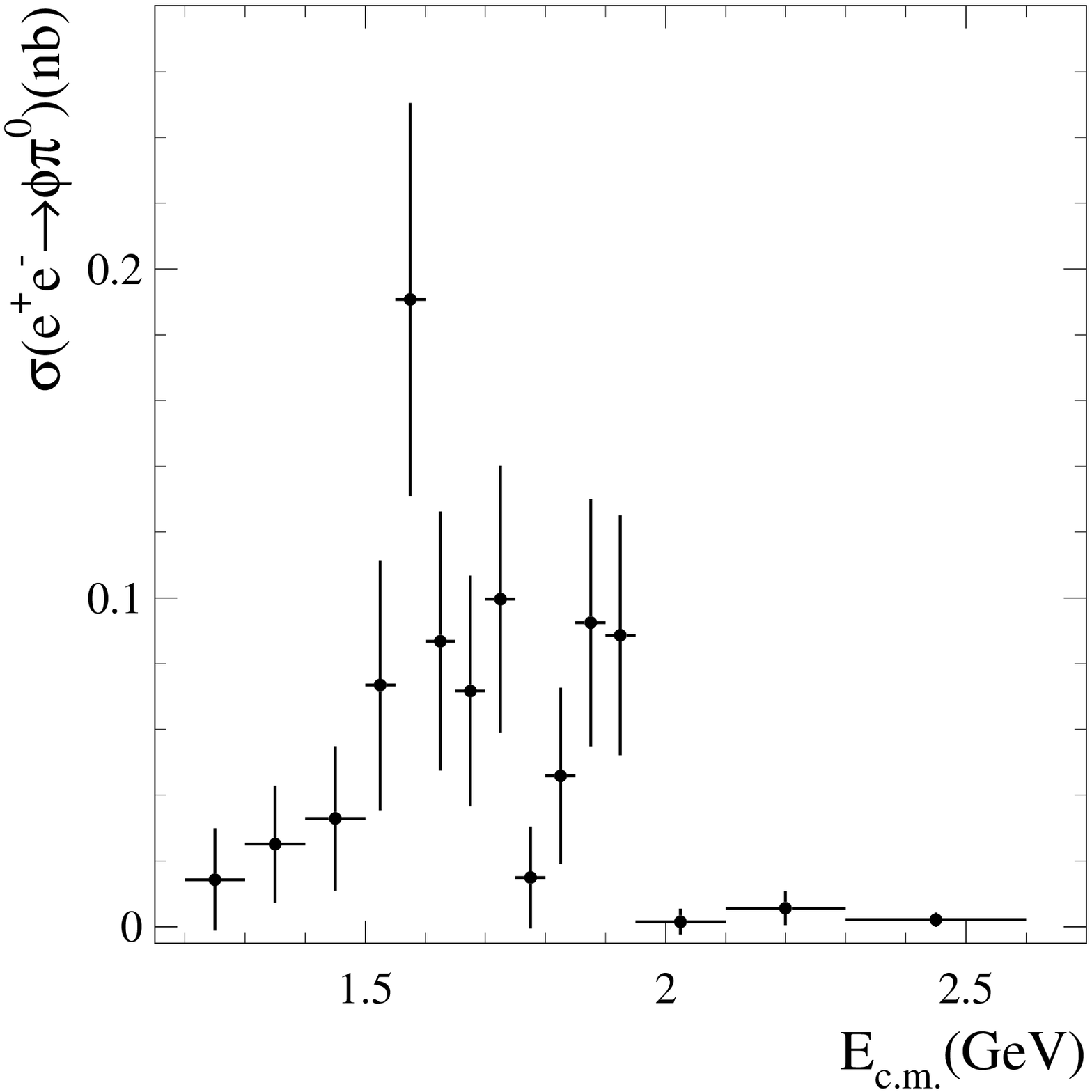}
\caption{The $\epem\to\phipiz$ cross section.}
\label{fig:phipi0_cs}
\end{figure}

The efficiency corrections and the 
systematic uncertainties which affect the \phipiz cross section measurement
are similar to those of the \kkpiz reaction. However, the uncertainties on 
background subtraction are larger, and we have to account also for the 
1.2\% uncertainty on the $\phi\to\kk$ branching fraction.
The total systematic error associated to the cross section measurement 
is therefore $\pm8.7\%$.


\section{The \kketa FINAL STATE}
\label{sec:phieta}

\begin{table*}
\caption{Measurement of the $\epem\to\eta \Kp\Km$(NOT$\phi$) cross section as a function of \Ecm. 
Errors are statistical only.}
\label{phieta_notphi_tab}
\begin{ruledtabular}
\begin{tabular}{ c c c c c c c c}
$\Ecm (\gev)$ & $\sigma$ (nb)& 
$\Ecm (\gev)$ & $\sigma$ (nb)& 
$\Ecm (\gev)$ & $\sigma$ (nb)& 
$\Ecm (\gev)$ & $\sigma$ (nb)\\
\hline
 1.64$-$1.74 &  0.08 $\pm$  0.04 &  2.24$-$2.34 &  0.02 $\pm$  0.01 &  2.72$-$2.76 &  0.02 $\pm$  0.02 &  2.96$-$3.00 &  0.03 $\pm$  0.02 \\
 1.74$-$1.84 &  0.01 $\pm$  0.03 &  2.34$-$2.44 &  0.01 $\pm$  0.01 &  2.76$-$2.80 &  0.01 $\pm$  0.01 &  3.00$-$3.04 &  0.01 $\pm$  0.01 \\
 1.84$-$1.94 &  0.06 $\pm$  0.02 &  2.44$-$2.54 &  0.03 $\pm$  0.01 &  2.80$-$2.84 &  0.01 $\pm$  0.01 &  3.04$-$3.08 &  0.01 $\pm$  0.01 \\
 1.94$-$2.04 &  0.03 $\pm$  0.02 &  2.54$-$2.64 &  0.01 $\pm$  0.01 &  2.84$-$2.88 &  0.01 $\pm$  0.02 &  3.08$-$3.10 &  0.27 $\pm$  0.09 \\
 2.04$-$2.14 &  0.03 $\pm$  0.01 &  2.64$-$2.68 &  0.01 $\pm$  0.01 &  2.88$-$2.92 &  0.01 $\pm$  0.01 &  3.10$-$3.12 &  0.30 $\pm$  0.09 \\
 2.14$-$2.24 &  0.00 $\pm$  0.01 &  2.68$-$2.72 &  0.01 $\pm$  0.02 &  2.92$-$2.96 &  0.02 $\pm$  0.02 &  3.12$-$3.14 &  0.11 $\pm$  0.05 \\
\end{tabular}
\end{ruledtabular}
\end{table*}

We analyze \phieta and \kketa (with non-resonant \kk\ pairs) 
final states separately, selecting
samples  with an invariant mass of the \kk\ system below and above  
1.045~\gevcc: $\phi$ and NOT-$\phi$ events, respectively.
In both cases, because of the cleanliness  of the $\eta$ signal, 
no additional cuts on photons or extra neutral energy are required. 
The \kk\ and $\gamma\gamma$ invariant mass distributions for selected events are shown
in Fig.~\ref{fig:phieta_spectra}. 
\begin{figure}[htb]
\includegraphics[width=\columnwidth]{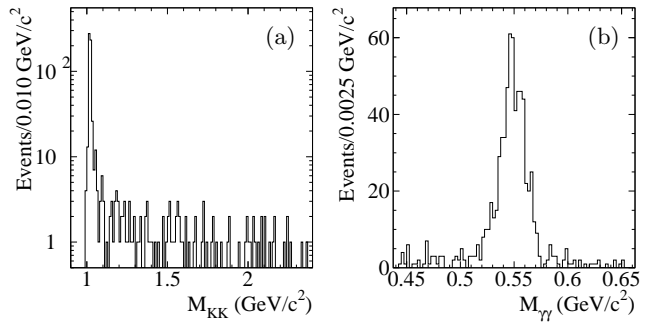}
\caption{The \kk\ (a) and $\gamma\gamma$ (b) invariant mass distributions 
  for selected events.} 
\label{fig:phieta_spectra}
\end{figure}

\subsection{\boldmath  $\epem\to\k\k\eta$}
\label{sec:kketa}

Following the same strategy as the analysis described in the previous sections, 
we select a total of 113 candidate events. 
We estimate $42\pm42$ \qqbar and $15\pm8$ ISR background events: $56\pm43$ 
events is then the overall signal yield. 
Figure~\ref{fig:kketa_ev} shows the $\epem\to\kketa$ mass spectrum 
from production threshold up to about 4~\gev, 
together with the estimated  backgrounds,
while the  relative cross section
is depicted in Fig.~\ref{fig:kketa_cs} and the results are listed
in Table~\ref{phieta_notphi_tab}.\\
\begin{figure}[htb]
\includegraphics[width=.9\columnwidth]{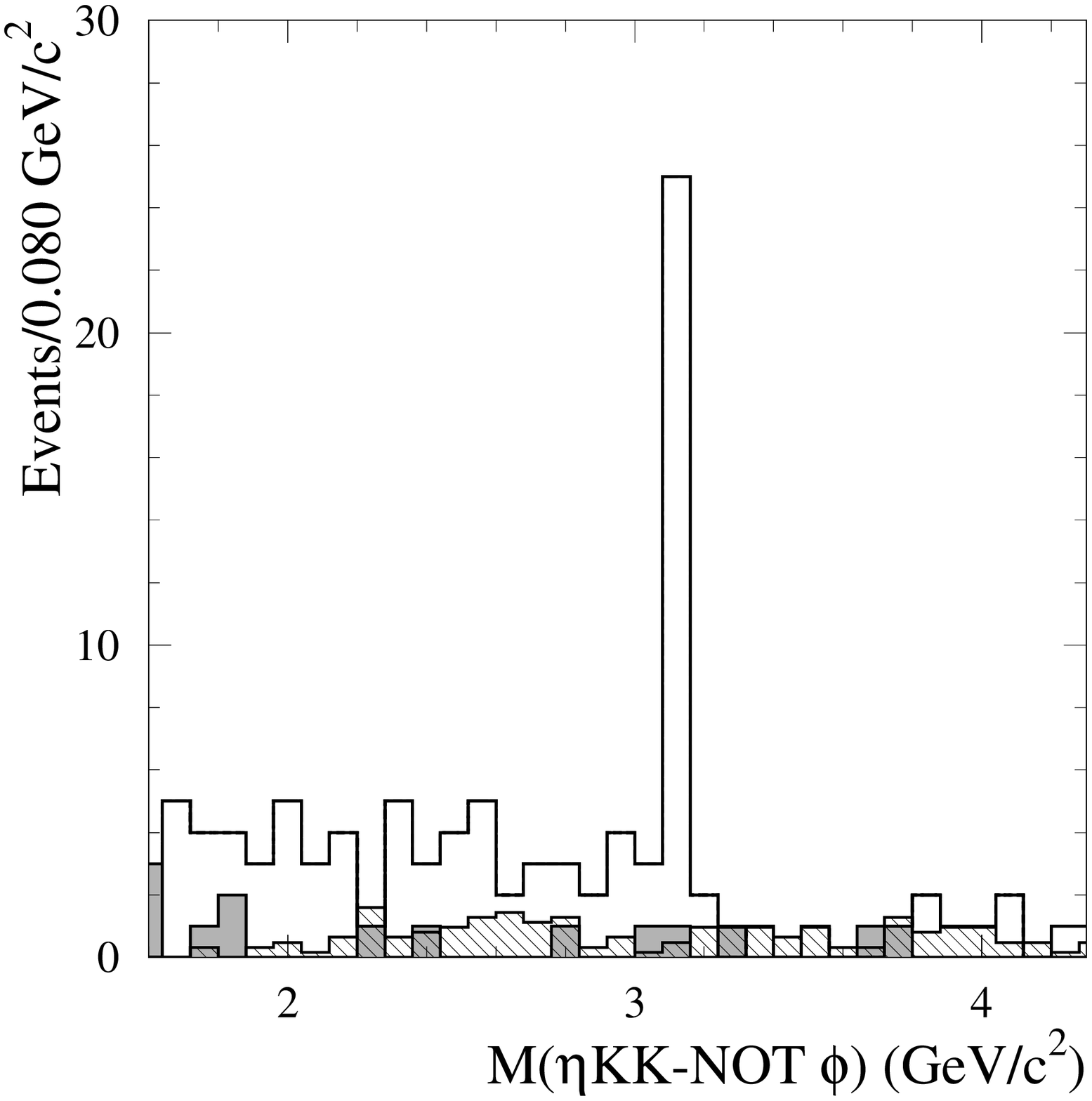}
\caption{The invariant mass distribution of the $\kketa$ system (open histogram).
The hatched and the gray histograms show the \qqbar and ISR backgrounds, respectively.}
\label{fig:kketa_ev}
\end{figure}
\begin{figure}[htb]
\includegraphics[width=.9\columnwidth]{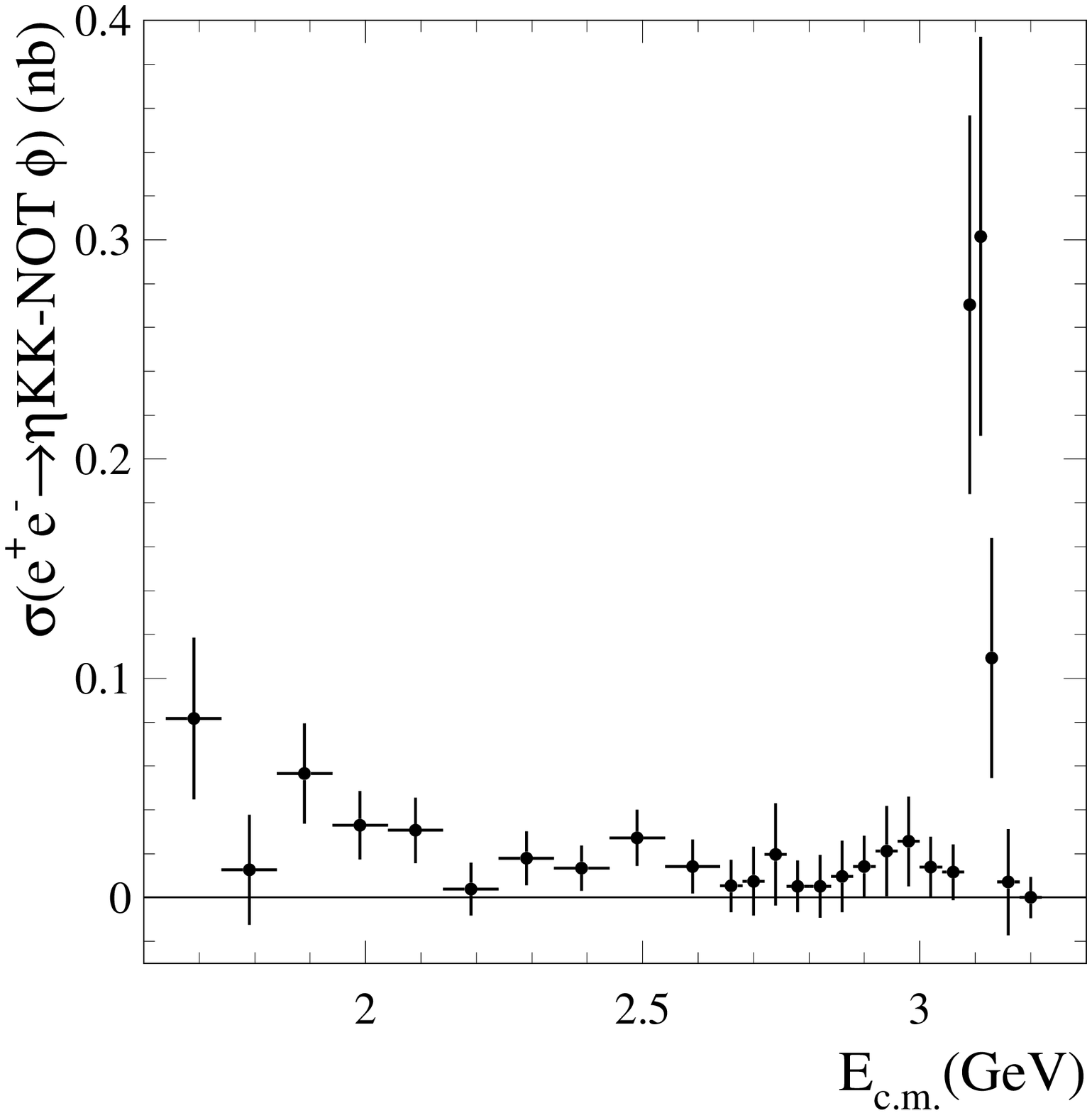}
\caption{The $\epem\to\kketa$  cross-section as a function of c.m. energy.}
\label{fig:kketa_cs}
\end{figure}
\indent
The uncertainties on background estimation, in particular 
the \qqbar contribution, completely dominate the systematic errors,
which sum up to about 80\%. 
On the other hand, the \jpsi region is 
essentially free from any background, 
as can be seen from the very clean $\eta$ 
and \jpsi signals shown in Fig.~\ref{fig:etakk_eta}. 
We select $21\pm3$ $\jpsi\to\Kp\Km\eta$ events, 
from which we measure
\begin{eqnarray}
    {\cal B}_{\jpsi\to\Kp\Km\eta}\cdot\Gamma_{ee}^\jpsi
    ~=~(4.8\pm 0.7\pm 0.3)\;\ev,
\end{eqnarray}
and extract the first 
evaluation of the $\jpsi\to\Kp\Km\eta$ branching fraction:
\begin{eqnarray}
    {\cal B}_{\jpsi\to\Kp\Km\eta}~=~(8.7\pm 1.3\pm 0.7)\;10^{-4}.
\end{eqnarray}

\begin{figure}[htb]
\includegraphics[width=\columnwidth]{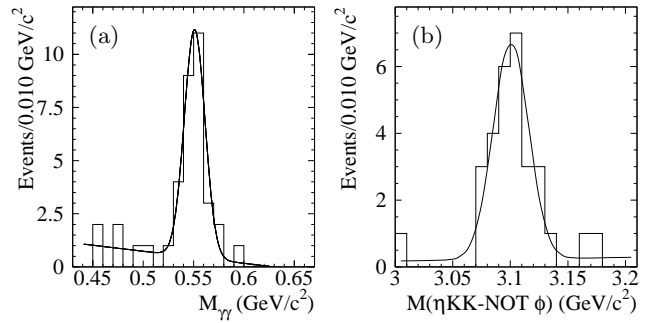} 
\caption{(a) the $\gamma\gamma$ invariant mass distribution in the $\eta$ region, 
  when a $\jpsi\to\kketa$ decay is selected.~~ 
  (b) the final \jpsi signal ($0.513<M_{\gamma\gamma}<0.583~\gevcc$ required).}
\label{fig:etakk_eta}
\end{figure}

\subsection{\boldmath $\epem\to\phi\eta$} 
\label{sec:phieta2}

The extremely clean $\phi$ signal makes this channel almost background-free.
Following the background-subtraction 
procedure as for the \phipiz final state  
we get  483 signal and 9 background events.
The $\phi$ signal, once a $\eta$ is selected, and the $\eta$ signal, 
after the $\phi$ selection, are shown in  Fig.~\ref{fig:metaphi_phieta_20_feb}.\\
\indent
We simulate several specific reactions with real $\phi$ and/or $\eta$ in the
final state,   
namely different ISR processes and \qqbar, $\tau\tau$ production, 
finding no significant contribution.  
The mass spectrum for signal events is shown in Fig.~\ref{fig:phieta_mass},
together with the subtracted distribution for the background events
obtained  from the side bands.

\begin{figure}[htb]
\includegraphics[width=\columnwidth]{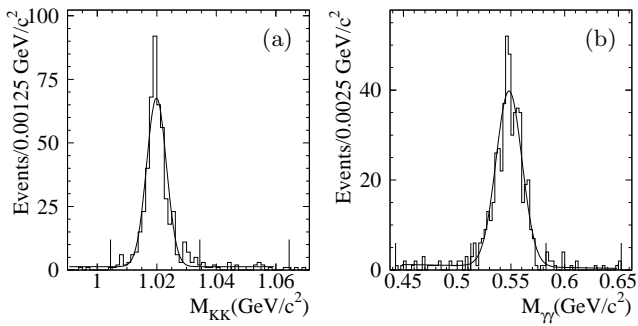}
\caption{
The \Kp\Km (a) and $\gamma\gamma$ (b) invariant mass distributions.
Signal and side-band regions are shown as vertical lines in the plots.
}
\label{fig:metaphi_phieta_20_feb}
\end{figure}

\begin{figure}[htb]
\includegraphics[width=.9\columnwidth]{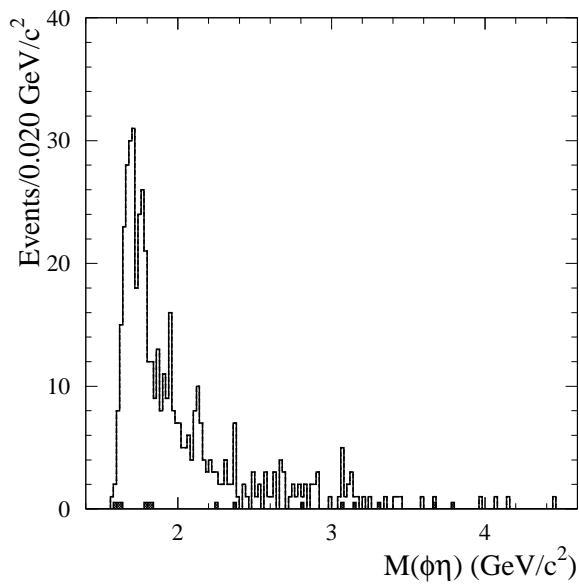}
\caption{The invariant mass distribution of the \phieta system. The shaded histogram
refers to the small subtracted background component.}
\label{fig:phieta_mass}
\end{figure}

\begin{figure}[t]
\includegraphics[width=.9\columnwidth]{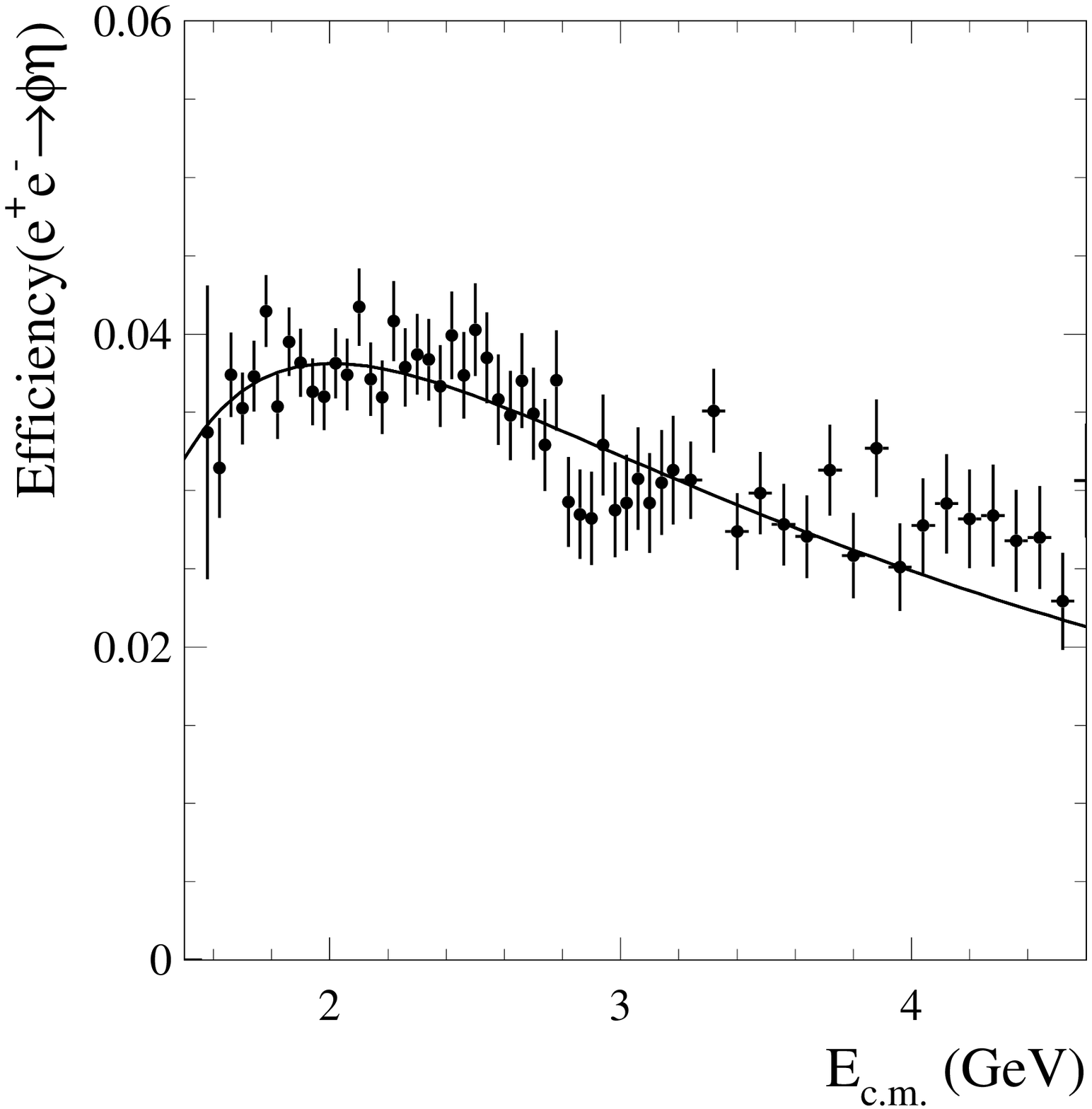}
\caption{Detection and reconstruction efficiency as a function 
  of the c.m. energy for the \phieta final state.
  The solid line is the result of a fit to the data with the function 
  $a\left( 1 - (x/b)^p\cdot e^{\xi (b-x)}\right)$.}
\label{fig:phieta_acc}
\end{figure}

\begin{figure}[htb]
\includegraphics[width=.9\columnwidth]{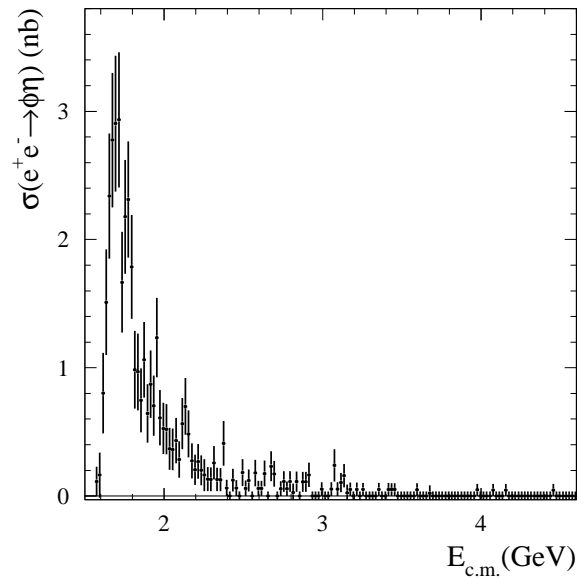}
\caption{$\epem\to\phieta$ cross section distribution as a function
  of c.m. energy.}
\label{fig:phieta_cs}
\end{figure}

The efficiency corrections and the 
systematic uncertainties that affect the \phieta cross section measurement
are very similar to those of the \kkpiz reaction (see Table~\ref{tab:error-kkpi0}). 
However, the background level is so low for the \phieta final state that
the associated errors are negligible and do not affect the background subtraction procedures. 
The systematic uncertainty  associated with $\eta$ reconstruction is equivalent to that 
for photon reconstruction, and the uncertainty on the $\phi\to\kk$ decay rate is 1.2\%. 
The total systematic error is thus $\pm 5.3\%$, 
while the efficiency is overestimated by 4.6\%.
The detection efficiency shown in Fig.~\ref{fig:phieta_acc} as a 
function of c.m. energy, is obtained from MC and accounts also for
$\eta\to\gaga$ decay rate.
The corresponding values for the $\epem\to \phieta$ cross section, 
obtained with an expression analogous to Eq.~\ref{eq:phipi0}, 
are reported in Table~\ref{phieta_tab}
and in Fig.~\ref{fig:phieta_cs}.\\
\indent The cross section peaks at about 3.0~nb around
1.7~\gev and also hints at a possible resonant behavior 
around 2.1~\gev. 
Above the \jpsi mass the cross section is very small,
as shown in the inset of Fig.~\ref{fig:phieta_cs}.

\begin{table*}[htb]
\caption{Measurement of the $\epem\to \phi\eta$ 
cross section as a function of \Ecm. Errors are statistical only.}
\label{phieta_tab}
\begin{ruledtabular}
\begin{tabular}{ c c c c c c c c}
$\Ecm (\gev)$ & $\sigma$ (nb)& 
$\Ecm (\gev)$ & $\sigma$ (nb)& 
$\Ecm (\gev)$ & $\sigma$ (nb)& 
$\Ecm (\gev)$ & $\sigma$ (nb)\\
\hline
 1.56$-$1.58 &  0.11 $\pm$  0.11 &  1.88$-$1.90 &  0.64 $\pm$  0.23 &  2.20$-$2.22 &  0.27 $\pm$  0.13 &  2.64$-$2.68 &  0.12 $\pm$  0.06 \\
 1.58$-$1.60 &  0.16 $\pm$  0.18 &  1.90$-$1.92 &  0.87 $\pm$  0.26 &  2.22$-$2.24 &  0.20 $\pm$  0.12 &  2.68$-$2.72 &  0.09 $\pm$  0.05 \\
 1.60$-$1.62 &  0.80 $\pm$  0.31 &  1.92$-$1.94 &  0.71 $\pm$  0.24 &  2.24$-$2.26 &  0.16 $\pm$  0.12 &  2.72$-$2.76 &  0.09 $\pm$  0.05 \\
 1.62$-$1.64 &  1.51 $\pm$  0.41 &  1.94$-$1.96 &  1.24 $\pm$  0.31 &  2.26$-$2.28 &  0.13 $\pm$  0.09 &  2.76$-$2.80 &  0.08 $\pm$  0.05 \\
 1.64$-$1.66 &  2.34 $\pm$  0.49 &  1.96$-$1.98 &  0.61 $\pm$  0.22 &  2.28$-$2.30 &  0.13 $\pm$  0.09 &  2.80$-$2.84 &  0.07 $\pm$  0.05 \\
 1.66$-$1.68 &  2.78 $\pm$  0.53 &  1.98$-$2.00 &  0.53 $\pm$  0.20 &  2.30$-$2.32 &  0.26 $\pm$  0.13 &  2.84$-$2.88 &  0.06 $\pm$  0.04 \\
 1.68$-$1.70 &  2.91 $\pm$  0.53 &  2.00$-$2.02 &  0.52 $\pm$  0.20 &  2.32$-$2.34 &  0.13 $\pm$  0.09 &  2.88$-$2.94 &  0.09 $\pm$  0.04 \\
 1.70$-$1.72 &  2.94 $\pm$  0.53 &  2.02$-$2.04 &  0.37 $\pm$  0.16 &  2.34$-$2.36 &  0.13 $\pm$  0.09 &  2.94$-$3.00 &  0.02 $\pm$  0.02 \\
 1.72$-$1.74 &  1.67 $\pm$  0.39 &  2.04$-$2.06 &  0.36 $\pm$  0.16 &  2.36$-$2.38 &  0.41 $\pm$  0.17 &  3.00$-$3.06 &  0.02 $\pm$  0.02 \\
 1.74$-$1.76 &  2.18 $\pm$  0.45 &  2.06$-$2.08 &  0.43 $\pm$  0.18 &  2.38$-$2.40 &  0.06 $\pm$  0.06 &  3.06$-$3.12 &  0.13 $\pm$  0.05 \\
 1.76$-$1.78 &  2.32 $\pm$  0.45 &  2.08$-$2.10 &  0.29 $\pm$  0.14 &  2.40$-$2.44 &  0.06 $\pm$  0.04 &  3.12$-$3.18 &  0.08 $\pm$  0.04 \\
 1.78$-$1.80 &  1.79 $\pm$  0.41 &  2.10$-$2.12 &  0.56 $\pm$  0.20 &  2.44$-$2.48 &  0.03 $\pm$  0.03 &  3.18$-$3.24 &  0.02 $\pm$  0.02 \\
 1.80$-$1.82 &  0.99 $\pm$  0.30 &  2.12$-$2.14 &  0.70 $\pm$  0.22 &  2.48$-$2.52 &  0.12 $\pm$  0.06 &  3.24$-$3.30 &  0.02 $\pm$  0.02 \\
 1.82$-$1.84 &  0.97 $\pm$  0.30 &  2.14$-$2.16 &  0.48 $\pm$  0.18 &  2.52$-$2.56 &  0.06 $\pm$  0.04 &  3.30$-$3.36 &  0.01 $\pm$  0.02 \\
 1.84$-$1.86 &  0.75 $\pm$  0.25 &  2.16$-$2.18 &  0.27 $\pm$  0.14 &  2.56$-$2.60 &  0.12 $\pm$  0.06 &  3.36$-$3.42 &  0.02 $\pm$  0.02 \\
 1.86$-$1.88 &  1.06 $\pm$  0.30 &  2.18$-$2.20 &  0.20 $\pm$  0.12 &  2.60$-$2.64 &  0.12 $\pm$  0.06 &  3.42$-$3.48 &  0.03 $\pm$  0.02 \\
\end{tabular}
\end{ruledtabular}
\end{table*}


\section{Cross section Fits}
\label{sec:fitting}
The cross section parametrization used to fit the data 
leading to the final state $f$,  $\epem\to f$,  at  $s=\Ecm^2$ is
\begin{eqnarray}
\sigma_f(s)&\!\!\!=\!\!\!&12\pi \mathcal{P}_f(s)\Bigg|
A_f^{\rm n.r.}(s)+\Bigg.\no\\
\Bigg. \!\!\!&&\!\!\!+\sum_R \sqrt{\mathcal{B}^R_{f}\Gamma^R_{ee}}
\frac{\sqrt{\Gamma_R/\mathcal{P}_f(M_R^2)}e^{i\Psi_R}}{M_R^2-s-i\sqrt{s}\Gamma_R(s)}
\Bigg|^2,
\label{general-param}
\end{eqnarray}
where $\mathcal{P}_f(s)$ is the phase space of the $f$ final state, $A_f^{\rm n.r.}(s)$
describes the non-resonant background, mainly due to the tails of resonances below
threshold, and the sum runs over all the vector resonances, with mass $M_R$, width
$\Gamma_R$ and relative phase $\Psi_R$, assumed to contribute to the cross section.
All of the final states analyzed contain a vector and a pseudoscalar meson.
The phase space for $f=VP$ has the form:
\begin{equation}
\mathcal{P}_{VP}(s)=\left[\frac{(s+M_V^2-M_P^2)^2-4M_V^2 s}
{s}\right]^\frac{3}{2}.
\label{pahase-space}
\end{equation}
%
%
\subsection{Fitting $K\ov{K}\pi$ and \phieta cross sections}
\label{subsec:kkstar}
The statistics are large enough to attempt a complete description of
the \k\kst channel. The \kskpi Dalitz plot
yields not only  the isoscalar $\sigma_0$ and isovector $\sigma_1$
components, but also their relative phase $\Delta\phi$
(Figs.~\ref{kskpi-i0},~\ref{kskpi-i1} and~\ref{kskpi-dph}). 
In addition, the $\epem\to\Kpm K^{*\mp}(892)$ cross section, 
obtained by integrating the \kkpiz Dalitz plot multiplied by the isospin factor of three, 
and taking into account corrections due to the interference relevant for $K^\pm\pi^0$  
in the \kst region, provides a further constraint on $\sigma_0$, $\sigma_1$ 
and $\Delta\phi$. Indeed, both the cross sections for $\epem\to K^\pm K^{*\mp}(892)$ and 
$\epem\to K^0 \ov{K}^{*0}(892)$+c.c. can be defined in terms of opposite combinations 
of the same isospin components:
\begin{eqnarray}
\sigma_{K^0 \ov{K}^{*0}(892)+\rm c.c}\!\!&=&\!\!
\left|\sqrt{\sigma_0}+\sqrt{\sigma_1} e^{i\Delta\phi}\right|^2
\label{eq:kkpi0}\\
&=&\sigma_0+\sigma_1+2\sqrt{\sigma_0\sigma_1}\cos\Delta\phi\no\\
\sigma_{K^\pm K^{*\mp}(892)}\!\!&=&\!\!
\left|\sqrt{\sigma_0}-\sqrt{\sigma_1} e^{i\Delta\phi}\right|^2\no\\
&=&
\sigma_0+\sigma_1-2\sqrt{\sigma_0\sigma_1}\cos\Delta\phi.\no
\end{eqnarray} 
We note that the square root of the cross sections
$\sigma_0$ and $\sigma_1$, having roughly the same phase space 
(the mass difference between neutral and charged \kst can be neglected here and 
in the following), are proportional to the corresponding amplitudes.
The isoscalar $K\kst$ cross section $\sigma_0$ and $\sigma_{\phieta}$
appear to be dominated by the same structure, a broad peak 
with $M\approx 1.7\;\gev/c^2$.
This behavior is theoretically expected because the two final states 
have the same quantum numbers $I^G(J^{PC})=0^-(1^{--})$ and similar
strangeness content.
In principle, quantum number conservation
should allow  $\omega$-like and $\phi$-like contributions for both 
cross sections. However, in the case of the \phieta final state, the 
OZI rule~\cite{ozi} strongly suppresses any intermediate $\omega$-recurrence.
On the other hand, for the channel $K\kst$, where the OZI rule does not apply, 
there are also suppression factors. 
Suppression of $\omega$-like versus $\phi$-like coupling could account for as much as
a factor of four~\cite{pdg,achasov}; an additional price has to be paid if
a strange quark pair has to be created from the vacuum.
Hence, in the following 
we will consider all the structures contributing to the isoscalar amplitudes
as $\phi$-like resonances. In particular, the peak with $M\approx 1.7\;\gevcc$
matches the first $\phi$-like excited state, so
we will assume it to be  simply the $\phi'$. 

The first result that can be drawn comparing the cross section 
data on $\sigma_0$ and $\sigma_{\phieta}$ (Figs.~\ref{kskpi-i0} 
and~\ref{fig:phieta_cs}) is an estimate for the ratio between the decay 
fractions of the $\phi'$ into $K\kst$ and \phieta. Indeed, at the $\phi'$ peak:
\begin{eqnarray}
\frac{\sigma_{\phieta}(M_{\phi'}^2)}{\sigma_{0}(M_{\phi'}^2)}\approx
\frac{\mathcal{B}^{\phi'}_{\phieta}}{\mathcal{B}^{\phi'}_{K\kst}}\approx \frac{1}{3},
\label{eq:ratiobw}
\end{eqnarray}
where $\mathcal{B}^R_f$ is the branching ratio for the decay $R\to f$.
From Eq.~\ref{eq:ratiobw} one can infer that both the $K\kst$ and the 
\phieta channel represent an important fraction of the $\phi'$ total width. 

In order to make optimal use of the experimental data,
we  perform a global fit exploiting all available information from the 
$K\ov{K}\pi$ final states (four sets of data: $\sigma_{0}$, $\sigma_{1}$, 
$\Delta\phi$ and  $\sigma_{K^\pm K^{*\mp}(892)}$) as well as the \phieta 
cross section.

We have considered also a recently published \babar\ measurement on the \phieta 
cross section~\cite{phieta-solodov} in the $\epem\to\Kp\Km\pip\pim\piz$ channel. 
These data, even though with a lower statistics, are in agreement with our 
measurement.

We have performed two fits, in the first case we used only our samples 
(five sets of data: $\sigma_{0}$, $\sigma_{1}$, $\Delta\phi$, 
$\sigma_{K^\pm K^{*\mp}(892)}$ and $\sigma_{\phieta}$), in the second case
we included the \phieta cross section from Ref.~\cite{phieta-solodov}.

We parametrize the $K\ov{K}\pi$ isoscalar and the isovector cross sections
using Eq.~\ref{general-param} with one resonance each 
and a non-resonant background,  assumed to be the $\phi$ tail.
The isoscalar cross section shows a clearly resonant behavior, the 
shape of the sub-dominant isovector cross section instead appears compatible
with a pure phase space. However, using the information  
of the relative phase, $\Delta\phi$ (Fig.~\ref{kskpi-dph}), it can be shown that
such a quantity is not consistent with 
a real isovector amplitude, as 
expected in case of a non-resonant 
behavior. 
If this were indeed the case, the relative
phase should be as shown by the 
light gray band in Fig.~\ref{fig:dfase}. This is clearly not compatible
with the data (solid circles). 
The isovector cross section $\sigma_1$, then, has to be described in terms of 
a $\rho$-like excited state, the $\rho'$ in the following.

Finally, the \phieta cross section is parametrized  using two resonances:
 in addition to the dominant $\phi'$ just above threshold, we include
a small contribution, probably due to an excited $s\ov{s}$ or a more exotic
state (tetraquark, hybrid) with a mass 
around $2.13\,\gevcc$, the $\phi''$, that has been observed decaying into  
 $\phi f_0(980)$~\cite{solodov}. 

A good parametrization of the non-resonant background, the
function $A^{\rm n.r.}_{f}(s)$ ($f=K\kst$, \phieta), can be obtained 
using a $1/s$ scaling
($A^{\rm n.r.}_{f}(s)\stackrel{\rm def}{=}  A^{\rm n.r.}_{f}/s$).
As a result, the shape of the non-resonant part, Eq.~\ref{general-param}, 
is given directly by
the phase space, and we may define the non-resonant cross section as:
\begin{equation}
\sigma_f^{\rm bkg}(s)=12\pi \mathcal{P}_f(s)\left[A_f^{\rm n.r.}\right /s]^2.
\label{sigma0}
\end{equation}
We stress that, as shown in Table~\ref{tab:multifit}, this cross section 
represents only a small fraction of the resonant one.

We describe wide  resonances  using
energy-dependent widths, $\Gamma_R(s)$ (see Eq.~\ref{general-param}),
parametrized as:
\begin{equation}
\Gamma_R(s)=\Gamma_R\sum_k \frac{\mathcal{P}_k(s)}{\mathcal{P}_k(M_R^2)}\mathcal{B}^R_{k},
\label{eq:larghezza}
\end{equation}
where $\Gamma_R$ is the constant total width, and $\mathcal{P}_k(s)$ and 
$\mathcal{B}^R_{k}$ are the phase space and the branching fraction 
for the transition of $R$ into the final state $k$ respectively
($\sum_k\mathcal{B}^R_{k}=1$).
In case of narrow peaks we use fixed widths.

\begin{figure}[htb]
\includegraphics[width=.9\columnwidth]{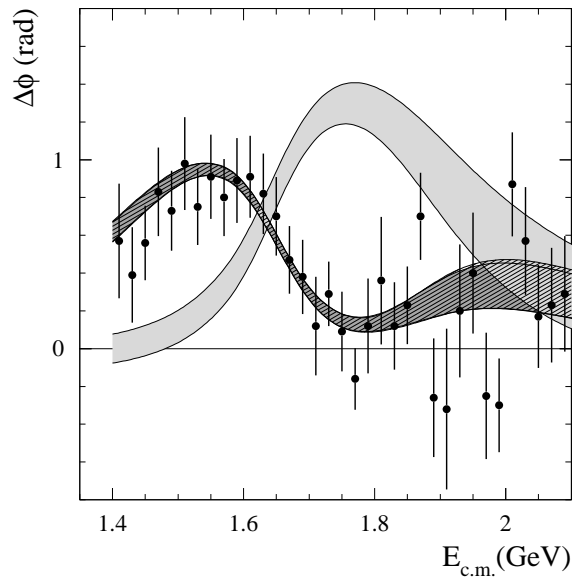}
\caption{\label{fig:dfase}%
  Phase difference between isovector and isoscalar
  $K\kst$ amplitudes: data and fits superimposed (dark-gray and hatched bands).
  In particular the dark-gray band is obtained by using only our \phieta cross
  section, while the hatched band including also the data from Ref.~\cite{phieta-solodov}.
  The light-gray band represents the relative phase expected 
  in case of non-resonant isovector component.}
\end{figure}
\begin{figure}[htb!]
\includegraphics[width=.9\columnwidth,height=15.cm]{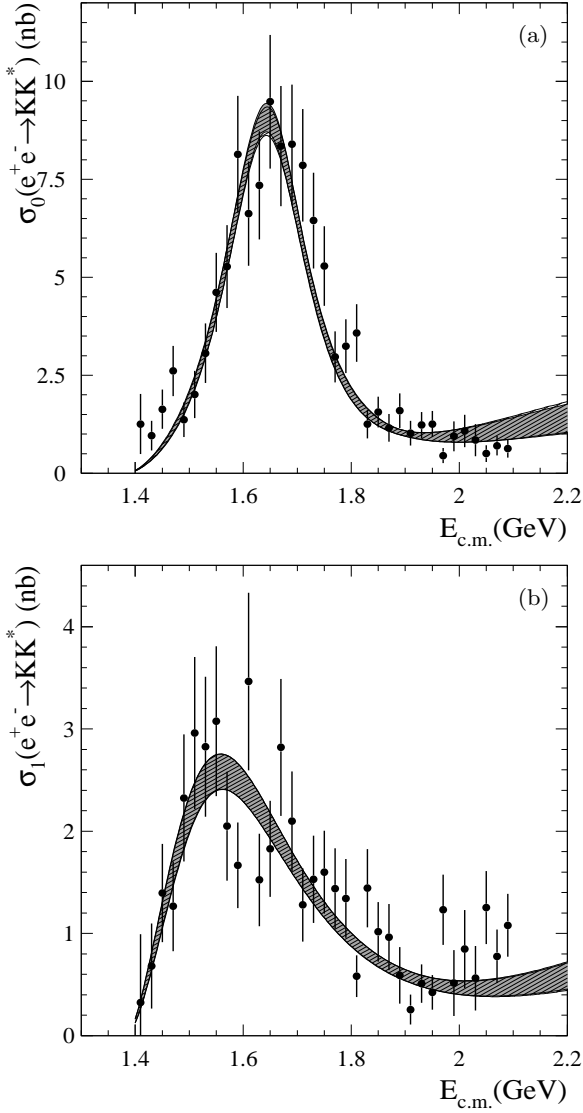}
\caption{\label{s0-s1}
  Isoscalar (a) and isovector (b) components of the $\kskpi$ 
  cross section. The solid points are the data and the 
  bands represent the fits. The hatched band is obtained 
  including the \phieta cross section from Ref.~\cite{phieta-solodov}.}
\end{figure}\vspace{-0mm}

As already mentioned, the most important resonance in these processes 
is the $\phi'$: it represents the main contribution for both $\sigma_0$
and $\sigma_{\phieta}$. It follows that the total width 
$\Gamma_{\phi'}(s)$ will
account for both these $VP$-like final states: the isoscalar $K\kst$ and 
the \phieta. Other final states with small branching ratios are
parametrized by adding  a constant term to the width 
\begin{eqnarray}
\Gamma_{\phi'}(s)\!\!&=\Gamma_{\phi'}\Bigg[&\!\!%
\frac{\mathcal{P}_{K\kst}(s)}{\mathcal{P}_{K\kst}(M_{\phi'}^2)}\mathcal{B}^{\phi'}_{K\kst}+\Bigg.\\
\!\!&&\!\!\Bigg.
\frac{\mathcal{P}_{\phieta}(s)}{\mathcal{P}_{\phieta}(M_{\phi'}^2)}\mathcal{B}^{\phi'}_{\phieta}+
\left(1\!-\!\mathcal{B}^{\phi'}_{K\kst}\!-\!\mathcal{B}^{\phi'}_{\phieta}\right)
\Bigg].\no
\end{eqnarray}
The $\phi''$, with a constant width, is included to model the \phieta cross section. 

The main contribution to the cross section $\sigma_1$ comes from
the $\rho'$ resonance. We assume the $\rho'$ to decay mainly
in four pions, and again include a constant term to account for 
other channels:
\begin{eqnarray}
\Gamma_{\rho'}(s)\!\!&=\Gamma_{\rho'}\Bigg[&\!\!%
\frac{\mathcal{P}_{4\pi}(s)}{\mathcal{P}_{4\pi}(M_{\rho'}^2)}\mathcal{B}^{\rho'}_{4\pi}+
\left(1\!-\!\mathcal{B}^{\rho'}_{4\pi}\right)\Bigg],
\end{eqnarray}
where, for $\mathcal{P}_{4\pi}(s)$ we use:
\begin{eqnarray}
\mathcal{P}_{4\pi}(s)=\frac{\left(s-16M_\pi^2\right)^n}{s},
\label{eq:4pi}
\end{eqnarray}
with $n=3/2$. Note that the fit is not sensitive to this power, as higher
values (e.g. $n=5/2$) give similar parameters.

%
%
%
%
\begin{table*}[htb!]\vspace{0mm}
\caption{\label{tab:multifit}
  Parameters for resonances $\phi'$
  and $\rho'$ obtained by fitting simultaneously
  the isoscalar and isovector component of the 
  $KK^*(892)$ + c.c. cross section, their relative phase,
  the \kkpiz and the \phieta cross sections
  (only our samples on the left, including also data from
   Ref.~\cite{phieta-solodov} on the right). 
  Real backgrounds are defined through Eq.~\ref{sigma0}.
  All phases $\Psi_R$ (see Eq.~\ref{general-param}) are set to zero.}
\vspace{2mm}
\renewcommand{\arraystretch}{1.5}
\begin{tabular}{lccccc}
\cline{1-3}\cline{5-6}
&\multicolumn{2}{c}{$\frac{\chi^2}{\rm n.d.f.}=\frac{184.9}{160-16}=1.28$}
&\hspace{5mm} &\multicolumn{2}{c}{$\frac{\chi^2}{\rm n.d.f.}=\frac{205.4}{187-16}=1.20$}\\
\cline{1-3}\cline{5-6}
\multicolumn{2}{c}{\vspace{-5mm}}\\
\cline{1-3}\cline{5-6}
$R$ with $I=0\hspace{10mm}$ & $\hspace{10mm}\phi'\hspace{10mm}$ & $\hspace{10mm}\phi''\hspace{10mm}$ 
&& $\hspace{10mm}\phi'\hspace{10mm}$ & $\hspace{10mm}\phi''\hspace{10mm}$ \\
\cline{1-3}\cline{5-6}
$\Gamma^R_{ee}\mathcal{B}^{R}_{K\kst}(\ev)$  & $367\!\pm\! 47$& - && $369\!\pm\! 53$& - 
\\
$\Gamma^R_{ee}\mathcal{B}^{R}_{\phieta}(\ev)$  & $154\!\pm\! 32$& $1.7\!\pm\!0.8$
&& $138\!\pm\! 33$& $1.7\!\pm\! 0.7$
\\
$1-\mathcal{B}^{R}_{K\kst}-\mathcal{B}^{R}_{\phieta}$  & $0.33\!\pm\! 0.14$& - 
&& $0.36\!\pm\! 0.15$& - 
\\
$M_R(\mev)$ &  $1709\!\pm\! 19 $& $ 2127 \!\pm\! 24$
&&  $1709\!\pm\! 20 $& $ 2125 \!\pm\! 22$
\\
$\Gamma_R(\mev)$ & $325\!\pm\! 68$ & $60\!\pm\!50$
&& $322\!\pm\! 77$ & $61\!\pm\!50$
\\ 
$\sigma_{K\kst}^{\rm bkg}(M_{\phi'}^2)$(nb) & $0.8\!\pm\! 0.3$ & -
&& $0.9\!\pm\! 0.3$ & -
\\
$\sigma_{\phieta}^{\rm bkg}(M_{\phi'}^2)$(nb) &  \multicolumn{2}{c}{$(4.7\!\pm\! 1.4)\times 10^{-3}$} 
&&  \multicolumn{2}{c}{$(4.7\!\pm\! 1.8)\times 10^{-3}$} 
\\
\cline{1-3}\cline{5-6}
\cline{1-3}\cline{5-6}
\multicolumn{5}{c}{\vspace{-5mm}}\\
\cline{1-3}\cline{5-6}
$R$ with $I=1\hspace{10mm}$ & \multicolumn{2}{c}{$\hspace{10mm}\rho'\hspace{10mm}$}
&& \multicolumn{2}{c}{$\hspace{10mm}\rho'\hspace{10mm}$} \\
\cline{1-3}\cline{5-6}
$\Gamma^R_{ee}\mathcal{B}^{R}_{K\kst}(\ev)$ & \multicolumn{2}{c}{$129\!\pm\! 15$}
&& \multicolumn{2}{c}{$127\!\pm\! 15$} \\
$1-\mathcal{B}^{R}_{4\pi}$ & \multicolumn{2}{c}{$0.36\!\pm\! 0.22$}
&& \multicolumn{2}{c}{$0.35\!\pm\! 0.23$} \\
$M_R(\mev)$ &  \multicolumn{2}{c}{$1508\!\pm\! 19$} 
&&  \multicolumn{2}{c}{$1505\!\pm\! 19$} \\
$\Gamma_R(\mev)$ & \multicolumn{2}{c}{$418\!\pm\! 26$} 
&& \multicolumn{2}{c}{$418\!\pm\! 25$} \\ 
$\sigma_{K\kst}^{\rm bkg}(M_{\phi'}^2)$(nb) & \multicolumn{2}{c}{$0.9\!\pm\! 0.2$}
&& \multicolumn{2}{c}{$0.9\!\pm\! 0.2$}\\
\cline{1-3}\cline{5-6}
\end{tabular}
%
%
\end{table*}\vspace{-0mm}

\begin{figure}[htb!]
\includegraphics[width=.9\columnwidth]{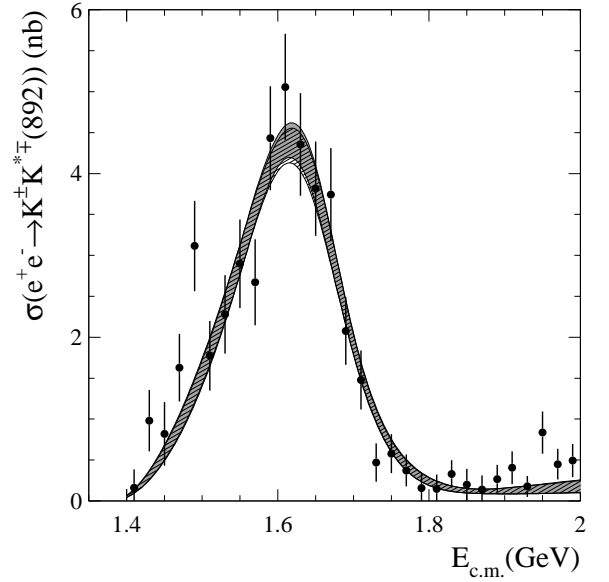}
\caption{\label{fig:kkpi0-cs} 
  Data on the annihilation cross section $\sigma_{\Kpm K^{*\mp}(892)}$, see Eq.~\ref{eq:kkpi0},
  and descriptions (solid and hatched bands) in terms of the parameterization of 
  Eq.~\ref{general-param}, with parameters reported in \hbox{Table~\ref{tab:multifit}}. 
  The hatched band is obtained including the \phieta cross section from Ref.~\cite{phieta-solodov}.}
\end{figure}
\begin{figure}[htb!]
\includegraphics[width=.9\columnwidth]{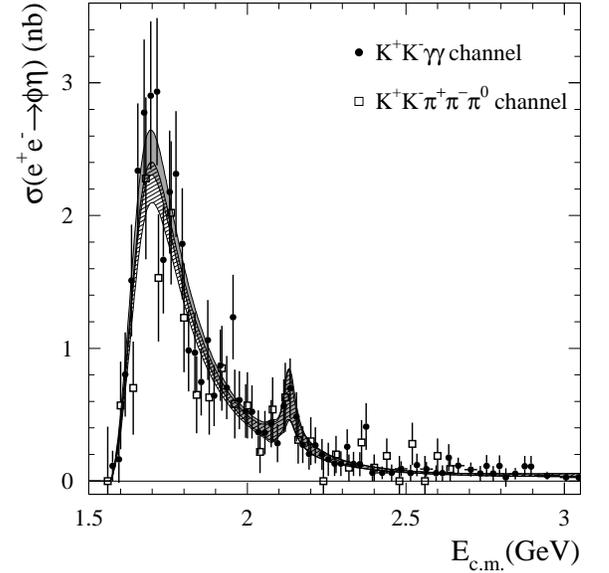}
\caption{\label{fig:phieta-cs}
  Data on the annihilation cross section $\sigma_{\phieta}$
  (the solid circles are from this analysis, while the empty squares 
  are from Ref.~\cite{phieta-solodov})
  and descriptions (solid and hatched bands) in terms of the parameterization of 
  Eq.~\ref{general-param}, with parameters reported in Table~\ref{tab:multifit}.
  The hatched band is obtained including the \phieta cross section from 
  Ref.~\cite{phieta-solodov}.}
\end{figure}

Our global fit results are summarized in Table~\ref{tab:multifit} and
shown in Figs.~\ref{fig:dfase}-\ref{fig:phieta-cs}, where the errors of data points 
include both statistical and systematic uncertainties, and the bands represent the
convolution of the one-sigma statistical errors of the fit.
The solid and the hatched fill refer to the two considered cases.
In particular, Figs.~\ref{s0-s1} and~\ref{fig:dfase} show data and fits for the isospin 
components and the relative phase from the \kskpi Dalitz plot analysis,
Fig.~\ref{fig:kkpi0-cs} shows the cross section $\sigma_{\Kpm K^{*\mp}(892)}$,
and Fig.~\ref{fig:phieta-cs} shows $\sigma_{\phieta}$, as obtained in this
analysis and in Ref.~\cite{phieta-solodov}.

All these data sets support the hypothesis that the main contribution in the 
isoscalar amplitudes is the resonance $\phi'$, while the isovector
component is almost fully described by a broad excited $\rho$ state. However, 
the normalized $\frac{\chi^2}{\rm n.d.f.}$ values, 1.28 in the first case and  
1.20 in the second case when the data on the \phieta cross section from 
Ref.~\cite{phieta-solodov} are used,
do not exclude the presence of other possible structures but, on the other
hand, the accuracy of the data does not allow any clear identification.

As mentioned above, the
mass and width of the $\phi'$ are compatible with the 
\phip~\cite{pdg}. The parameters obtained for the $\phi''$, 
the second isoscalar excited state, 
are consistent with those reported in Ref.~\cite{solodov}.
A clear evidence for this decay channel cannot be established
with the available data sample, however, the probability to
be a statistical fluctuation is 6$\times10^{-4}$ (25 events
under the peak with an expectation of 11 evaluated from 
the mass spectrum in Fig.~\ref{fig:phieta_cs}).  
The $\rho'$ from the isovector component of the 
$K\kst$ cross section has a mass value in agreement with Ref.~\cite{pdg},
although this agreement is not particularly striking given the width of this resonance.

It is worth stressing that  at  \Ecm  above 1.8 \gev,  
contributions coming from any higher mass $K^*$ 
certainly affect the fit in the $\K\kst$ channel. 
At low \Ecm 
the fit is very sensitive to the $K^*(892)$ width, when
the \k$\pi$ invariant mass is  close  to the \kst threshold. However, analyticity 
relates $KK^*(892)$ and $\phi\eta$ amplitudes to the
$K^*(892)$ and $\phi$ radiative widths (that is $KK^*(892)$ and $\phi\eta$
 at $s=0$) via dispersion relations. Exploiting these relations might lead to an 
improved determination of the threshold  cross section.  

The systematic effects due to the choice of fit model were evaluated by
repeating the fits using different values for the relative phases $\Psi_R$,
and by choosing different parameterizations for the energy-dependent
widths, as appropriate.

\subsection{\mathversion{bold} Fitting the $\phipiz$ cross section}
\label{subsec:phipi0fit}

The \phipiz channel is a pure isospin one final state, hence, using
Eq.~\ref{general-param} to describe the corresponding cross
section, only isovector contributions must be included. 
This channel has been selected as a possible candidate 
for multiquark intermediate states:  contributions are expected from
 $\rho$ recurrences, even though they are OZI-suppressed.
We  label any resonance in this channel as 
a $\rho$-like excited state.

Two cases have been studied in detail:
\begin{itemize}
\item only one $\rho$-like excited state, a
      $\rho''$ (to distinguish it from the $\rho'$ of the previous case);
\item two $\rho$ recurrences, {\it i.e.}, in addition to $\rho''$,
      a narrow resonance to better represent the data  
      around $\Ecm \approx 1900\;\mev$ (Fig.~\ref{phipi0-cs-fig}),
 labelled in the following  as $\rho(1900)$.
\end{itemize}
The $\rho(1900)$ width is assumed constant,  
while for the $\rho''$ we use the energy-dependent width
\begin{eqnarray}
\Gamma_{\rho''}(s)\!\!&=\Gamma_{\rho''}\Bigg[&\!\!%
\frac{\mathcal{P}_{4\pi}(s)}{\mathcal{P}_{4\pi}(M_{\rho''}^2)}\mathcal{B}^{\rho''}_{4\pi}+
\left(1\!-\!\mathcal{B}^{\rho''}_{4\pi}\right)
\Bigg],\no
\end{eqnarray}
where $\mathcal{P}_{4\pi}(s)$ is the four-pion phase space
defined in Eq.~\ref{eq:4pi}, and $\mathcal{B}^{\rho''}_{4\pi}$ is
the branching fraction for $\rho''\to4\pi$.

The results of the fit with only the $\rho''$ are reported in the first column
of the Table~\ref{phipi0-partot}.

In the second case, where the $\rho(1900)$ is also included,
a two step procedure is used to determine the 
phases of the two quasi-real amplitudes. 
We find $(\Psi_{\rho''},\Psi_{\rho(1900)})=(0,\pi)$ to be the best combination.
With these phases we get the results reported in the
second and third columns of the Table~\ref{phipi0-partot}. 

\begin{figure}
\includegraphics[width=.9\columnwidth]{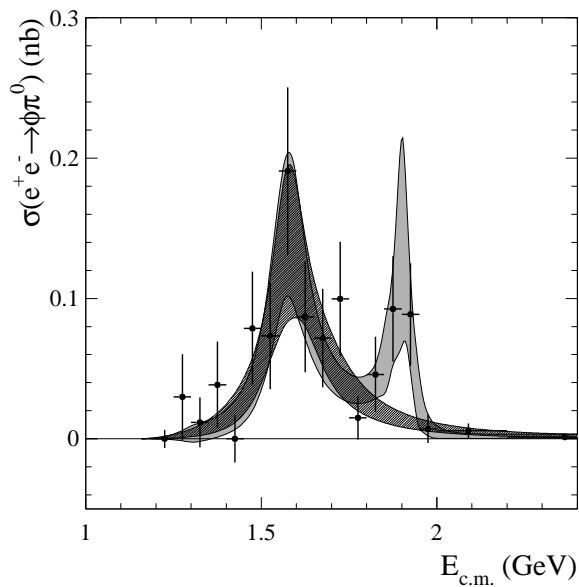}
\caption{\label{phipi0-cs-fig}
  Data on the annihilation cross section $\sigma_{\phipiz}$
  and descriptions in terms of the parameterization of 
  Eq.~\ref{general-param} with the only $\rho''$, hatched band, and including 
  also the $\rho(1900)$, gray band.}
\end{figure}
%
%
%
%
%
\begin{table}[h]\vspace{0mm}
\caption{\label{phipi0-partot}%
Parameters obtained for the \phipiz cross section.
First column: with the only $\rho''$ resonance, second and third
columns: including also the $\rho(1900)$.
The normalized $\chi^2$ and corresponding C.L. are reported
in each case.}
\vspace{2mm}
\renewcommand{\arraystretch}{1.5}
\begin{tabular}{lcccc}
\cline{1-2}\cline{4-5}
($\frac{\chi^2}{\rm n.d.f.}$,C.L.) & ($\frac{14.36}{16-5},0.31$) &&
\multicolumn{2}{c}{($\frac{7.37}{16-8},0.50$)} \\
\cline{1-2}\cline{4-5}
\multicolumn{4}{c}{\vspace{-5mm}}\\
\cline{1-2}\cline{4-5}
 $R$ &$\rho''$ &&$\rho''$ & $\rho(1900)$ \\
\cline{1-2}\cline{4-5}
 $\Gamma^R_{ee}\mathcal{B}^{R}_{\phipiz}(\ev)$ 
&$4.4\!\pm\!1.0$ &&$3.5\!\pm\! 0.9$  & $2.0\!\pm\! 0.6$ \\
$(1-\mathcal{B}^{R}_{4\pi})$ & $0.67\!\pm\!0.43$ &&$0.44\!\pm\! 0.49$ & -\\
$M_R(\mev)$ & $1593\!\pm\!32$&& $1570\!\pm\! 36$ & $1909\!\pm\! 17$ \\
$\Gamma_R(\mev)$ &$203\!\pm\!97$&& $144\!\pm\! 75$ & $48\!\pm\!17$  \\ 
$\Psi_R$(rad) &$0$ &&$0$ & $\pi$  \\ 
$\sigma^{\rm bkg}_{\phipiz}(M_{\rho(1900)}^2)$(nb) 
&$(\!0.4\!\pm\!0.2\!)\!\!\times\!\! 10^{-3}$ &&
\multicolumn{2}{c}{$(\!0.5\!\pm\!1.5\!)\!\!\times\!\! 10^{-3}$}\\
\hline
\end{tabular}
\end{table}

The results of the two fits are shown in Fig.~\ref{phipi0-cs-fig},
superimposed on the cross section data.

A slightly better $\chi^2$ is obtained by adding this extra resonance.
We can not however exclude that the  observed accumulation of events 
at $\Ecm\approx 1.9$ \gev is produced by a statistical fluctuation.
In fact, we observe 18 events to be compared to an expectation of 8 events,
taking into account both the background and the tail of the $\rho''$,
and this translates in a Poisson probability of $2\times10^{-3}$. 

Mass, width, and quantum numbers [$I^G(J^{PC})=1^+(1^{--})$] obtained for
the $\rho(1900)$ are compatible with those of the 
so-called ``dip'' observed in other channels, primarily multi-pion final 
states~\cite{dip}.


%
\begin{table*}
\caption{\label{tab:summary} Summary of parameters of resonances in the studied final states,
obtained including also the data on the \phieta cross section from 
Ref.~\cite{phieta-solodov}.
The parameters for the $\rho''$ are taken from the fit to 
the \phipiz cross section with two resonances.
}
\small
\begin{ruledtabular}
\renewcommand{\arraystretch}{1.5}
\begin{tabular}{cccccc}
Isospin &$R$ & $\Gamma^R_{ee}\mathcal{B}^R_{K\kst}$(\ev) & 
$\Gamma^R_{ee}\mathcal{B}^R_{\phieta}$(\ev) & $M_R$(\mev) & $\Gamma_R$(\mev) \\
\hline
%
%
\multirow{2}{*}{0} & $\phi'$ & $369\!\pm\!53\!\pm\!1$ & $138\!\pm\!33\!\pm\!28$ & 
$1709\!\pm\!20\!\pm\!43$ &  $322\!\pm\!77\!\pm\!160$ \\
 & $\phi''$ & $-$ & $1.7\!\pm\!0.7\!\pm\!1.3$ & $2125\!\pm\!22\!\pm\!10$ &  
$61\!\pm\!50\!\pm\!13$ \\
\hline\hline
isospin & $R$ & $\Gamma^R_{ee}\mathcal{B}^R_{K\kst}$(\ev) & 
$\Gamma^R_{ee}\mathcal{B}^R_{\phipiz}$(\ev) & $M_R$(\mev) & $\Gamma_R$(\mev) \\
\hline
\multirow{3}{*}{1} & $\rho'$ & $127\!\pm\!15\!\pm\!6$ & $-$ & $1505\!\pm\!19\!\pm\!7$ &  $418\!\pm\!25\!\pm\!4$ \\
 & $\rho''$     & $-$ & $3.5\!\pm\!0.9\!\pm\!0.3$ & $1570\!\pm\!36\!\pm\!62$ &  $144\!\pm\!75\!\pm\!43$ \\
 & $\rho(1900)$ & $-$ & $2.0\!\pm\!0.6\!\pm\!0.4$ & $1909\!\pm\!17\!\pm\!25$ &  $48\!\pm\!17\!\pm\!2$   \\
\end{tabular}
\end{ruledtabular}
\end{table*}
\section{ {\boldmath Summary and conclusions}}
We have studied the \kskpi, \kkpiz, \phipiz, \kketa, and \phieta
final states, produced in \babar\ via ISR.

We have measured the production cross section from threshold up to $\approx$4.6\gev
for all of these processes to an unprecedented accuracy.
 
By studying the asymmetric \kskpi Dalitz plot, we have obtained the moduli 
and relative phase of the isospin components for the $K\kst$ 
cross section. The knowledge of the isoscalar and isovector
cross sections, $\sigma_0$ and $\sigma_1$, allows a simple description in terms of
vector meson resonances, where $\rho$- and $\phi$-like
mesons do not mix. In addition, the relative phase gives
unique information on the sub-dominant isovector component, 
which, although compatible with a pure phase space behavior, 
reveals a resonant structure. 

Two global fits, that benefit from five interconnected sources
of information ($\sigma_0$, $\sigma_1$, $\sigma_{\phieta}$, 
$\sigma_{\kkpiz}$ and $\Delta\phi$), have been performed.
In a first case we have used only the data samples coming from this
analysis, while in the second case recent \babar\ data on $\sigma_{\phieta}$ 
in the $\Kp\Km\pip\pim\piz$ channel have been included. 

Both \phieta and the isoscalar $K\kst$ cross sections have
been parametrized with the same dominant resonance $\phi'$;
the suitable combination of $\sigma_0$, $\sigma_1$
and $\Delta\phi$, obtained from the \kskpi channel,
is used to describe the \kkpiz cross section. 
The mass and width for the $\phi'$, the main resonant contribution in
the isoscalar channels, are compatible with those of the 
\phip meson reported in Ref.~\cite{pdg}. Concerning the $\phi''$ resonance, 
included only in the \phieta channel, we find parameters compatible with 
those of the first observation in the $\phi f_0(980)$ final state~\cite{solodov}, 
confirming  the quantum numbers of this resonance,
$I^G(J^{PC})=0^-(1^{--})$. 

The isovector component of the $K\kst$ cross section is
described by one broad resonance, $\rho'$, whose mass is
compatible with the $\rho(1450)$, the first  $\rho$ recurrence.
The slight inconsistency for the width is probably due to the fact that a single
structure is not sufficient to describe this cross section and
that the fit tends to broaden the $\rho'$ to mimic a more complex structure.
The present statistics does not allow the identification of 
multiple structures.

The \phipiz cross section has been fit  with 
$\rho$ recurrences, whose coupling with the final state is, however, 
OZI-suppressed. 
Two possible descriptions have been considered, with one and two resonances.
Their parameters are reported in Table~\ref{phipi0-partot}.

The resonance labeled $\rho''$, might be the so called $C(1480)$, observed in 
$\pim p\to\phipiz n$ charge-exchange reactions~\cite{landsberg1}, in the same 
\phipiz final state. However, a firm conclusion cannot be drawn at the
moment; an OZI-violating decay of the meson $\rho(1700)$~\cite{pdg} 
cannot be excluded.
The second structure, $\rho(1900)$, is compatible with the ``dip'' already observed
in other experiments, predominantly in multi-pion final states~\cite{dip}.

In Table~\ref{tab:summary} we summarize the whole set of parameters obtained
from the fit to the cross sections, with the inclusion of model related 
systematic errors. 
%
%

As already pointed out, the normalized $\chi^2$'s
obtained from this global-fit procedure, are still compatible with
other possible minor contributions that, however, cannot be precisely
identified because of the accuracy of the data.

In the $(2\-- 3)\;\gev$ c.m. energy region the  
$\kskpi$ final state has been analyzed, adding the
$K_2^{*}(1430)K$  to the $\kst K$ intermediate state. 
The strong asymmetry between neutral and charged channels of the Dalitz plot 
($M^2_{K_S\pi^{\!\pm\!}}$ vs. $M^2_{K^{\!\pm\!}\pi^{\mp}}$) can be explained in
terms of constructive and destructive interference between
different  isospin components. This asymmetry might be connected 
to a similar effect observed in the radiative decay rates of the
neutral and charged $K^*_2(1430)$~\cite{pdg}.

Very clean \jpsi signals have been observed in most of the studied final states,
allowing the measurement of the corresponding branching fractions.

We have no evidence of the $Y(4260)$ decaying into \kskpi and \kkpiz. 
We determine $\Gamma^{Y(4260)}_{ee}\mathcal{B}^{Y(4260)}_{\kskpi}< 0.5$ \ev 
and $\Gamma^{Y(4260)}_{ee}\mathcal{B}^{Y(4260)}_{\kkpiz}< 0.6$ \ev 
at 90\% C.L..
We observe also (inset of Fig.~\ref{fig:sigma_kskpi_all}) a $3.5 \;\sigma$
fluctuation in the \epem$\to$\kskpi cross section at \Ecm$\approx4.20$ \gev. 


\section*{ACKNOWLEDGEMENTS}
We warmly acknowledge Stanley Brodsky for very fruitful discussions
and suggestions on many aspects of this work.\\
\indent 
We are grateful for the 
extraordinary contributions of our \pep2\ colleagues in
achieving the excellent luminosity and machine conditions
that have made this work possible.
The success of this project also relies critically on the 
expertise and dedication of the computing organizations that 
support \babar.
The collaborating institutions wish to thank 
SLAC for its support and the kind hospitality extended to them. 
This work is supported by the
US Department of Energy
and National Science Foundation, the
Natural Sciences and Engineering Research Council (Canada),
the Commissariat \`a l'Energie Atomique and
Institut National de Physique Nucl\'eaire et de Physique des Particules
(France), the
Bundesministerium f\"ur Bildung und Forschung and
Deutsche Forschungsgemeinschaft
(Germany), the
Istituto Nazionale di Fisica Nucleare (Italy),
the Foundation for Fundamental Research on Matter (The Netherlands),
the Research Council of Norway, the
Ministry of Science and Technology of the Russian Federation, 
Ministerio de Educaci\'on y Ciencia (Spain), and the
Science and Technology Facilities Council (United Kingdom).
Individuals have received support from 
the Marie-Curie IEF program (European Union) and
the A. P. Sloan Foundation.



%
\end{document}